%% file: string_target.tex
\documentclass[11pt]{article}
\usepackage[pdftex]{graphicx,color} 
\usepackage{jheppub}
\usepackage{amsmath}
\usepackage{amssymb}

\usepackage{comment}
\usepackage{shuffle}

\newcommand{\ii}{\mathrm{i}}

\newcommand{\Tr}{\textrm{Tr}}

\newcommand\dd{\mathrm{d}}
\newcommand{\be}{\begin{equation}}
\newcommand{\ee}{\end{equation}}
\newcommand{\bali}{\begin{align}}
\newcommand{\eali}{\end{align}}
\newcommand{\bea}{\begin{equation}\begin{aligned}}
\newcommand{\eea}[1]{\label{#1}\end{aligned}\end{equation}}
\newcommand\Res{\mathrm{Res}}
\newcommand\Amp{\mathcal{A}}

\renewcommand\Im{\operatorname{Im}}

\newcommand{\bt}[1]{{\mathord{\vcenter{\hbox{\includegraphics[scale=0.3]{{{#1}}}}}}}}
\newcommand{\bts}[1]{{\mathord{\vcenter{\hbox{\includegraphics[scale=0.22]{{{#1}}}}}}}}

\graphicspath{ {./bt/} }

\newcommand{\boxit}[1]{%
  \[\fbox{%
      \addtolength{\linewidth}{-2\fboxsep}%
      \addtolength{\linewidth}{-2\fboxrule}%
      \begin{minipage}{\linewidth}%
      #1%
      \end{minipage}%
    } \nonumber \]%
}

\title{String theory in target space}
\author{Rutger H. Boels}
\author{and Tobias Hansen}
\affiliation{II. Institut f\"ur Theoretische Physik, Universit\"at Hamburg\\ Luruper Chaussee 149, D- 22761 Hamburg, Germany }
\emailAdd{Rutger.Boels@desy.de}
\emailAdd{Tobias.Hansen@desy.de}


\abstract{It is argued that the complete S-matrix of string theory at tree level in a flat background can be obtained from a small set of target space properties, without recourse to the worldsheet description. The main non-standard inputs are (generalised) Britto-Cachazo-Feng-Witten shifts, as well as the monodromy relations for open string theory and the Kawai-Lewellen-Tye relations for closed string theory. The roots of the scattering amplitudes and especially their appearance in the residues at the kinematic poles are central to the story. These residues determine the amplitudes through on-shell recursion relations. Several checks of the formalism are presented, including a computation of the Koba-Nielsen amplitude in the bosonic string. Furthermore the question of target space unitarity is \mbox{(re-)investigated}. For the Veneziano amplitude this question is reduced by Poincar\'e invariance, unitarity and locality to that of positivity of a particular numerical sum. Interestingly, this analysis produces the main conditions of the no-ghost theorem on dimension and intercept from the first three poles of this amplitude.  }

\begin{document}
\maketitle

\section{Introduction}

The birth of string theory is usually taken to be the publication of Veneziano's paper \cite{Veneziano:1968yb}.  
As is manifest in its title the motivation behind this paper was to find a four point scattering amplitude for pions which 
had properties which seemed desirable at the time from experiment: crossing-symmetry and Regge-behaviour. This makes 
it a pure exponent of the analytic S-matrix philosophy, which attempts to obtain scattering amplitudes as classes 
of constrained functions. Veneziano's paper inspired a series of developments which led eventually to string theory
 in its worldsheet formulation as it is taught in modern textbooks, see e.g.\ \cite{DiVecchia:2007vd} for an account 
 of this early period. The development of string theory itself has led further and further away from its S-matrix based roots. 

The analytic S-matrix programme has recently made a triumphant return in field theory. Inspired by Witten's \cite{Witten:2003nn} twistor string proposal many new techniques have been developed to calculate scattering amplitudes without the use of Feynman graphs (see \cite{Elvang:2013cua} for a recent overview and gateway to the literature). This leads to the question if these new analytic S-matrix developments can yield new insight into string theory. So far most work in this direction has relied in one form or another on the worldsheet picture. In this article it is shown that worldsheets can, at least in principle, be avoided altogether. We hope that this could lead to a different and potentially very powerful perspective on the foundations of string theory. One concrete motivation for the medium-to-longer term is to circumvent the difficulties in generalising string theory to curved backgrounds in the worldsheet approach. In this article however the more modest main goal is the study of tree level string theory amplitudes in a flat background (``textbook strings''). 

Of particular importance for the current article is the concept of on-shell
recursion relations \cite{Britto:2004ap, Britto:2005fq} which in field theory allow
the computation of higher point amplitudes from lower point amplitudes. Proving on-shell
recursion relations for a specific field theory involves a study of the behaviour
of a scattering amplitude when the momenta of two of the legs tend to infinity in
a particular way. This is closely related, but generically distinct from Regge behaviour \cite{Boels:2008fc, Fu:2013cza}. The relations can then 
be used to argue that an amplitude can be reconstructed from residues at poles.
These residues are in principle determined by perturbative unitarity: they are products of lower
point amplitudes with one leg shared between the amplitudes, summed over all
possible states in the theory at the mass level set by the specific pole. See \cite{Feng:2011np}
for a dedicated review of on-shell recursion relations in field theory.

On-shell recursion relations in string theory were first discussed in
\cite{Boels:2008fc}. A generic proof that on-shell recursion relations hold
in string theory appeared in \cite{Cheung:2010vn, Boels:2010bv}.
The underlying analysis shows that the large momentum behaviour needed for
on-shell recursion follows from a suitable extension of the Regge behaviour
of string amplitudes, inspired by a computation in \cite{Brower:2006ea}.
The recursion relations are however not immediately useful in string theory.
The main problem is that they involve sums
over the complete tower of levels appearing in the string spectrum as well
as a sum over all possible polarisations at a fixed level. So even if one
starts with, say, a purely tachyonic amplitude in the open bosonic string,
to compute this one needs three point scattering amplitudes for the complete set of
string states. Whereas in field theory Poincar\'e invariance and locality fix the
needed three point amplitudes typically up to a single (coupling) constant, this is no longer true in string theory. 
Expressions have been derived from the worldsheet, see e.g.\
\cite{DelGiudice:1971fp}, but they are unwieldy even before taking
 sums over products of them. Some headway on this problem was
made in \cite{Chang:2012qs} by summing over spurious states, but to our
knowledge no complete solution exists. 

A hint that more may be possible was given in \cite{Cheung:2010vn} where
`internal' recursion relations for Koba-Nielsen amplitudes were derived
from their integral representation.
These relations express tachyon amplitudes in terms of certain sums over
products of tachyon amplitudes. Generalisation of the worldsheet based
methods used
there to other amplitudes seems prohibitively complicated using known techniques. In this article an
additional ingredient is introduced which allows us to bypass the sums
over all particles and their polarisations at a fixed mass level. The
clue to this ingredient comes from a seemingly unrelated question: 

\subsubsection*{Where do the roots of string theory amplitudes come from?}
The general form of the (colour-ordered) Veneziano amplitude describing four tachyon scattering in open string theory reads
\begin{equation}\label{eq:venezianoamp}
\Amp(s_{12},s_{23}) \propto \frac{\Gamma[\alpha(s_{12})]\Gamma[\alpha(s_{23})]} {\Gamma[\alpha(s_{12}) + \alpha(s_{23})]} \, ,
\end{equation}
where $\alpha$ is a linear function of the usual Mandelstam invariants $s_{12}$ and $s_{23}$. Every string textbook points out that the two $\Gamma$ functions in the numerator have an infinite series of poles at negative integer arguments.
These poles acquire a meaning through perturbative unitarity as they display a part of the infinite tower of states in the string spectrum. The crucial question for this article is: do the \emph{roots} of the amplitude in equation \eqref{eq:venezianoamp} from the $\Gamma$ function in the denominator have a physical meaning too? Moreover, can they be predicted? This turns out to have been studied long ago \cite{D'Adda:1971te} from an argument based on the monodromy relations found in \cite{Plahte:1970wy}. A modern derivation of these relations based on CFT methods can be found in \cite{Boels:2010bv}. 

The existence of roots in amplitudes has shown up in a different context before in Yang-Mills theories \cite{Zhu:1980sz}. The roots there are for the total (not colour-ordered) amplitude. In this case these have eventually been understood as a consequence of the Bern-Carrasco-Johansson (BCJ) \cite{Bern:2008qj} relations. The BCJ relations arise in the field theory expansion of the string monodromy relations and were in fact first proven this way \cite{BjerrumBohr:2009rd, Stieberger:2009hq}. Here it will be shown that the monodromy relations can be used to not only predict the location of the roots of the \emph{residues} at kinematic poles but even that with some additional work they fix the residue of at least the Koba-Nielsen amplitudes completely. 

This article is structured as follows: Section \ref{sec:review} contains a brief
review of relevant background material. The reviewed techniques are then used to
derive the residues at poles of mainly bosonic string amplitudes in Section \ref{sec:ampsfrommonod}.
By on-shell recursion this provides the complete scattering amplitudes in the string
theory. It is an interesting question how results which are usually derived through
the worldsheet picture find a place in a target space approach. As a prime example
of this we re-initiate the systematic study of perturbative unitarity in the target
space in Section \ref{sec:unit}. In particular a complete $SO(D-1)$ covariant expression is derived for the 
two tachyon-anything three point amplitudes, including the numerical constant. Using the same techniques a purely 
target space based derivation of the no-ghost theorem conditions is presented. The results are then gathered into a target space definition
of string theory in \ref{sec:stringdef}. Various ways in which the set of constraints for string amplitudes could 
conceivably be improved further are discussed.  As a further example, it is argued from analysis of the
five point case that this definition reproduces the full open superstring S-matrix.
The discussion section sketches several broad classes
of possible applications and directions for further research. Several
appendices contain details of calculations.

%%%%%%%%%%%%%%%%%%%%%%%%%%%%%%%%%%%%%%%%%%%%%%%%%%%%%%%%%%
%%%%%%%%%%%%%%%%%%%%%%%%%%%%%%%%%%%%%%%%%%%%%%%%%%%%%%%%%%

\section{Review}\label{sec:review}

The full string amplitude is given as a sum over all non-cyclic permutations of so-called colour-ordered amplitudes times single trace factors,
\begin{equation}\label{eq:stdcolorordering}
\Amp_n = \sum_{\sigma \in \mathbb{P}_n \slash \mathbb{Z}_n} \Amp_{\textrm{colour-ordered}}(\sigma_1, \ldots, \sigma_n) \Tr \left(T^{\sigma_1} \ldots T^{\sigma_n}\right) .
\end{equation}
Throughout open string amplitudes will be assumed to be colour-ordered amplitudes. This decomposition is natural in string theory: the traces of matrices $T$ in the fundamental representation of $U(N)$ are simply the Chan-Patton factors. The string theory picture played a large \cite{Mangano:1987xk} but not exclusive \cite{Berends:1987cv} role in introducing the concept of colour-ordering in field theory. See \cite{Boels:2010mj} for a derivation of colour-ordering from the more modern D-brane picture of string theory. General properties of colour-ordered amplitudes are well-known  \cite{Mangano:1987xk} and will not be reviewed here.

\subsection{Overview of conventions}\label{sec:conventions}
The metric will have signature $(-++\ldots +)$ so that the mass of a tachyon with momentum $k_i$ is
\begin{equation}
- \alpha' k_i^2 = \alpha' m^2 = -1 .
\end{equation}
Define for the product of two momenta
\be
k_{ij} = 2 \alpha' k_i \cdot k_j ,
\ee
and the Mandelstam invariants
\begin{align}
s_{ij} &= - \alpha'(k_i + k_{j})^2 = \alpha' (m_i^2 + m_j^2) - k_{ij} , \\
s_{1 \ldots a} &= - \alpha'(k_1 + \ldots + k_a)^2 = \sum\limits_{i=1}^a \alpha' m_i^2 - \sum\limits_{i = 1}^{a-1} \sum\limits_{j = i+1}^{a}k_{ij} .
\label{eq:mandelstams}
\end{align}
Mass levels such as $A$ are always defined in terms of a Mandelstam invariant as $s = A-1$ for
the bosonic and $s = A$ for the superstring. Hence the lowest mass particle is always the one at level $A=0$.

\subsection{On-shell recursion in string theory}
The main idea of on-shell recursion as introduced in \cite{Britto:2005fq} is to introduce a
single auxiliary complex parameter into scattering amplitudes, while keeping the amplitudes
physical. To this end, one picks two legs and deforms their momenta as
\begin{equation}\label{eq:bcfwshift}
k_{i} \rightarrow \hat{k}_i \equiv k_i + q \, z , \qquad k_{j} \rightarrow \hat{k}_j \equiv k_j - q \, z .
\end{equation}
This automatically satisfies momentum conservation. If one then also imposes 
\begin{equation}
q^2=q\cdot k_i = q \cdot k_j = 0 ,
\end{equation}
the two singled-out legs remain on their original mass-shell. These equations can always be solved in four or more dimensions. In four dimensions, two solutions exist (this is easily verified in the centre-of-mass frame \cite{ArkaniHamed:2008yf}). The deformations in equation \eqref{eq:bcfwshift} are collectively known as a BCFW-shift. Note that this shift makes momenta automatically complex.  

The point of single complex variables in physics is invariably the possibility to use Cauchy's theorem. In the present context, one would like to compute the original amplitude, $\Amp(0)$, which may be computed as
\begin{equation}
\Amp(0) = \oint_{z=0} \frac{\Amp(z)}{z} .
\end{equation}
Here and in the following, all residue-type integrals contain $\frac{1}{2 \pi \ii}$ factors. 

\subsubsection*{Diversion: the pole structure of $\Amp(z)$}
As a function of $z$, the amplitude $\Amp(z)$ can have physical poles\footnote{In this article it will be implicitly assumed that amplitudes do not have un-physical poles. Moreover, it is assumed that the poles originate in nothing more exotic than Feynman-type propagators going on mass-shell.}. In fact, for generic external momenta it will only have single poles. The residues at these poles have a physical interpretation from perturbative unitarity. Say one takes a certain channel defined by a set $\sigma$ of adjacent particles which includes the shifted particle $i$ but not the other one $j$. The pole in $z$ in this channel occurs when the associated internal propagator goes on-mass-shell, i.e.
\begin{equation}\label{eq:solveforz}
-(\hat{k}_i + \sum_{l\in \sigma \backslash \{i\} } k_l)^2 = m^2 ,
\end{equation}
for some mass of a particle in the particular theory under study. Note that the location of this pole is at a finite value of $z$. The residue of the amplitude at this pole is predicted by perturbative unitarity\footnote{Here convention is followed by calling this perturbative unitarity. As pointed out in \cite{Schuster:2008nh}, these equations are somewhat stronger when complex momenta are considered. It is this stronger sense which will be needed below. } to be
\begin{equation}
\lim_{-(\hat{k}_i + \sum_{l\in \sigma \backslash \{i\} } k_l)^2 \rightarrow m^2} \left[\left((\hat{k}_i + \sum_{l\in \sigma \backslash \{i\} } k_l)^2 + m^2 \right) \Amp(z) \right] = \sum_{\textrm{spectrum with mass } m } \Amp_{L} \Amp_{R} .
\label{eq:unitarity_recursion}
\end{equation}
Here the amplitudes $\Amp_{L}$ and $\Amp_{R}$ contain  the set $\sigma$
and the complement of this set respectively, as well as one `interchanged' particle in addition on both sides. The order of the legs simply derives from the order on the parent amplitude. 
Note that both these amplitudes have a strictly lower number of particles than the
original amplitude $\Amp(0)$. The sum ranges over the complete spectrum of the theory
at a fixed mass level. This is a double sum: every physical particle in a Poincar\'e
invariant theory transforms as an irreducible representation (irrep) of the appropriate little group. Hence one
first needs to sum over all irreps at mass $m$. Then within these irreps one
needs to sum over all states: the (higher dimensional analog of) spin or helicity
states. Limits like the one above will usually be denoted as residues in Mandelstams.
Written in this fashion \eqref{eq:unitarity_recursion} reads
\be
- \Res_{s_{\hat i,  \{l\in \sigma \backslash \{i\}\}} \to \alpha' m^2} \Amp(z)= \sum_{\textrm{spectrum with mass } m } \Amp_{L} \Amp_{R}  \ .
\ee
where the right-hand side must be evaluated at the value of $z$ for which \eqref{eq:solveforz} holds.

\subsubsection*{Back to on-shell recursion}
By interpreting $z$ as a coordinate on the Riemann sphere $\mathbb{CP}^1$ one can pull the contour to infinity and obtain
\be
\Amp(0) = \oint\limits_{z=0} \frac{\Amp(z)}{z} = 
- \sum\limits_{z_I \ \text{finite}} \Res_{z=z_I} \frac{\Amp(z)}{z}
- \Res_{z=\infty} \frac{\Amp(z)}{z} .
\ee
As just explained, the residues at finite values of $z$ have an interpretation through perturbative unitarity in terms of products of lower point scattering amplitudes. If therefore the residue at infinity is absent the just derived schematic equation constitutes an explicit on-shell recursion relation. The crucial question is therefore to obtain the residue at infinity, i.e.
\be
\Res_{z=\infty}\frac{ \Amp(z)}{z}  = \, ? \ .
\ee
A sufficient condition for this residue to vanish is that $\Amp(z) \rightarrow 0$ for $z \rightarrow \infty$. In principle if one can compute the residue at infinity explicitly there is also an effective recursion relation, but examples of this type tend to be quite involved. Hence vanishing residues will be aimed at henceforth. 

Up to now the discussion has been completely general. There is however a marked
difference in how efficient these recursion relations are in string or field theory.
In field theory the spectrum is finite, typically with just one or two (super-)particle
types. In string theory it is well-known that the theory contains an infinite tower of states,
labelled by the mass level. To get a feel for the matter content at fixed mass level in
terms of irreps of the little group, see \cite{Hanany:2010da}. The list of irreps grows
rather quickly with the mass level, but the number of tensor indices is always bounded
by the level (as defined in Section \ref{sec:conventions}) in the bosonic string and by the level $+$ 1 in the superstring.
Even if residues at infinity are
absent a naive application of the on-shell recursion relations requires knowledge of
\emph{all} three point amplitudes. 

Apart from effectiveness of the recursion relations, they of course also have to be proven. For this one needs to study the expansion of $\Amp(z)$ around $z=\infty$. In field theory a very direct analysis \cite{ArkaniHamed:2008yf} in $4$ or more dimensions yields
\begin{equation}
\Amp_{\textrm{ym}}(z) \sim \hat{\xi}_{1,\mu}  \hat{\xi}_{2,\nu} \Amp^{\mu\nu}(z) ,
\end{equation}
for the BCFW shift of two colour-adjacent gluons labelled one and two in a Yang-Mills amplitude (possibly minimally coupled to matter). Here the $\xi$ vectors are the polarisation vectors of the shifted gluons, whose large $z$ behaviour is easily analysed. The tensor $\Amp^{\mu\nu}$ is given as:
\begin{equation}
\Amp^{\mu\nu}(z) = z \left( \eta^{\mu\nu} f_0\left(\frac{1}{z}\right)  + \frac{1}{z} B^{\mu\nu} \left(\frac{1}{z}\right) + \mathcal{O} \left(\frac{1}{z}\right)^2 \right) ,
\end{equation}
where $f(w)$ and $B^{\mu\nu}(w)$ are polynomials in $w$ with generically non-zero constant term and the tensor $B^{\mu\nu}$ is anti-symmetric in its indices. Combining the $\Amp^{\mu\nu}$ tensor with the behaviour of the polarisation vectors then gives the result that for any choice of helicities of the singled-out two gluons a shift exists such that the amplitude may be computed through on-shell recursion. For this shift one obtains 
\be
\frac{\Amp(z)}{z} \sim \frac{1}{z^2} \textrm{  for  } z \rightarrow \infty \ .
\ee

In string theory the result for the large $z$ shift is very similar to the field theory result. As shown in \cite{Cheung:2010vn} and \cite{Boels:2010bv}, in the superstring
\begin{equation}
\Amp_{\textrm{open}, gg}(z) \sim \hat{\xi}_{1,\mu}  \hat{\xi}_{2,\nu}  z^{- 2 \alpha' k_1 \cdot k_2} \Amp^{\mu\nu}(z) ,
\end{equation}
holds for the shift of two colour-adjacent gluons, with arbitrary field content on the
other legs. The difference to the field theory is in the Regge-like prefactor. In the
bosonic string, this result for the BCFW shift of two colour-adjacent gluons is structurally the same, but the tensor $\Amp^{\mu\nu}$ is modified to
$\tilde{\Amp}^{\mu\nu}$ as
\begin{equation}
\tilde{\Amp}^{\mu\nu}(z) \equiv \Amp^{\mu\nu}(z)+ \, z \,\alpha' k^{\mu} k^{\nu} f_1\left(\frac{1}{z}\right) ,
\end{equation}
with $k^{\mu} = k_1^{\mu} + k_2^{\mu}$ and $f_1(w)$ a polynomial of $w$ with non-zero constant term. This particular term is forbidden in any supersymmetric field theory as it generates amplitudes with all helicities equal which is perturbatively impossible in a supersymmetric field theory \cite{Grisaru:1977px}. For shifts of two tachyons, the result reads
\begin{equation}\label{eq:largeztachyons}
\Amp_{\textrm{open}, TT}(z) \sim z^{ s_{12} + 1} \left(f_1\left(\frac{1}{z}\right) \right) ,
\end{equation}
again with arbitrary field content on the other legs. It is easy to show that the general structure of a BCFW shift for arbitrary choice of matter content on the two legs will always be a Regge-type factor times a polynomial in $1/z$. This can be computed directly from the OPE, see \cite{Cheung:2010vn} and \cite{Boels:2010bv} for details.

The shifts of colour-non-adjacent particles on an open string amplitude follow from the use of monodromy relations, see \cite{Boels:2010bv}. The BCFW shift of two particles on a closed string amplitude follows basically by either the same worldsheet based argument or from the use of the KLT relations \cite{Kawai:1985xq}.

\subsection{Monodromy relations}
Central to the discussion will be the monodromy relations first discussed in \cite{Plahte:1970wy}. The two basic monodromy relations for colour-ordered open string tree amplitudes in a flat background read 
\boxit{\be\label{eq:monodfund}
\Amp(\beta,1,2, \ldots,N) = - \sum_{i=1}^{N-2}    \exp\left[ \pm i\pi (\sum_{j=1}^i k_{\beta, j}) \right] \Amp(1,\ldots, i, \beta, i+1,\ldots ,N),
\ee}
for an amplitude involving $N$ bosonic particles. Basically the particle labelled $\beta$ is moved through the other colour-ordered particles, picking up a `sign' for every interchange. In string theory this follows from the braid relation for flat background vertex operators  \cite{Boels:2010bv}. Note there are two relations: one for each choice of sign in the exponent. For complex momenta these two relations are not complex conjugate. 

From the basic relations others may be derived \cite{Ma:2011um}. In modern language \cite{BjerrumBohr:2009rd, Stieberger:2009hq} the relations needed below can be written as
\be\label{eq:monod1}
\Amp(\beta^T,1,\alpha,N) = (-1)^{s}\sum\limits_{\sigma\in OP( \{\beta\}, \{\alpha\})} \mathcal{P}_{\{\beta^T,1,\alpha,N\},\{1,\sigma,N\}}\Amp(1,\sigma,N),
\ee
where $\beta = \{\beta_1,...,\beta_s\}$ is now an ordered set of particle labels and $\beta^T$ indicates the inversion of the ordered set $\beta$.  In the formula $\alpha = \{\alpha_1,...,\alpha_{N-s-2}\}$ is an ordered set of particle labels and $OP( \{\beta\}, \{\alpha\})$ are the ordered permutations of $\beta$ and $\alpha$ i.e.\ the permutations of the union $\beta \cup \alpha$ that preserve the order of both subsets. The sum over $OP( \{\beta\}, \{\alpha\})$ is known as the shuffle product $\beta \shuffle \alpha$.

The phase factor $\mathcal{P}$ can be neatly expressed in terms closely related to the so-called momentum kernel \cite{BjerrumBohr:2010hn}. In the notation of \cite{Ma:2011um}, it is given as a function of two permutations $\sigma, \tau$ as
\be\label{eq:momentum_kernel}
\mathcal{P}_{\{\sigma\},\{\tau\}}
=\exp\left[i\pi\sum\limits_{i,j}k_{ij}\theta(\sigma^{-1}(i)-\sigma^{-1}(j))\theta(\tau^{-1}(j)-\tau^{-1}(i))\right],
\ee
where
\be
\theta(x)=  \begin{cases}
              1 & (x>0) \\
              0 & (x\leq 0) \\
            \end{cases} .
\ee
The $\theta$'s are there to let any $k_{ij}$ appear in the exponent if and only if $i$ and $j$ appear in a different order in $\sigma$ and $\tau$. Some examples are
\be
\mathcal{P}_{\{\sigma\},\{\sigma\}} = 1 , \qquad \mathcal{P}_{\{1, 2, 3\},\{2, 1, 3\}} =\exp\left[i\pi k_{12} \right] ,
\qquad \mathcal{P}_{\{\sigma\},\{\sigma^T\}} = \exp\left[i\pi\sum\limits_{i<j}k_{ij}\right] .
\ee
For fermionic particles an additional minus sign appears every time a pair of fermions is interchanged, see \cite{Sondergaard:2009za} for more details. 

The relations are universal in that they do not depend on the particle content of the open string amplitude. Moreover, also the `conjugate' relations hold:
\be\label{eq:monod2}
\Amp(\beta^T,1,\alpha,N) = (-1)^{s}\sum\limits_{\sigma\in OP( \{\beta\}, \{\alpha\})} \mathcal{P}^*_{\{\beta^T,1,\alpha,N\},\{1,\sigma,N\}}\Amp(1,\sigma,N),
\ee
with only the sign of the exponent changed
\be\label{eq:momentum_kernel_conjugate}
\mathcal{P}^*_{\{\sigma\},\{\tau\}}
=\exp\left[-i\pi\sum\limits_{i,j}k_{ij}\theta(\sigma^{-1}(i)-\sigma^{-1}(j))\theta(\tau^{-1}(j)-\tau^{-1}(i))\right] .
\ee
These relations hold also for complex momenta: in the worldsheet derivation the exact phase simply corresponds to a choice of branch cut, while the amplitudes should be independent of this choice. Relations \eqref{eq:monod1} and \eqref{eq:monod2} can be subtracted to give
\be
\label{eq:monod_imaginary}
\sum\limits_{\sigma\in OP( \{\beta\}, \{\alpha\})} \mathcal{S}_{\{\beta^T,1,\alpha,N\},\{1,\sigma,N\}}
\Amp(1,\sigma,N)=0,
\ee
where
\be
\mathcal{S}_{\{\sigma\},\{\tau\}}=\Im \mathcal{P}_{\{\sigma\},\{\tau\}} = \sin \left[\pi\sum\limits_{i,j}k_{ij}\theta(\sigma^{-1}(i)-\sigma^{-1}(j))\theta(\tau^{-1}(j)-\tau^{-1}(i))\right].
\ee
In this article, equation \eqref{eq:monod_imaginary} will be used to study the residues of amplitudes $\Amp(123\ldots N)$ in the variables $s_{12}, s_{123}, \ldots, s_{1 \ldots N-2}$.
To this end it is useful to rewrite the expression in a way which exposes the pole in the $s_{1, \beta}$ channel
by splitting off the first element of $\alpha$ (this will be labeled $s+2$) and separating the sum over its positions.
Although not all particle labels will be specified in the following formulae, set $\beta = \{2, \ldots, s+1 \}, \alpha = \{s+3, \ldots, N-1 \}$ for the remainder of this paper. 

The relations in \eqref{eq:monod_imaginary} are graded by the size of the set $\beta$.
For instance, the relation that makes the pole in $s_{1 \beta_1}$ manifest  is
\bea
&\Amp(1,\beta_1,3,\alpha,N)& \\
= {}&-\frac{1}{\mathcal{S}_{\{\beta_1,1,3,\alpha,N\},\{1,\beta_1,3,\alpha,N\}} }\sum\limits_{\sigma\in OP(\{\beta_1\} , \{\alpha\})}
\mathcal{S}_{\{\beta_1,1,3,\alpha,N\},\{1,3,\sigma,N\}}\Amp(1,3,\sigma,N)\\
= {}& \frac{(-1)^{\alpha' m_{\beta_1}^2 + \alpha' m_1^2  }}{\sin(\pi s_{1,\beta_1}) }\sum\limits_{\sigma\in OP(\{\beta_1\} , \{\alpha\})}
\mathcal{S}_{\{\beta_1,1,3,\alpha,N\},\{1,3,\sigma,N\}}\Amp(1,3,\sigma,N).
\eea{eq:monod_im_lvl1}
where the definition of the Mandelstam variables in equation \eqref{eq:mandelstams} was used. Note that none of the amplitudes on the right-hand side has a pole in the $s_{1 \beta_1}$ channel. Since the sine functions in the numerator cannot cause poles, all poles must be captured by the sine in the denominator. Similarly, the pole in $s_{1 \beta_1 \beta_2}$ is manifest in
\bea
&\Amp(1,\beta_1,\beta_2,4,\alpha,N)\\
={}& \frac{(-1)^{\alpha' (m_{\beta_1}^2+m_{\beta_2}^2 +  m_1^2 ) }}{\sin(\pi s_{1,\beta_1,\beta_2}) }
\Biggl[ \sum\limits_{\sigma \in OP(\{\beta_1, \beta_2\} , \{\alpha\})} \!\! \mathcal{S}_{\{\beta_2,\beta_1,1,4,\alpha,N\},\{1,4,\sigma,N\}}\Amp(1,4,\sigma,N) \\
& + \sum\limits_{\sigma\in OP(\{\beta_2\} , \{\alpha\})}\!\!  \mathcal{S}_{\{\beta_2,\beta_1,1,4,\alpha,N\},\{1,\beta_1,4,\sigma,N\}}\Amp(1,\beta_1,4,\sigma,N) \Biggr].
\eea{eq:monod_im_lvl2}
The general form of this relation is
\bea
&\Amp(1,\beta_1,...,\beta_s,s+2,\alpha_1,...,\alpha_{N-s-3},N)\\
={}& \frac{(-1)^{\alpha' \left( m_1^2 +\sum\limits_{i=1}^{s} m_{\beta_i}^2 \right) }}{\sin(\pi s_{1\beta_1...\beta_s})}
\Biggl[\sum\limits_{\sigma\in OP(\{\beta_{1},...,\beta_s\}, \{\alpha\})} \!\!  \mathcal{S}_{\{\beta^T,1,s+2,\alpha,N\},\{1,s+2,\sigma,N\}}\Amp(1,s+2,\sigma,N)\\
&+\sum\limits_{l=1}^{s-1}\sum\limits_{\sigma\in OP(\{\beta_{l+1},...,\beta_s\}, \{\alpha\})}\!\!  \mathcal{S}_{\{\beta^T,1,s+2,\alpha,N\},\{1,\beta_1,...,\beta_l,s+2,\sigma,N\}}\Amp(1,\beta_1,...,\beta_l,s+2,\sigma,N)\Biggr].
\eea{eq:monod_im_lvl_s}
The sine in the denominator captures the complete pole in the $(1,\beta)$-channel. It should be clear these relations may be nested to uniquely
express a given open string amplitude in terms of a particular set of basis amplitudes with the positions of three particles fixed, e.g.\ $\Amp(1,2, \sigma, N)$. This particular form of the  monodromy relations has first appeared in \cite{Ma:2011um}, as far as we are aware.

\subsubsection*{Roots of amplitudes}
\label{sec:roots}

The monodromy relations can be used to find the roots of amplitudes as studied in \cite{D'Adda:1971te}.
Their argument to find the roots has to be slightly extended here to allow for complex momenta. 

In \eqref{eq:monod1} each factor $\mathcal{P}_{\{\beta^T,1,\alpha,N\},\{1,\sigma,N\}}$ depends on the
Mandelstam $s_{1,\beta}$ and additional momentum invariants
\begin{equation}
\begin{gathered}
\{k\}_{\sigma} = \{ k_{ij} \theta(\sigma^{-1}(i)-\sigma^{-1}(j))\theta(\tau^{-1}(j)-\tau^{-1}(i)) \ | \ i,j \in \{ \sigma \} \} ,\\
\text{where} \qquad \tau = \{ \beta \cup \alpha \} , \qquad  \sigma \in OP(\{ \beta \}, \{ \alpha \}).
\end{gathered}
\end{equation}
If all elements of all $\{k\}_{\sigma}$ are taken to non-negative integer values
\be
\{k\}_{\sigma} \subset \mathbb{N}_0 \quad \forall \sigma \in OP(\{ \beta \}, \{ \alpha \}) ,
\label{eq:amplitude_roots}
\ee
while $s_{1,\beta}$ is kept arbitrary the equations \eqref{eq:monod1}
and \eqref{eq:monod2} become
\be
\Amp(\beta,1,\alpha,N) = \exp (-i \pi s_{1,\beta}) F = \exp (i \pi s_{1,\beta}) F ,
\ee
for some function $F$. This can only be satisfied for generic $s_{1,\beta}$ if both $\Amp(\beta,1,\alpha,N)$ and $F$ vanish.
The restriction to non-negative integers was to avoid hitting poles in the amplitudes which appear in the 
monodromy relations.

A second remark is that using a more general form of the monodromy relations should allow us to obtain additional
sets of roots more straightforwardly. In \cite{D'Adda:1971te} only monodromy relations were used where $\beta$ has only 
one element which means there is one set of roots per amplitude that can trivially be read off as in
\eqref{eq:amplitude_roots}. Further sets of roots are obtained by combining monodromy relations and can 
contain conditions on multi-particle Mandelstams. For instance, a table in \cite{D'Adda:1971te} lists five sets of
roots of the $6$-point amplitude. Two of them are given by \eqref{eq:amplitude_roots} when $\beta$ has one or
two elements. The remaining sets of roots in the table involve conditions on multi-particle Mandelstams
and it still seems to be necessary to combine multiple monodromy relations to derive these. A general and simple way to derive all sets of roots is a worthwhile direction to explore, but will not be needed here. Below the form of the monodromy relations reviewed above will be used to study the roots.

The field theory limit\footnote{Loosely speaking, this is the ``$\alpha' \rightarrow 0$'' limit.
More correctly, this is the limit where $\alpha' s_{ij}\rightarrow 0$ for any $i,j$} of the
monodromy relations results in the BCJ-relations \cite{Bern:2008qj},
which can alternatively be derived using a non-adjacent BCFW shift \cite{Feng:2010my}.
It would be interesting to see if the string monodromy relations could also be derived from a non-adjacent BCFW shift.

%%%%%%%%%%%%%%%%%%%%%%%%%%%%%%%%%%%%%%%%%%%%%%%%%%%%%%%%%%
%%%%%%%%%%%%%%%%%%%%%%%%%%%%%%%%%%%%%%%%%%%%%%%%%%%%%%%%%%

\section{String amplitudes from monodromy relations}\label{sec:ampsfrommonod}

In this section it will be shown that the residues at kinematic poles can be derived from the monodromy relations. These
are then used in the on-shell recursion relations to construct the complete amplitude. 

Instrumental are the location of the roots of the residues of amplitudes.
Below it is shown that the form of the monodromy relations discovered more recently and reviewed above allow
for a more natural approach to studying roots than was possible in the original \cite{D'Adda:1971te} paper.
In their new form  the monodromy relations allow the systematic study of the the roots of the
\emph{residues} of the amplitude, a possibility that was not obvious from the original monodromy relations.
As inputs this section uses the behaviour under BCFW-shifts derived above. 

\subsection{Four point amplitudes}
To provide some orientation the four point amplitudes will be discussed extensively. At four points the monodromy relation \eqref{eq:monod_im_lvl1} can be written as
\begin{equation}\label{eq:monodrat4pt}
\Amp(1234) = (-1)^{\alpha' (m_1^2 + m_4^2)} \frac{\sin(\pi s_{13})}{\sin(\pi s_{12})} \Amp(1324).
\end{equation}
This relation is easily checked for the Veneziano amplitude in equation \eqref{eq:venezianoamp}. A simple consistency check is to consider the pole structure: the amplitude on the left-hand side has poles in the $s_{12}$ and $s_{23}$ channel, but not in the $s_{13}$ channel. Similarly, the amplitude on the right-hand side has poles in $s_{23}$ and $s_{13}$ channel, but not in the $s_{12}$ channel. This discrepancy is solved by the roots of the sine functions. 

In equation \eqref{eq:monodrat4pt} it is obvious that all poles in the $s_{12}$-channel of the left-hand side amplitude are contained in the sine-function in the denominator on the right-hand side. As a bonus, the equation also displays possible roots of the amplitudes. These are contained in the sine-function in the numerator. A restriction here is that for sufficiently large integer values of $s_{13}$ the amplitude on the right-hand side develops a pole, leading to a finite, non-vanishing result. In the bosonic string case for instance the amplitude $\Amp(1234)$ generically has a series of roots at
\begin{equation}
s_{13} \in \{-2,-3,-4, \ldots \} .
\end{equation} 
Comparing to the Veneziano amplitude in equation \eqref{eq:venezianoamp} it is seen that
\emph{all} roots of this particular amplitude arise this way. Note that the starting location
of the row of roots of the amplitude on the left-hand side is determined by the location
of the lowest mass pole of the amplitude on the right-hand side. The argument just given
applies to all possible choices of external states within the string spectrum and to the
superstring. The precise starting location of the roots depends on the external masses
and the spectrum, as some states will for instance not couple to two tachyons (see Section \ref{sec:unit}).

The previous reasoning can be extended to compute the residues at poles. For definiteness the focus will first be on the Veneziano amplitude with four external tachyons. From equation \eqref{eq:monodrat4pt} it follows that
\begin{equation}\label{eq:fixresidue4pt}
\textrm{Res}_{s_{12} \rightarrow A-1} \Amp(1234) =  \frac{(-1)^{A-1}}{\pi} \left[ \sin(\pi s_{13})  \Amp(1324) \right]_{s_{12} = A-1} ,
\end{equation}
for some non-negative integer $A$. By perturbative unitarity, Poincar\'e invariance and locality the left-hand side of this equation must be a polynomial in $s_{13}$. It is not manifest the right-hand side is. Note however that as a function of $s_{13}$ it no longer has an \emph{infinite} series of roots since by momentum conservation
\begin{equation}
(A-1) + s_{23} + s_{13} = \sum m_i^2 = -4  .
\end{equation}
Hence, if $s_{13}$ is $\in \{-1,0,1,\ldots\}$ it will hit the pole in the amplitude $\Amp(1324)$ in the $(1,3)$ channel while if  $s_{13}$ is $\in \{-2-A,-3-A,\ldots\}$ it will hit a pole in the $(2,3)$ channel. For four tachyons, this implies the residue is a polynomial of at least degree $A$, with roots at $\{-2,-3,\ldots, -1-A\}$. For $A=0$, the polynomial is a constant. The maximal degree of the polynomial in $s_{13}$ appearing in the residue at this pole is set by the maximal spin of the spectrum at mass level $A$ which is known to be $A$ itself. Actually, this can be demonstrated by studying a $(1,2)$ channel BCFW shift of the residue. By equation \eqref{eq:largeztachyons} one obtains for the residue under this shift in a cross-channel
\begin{equation}\label{eq:largeztachyonsatresidue}
\textrm{Res}_{s_{12} \rightarrow A-1} \Amp(1234)  \sim z^{A} \left(f_1\left(\frac{1}{z}\right) \right) .
\end{equation}
Note that technically, one should study a non-adjacent BCFW shift for the amplitude on the right-hand side of equation \eqref{eq:fixresidue4pt}. How to do this was explained in \cite{Boels:2010bv}, 
which in this particular case simply reduces to reading of the large $z$ shift from the left-hand side of equation \eqref{eq:fixresidue4pt}. It will be assumed the BCFW large z-limit in the $(1,2)$ channel and taking the residue in this channel commute\footnote{This can be proven from the worldsheet point of view using the full result for the large z-shift in \cite{Boels:2010bv}.}. Since the residue must be a function of $s_{13}$ only, the BCFW shift fixes the maximal spin of the spectrum at level $A$ to be $A$. 

By the main theorem of algebra, these observations fix the residue up to an overall constant
\begin{equation}\label{eq:ansatzres4ptfrommono}
\textrm{Res}_{s_{12} \rightarrow A-1} \Amp(1234) = c (s_{13}+2) \ldots (s_{13}+A+1) .
\end{equation}
This constant can be fixed by tuning $s_{13}$ to the value $-1$ in equation \eqref{eq:fixresidue4pt}. The right-hand side in this case does not vanish but factorises by unitarity into two $3$-tachyon amplitudes,
\begin{equation}
\textrm{lim}_{s_{13} \rightarrow -1} \left[ \frac{(-1)^{A-1}}{\pi} \sin(\pi s_{13})  \Amp(1324) \right]  = (-1)^{A-1} \Amp_3(T,T,T) \Amp_3(T,T,T) = (-1)^{A-1} g_o^2 ,
\end{equation}
these 3 point amplitudes are just the open string coupling constant $g_o$. Combining this expression for the right-hand side of equation \eqref{eq:fixresidue4pt}  with equation \eqref{eq:ansatzres4ptfrommono} for the left-hand side at $s_{13} = -1$ now fixes the constant $c$ to be
\begin{equation}
c= g_o^2 \frac{(-1)^{A-1}}{\Gamma[A+1]} .
\end{equation}
Note this computation has fixed the numerical coefficient of all the tachyon-tachyon-massive-state couplings in terms of the three tachyon coupling. As a result the complete residue is fixed by equation \eqref{eq:fixresidue4pt}, a combination of unitarity, locality, Poincar\'e invariance as well as Regge behaviour. The string coupling constants will mostly be suppressed in the following. 

\subsubsection*{The complete four point function through on-shell recursion}
The stage is now set for the derivation of the Veneziano amplitude through on-shell
recursion by assembling the above building blocks. Since the $s_{12}$ channel
poles have been worked out it is natural to study a shift on particles $2$ and $3$.
This will keep $s_{23}$ invariant. Hence it is advantageous to express the residues
in equation \eqref{eq:ansatzres4ptfrommono} in terms of $s_{23}$ instead of $s_{13}$,
\begin{equation}\label{eq:res4ptfrommono}
\textrm{Res}_{s_{12} \rightarrow A-1} \Amp(1234) = g_o^2 \frac{(-1)^{A-1}}{\Gamma[A+1]} (-s_{23}-A-1) \ldots (-s_{23}-2) .
\end{equation}
The on-shell recursive expression in this case simply gives (suppressing $g_o$)
\bea
\Amp(1,2,3,4)&  = - \sum_{A=0}^{\infty} \frac{1}{s_{12} - A+1} (-1)^A \frac{\Gamma[-s_{23}-1] }{\Gamma[A+1] \Gamma[-s_{23} - 1 - A]} \\ 
& = - \sum_{A=0}^{\infty} \frac{1}{s_{12} - A+1} (-1)^A \binom{ k_{23} }{ A } \\
& = \frac{\Gamma[-s_{12}-1] \Gamma[-s_{23}-1]}{\Gamma[-s_{12}-s_{23}-2]} ,
\eea{eq:A4_bcfw}
as the result of the in string theory very well-known summation formulae for the $\beta$ function. In the second line the binomial coefficient was used. 

In the rest of this article the Veneziano amplitude calculation will widely be extended. To motivate more general remarks further example computations will be presented first. 

\subsubsection{Example: three tachyons, one gluon}
In general string scattering amplitudes will involve particles with polarisation vectors. To show how this fits into the calculation first study the example of an amplitude with three tachyons and a gluon. Residues of the amplitude $\Amp(1,2,3,4_g)$ with three tachyons labelled $1$, $2$, $3$ and a gluon $4_g$ in the $(1,2)$ channel can depend on one momentum invariant, say $s_{23}$, and terms containing the polarisation $\xi_4 \cdot k_1$, $\xi_4 \cdot k_2$, $\xi_4 \cdot k_3$. Due to momentum conservation and orthogonality of the polarisation vector w.r.t.\ it's own momentum, one of these can be expressed in terms of the other two, e.g.
\be
\xi_4 \cdot k_2 = - \xi_4 \cdot (k_1 + k_3) \ .
\ee
Momentum conservation gives in this case
\be\label{3t1g_massless_relation}
s_{12} + s_{23} + s_{13} = \sum \alpha' m_i^2 = -3  .
\ee
By the same monodromy relation as before \eqref{eq:monodrat4pt}, repeated here for convenience,
\begin{equation}\label{eq:monod3T1g}
\Amp(1,2,3,4_g) = (-1)^{\alpha' (m_1^2 + m_4^2)} \frac{\sin(\pi s_{13})}{\sin(\pi s_{12})} \Amp(1,3,2,4_g) ,
\end{equation}
the amplitude $\Amp(1,3,2,4_g)$ has no poles (and thus $\Amp(1,2,3,4_g)$ has roots) for
\be
s_{13} \in \mathbb{Z} \quad \wedge \quad s_{23} \leq -2 \quad \wedge \quad s_{13} \leq -2 .
\ee
This becomes at the residue $s_{12} = A-1$, using \eqref{3t1g_massless_relation}
\be
s_{23} \in \mathbb{Z} \quad \wedge \quad s_{23} \leq -2 \quad \wedge \quad s_{23} \geq -A .
\ee
Hence there is, again, only a finite number of roots. This fixes a polynomial of degree $A-1$. Similar to the Veneziano example the residues have to be proportional to the following polynomials which exhibit all the required roots
\be\label{eq:ansatz3t4gpolyn}
\frac{\Gamma[-s_{23}-1] }{\Gamma[A] \Gamma[-s_{23} - A]} = \binom{k_{23}}{A-1} \qquad A>0.
\ee
The poles at $A=0$ and $A=1$ deserve special attention. For $A=0$ the exchanged particle in the $(1,2)$ channel is a tachyon. Hence the polarisation of the gluon can only be contracted to the momentum which belongs to the tachyon on the same $3$-point amplitude (up to momentum conservation). This gives
\be
\Res_{s_{12}\to -1} \Amp(1,2,3,4_g) = c_0 \, \xi_4 \cdot k_3 ,
\label{eq:ansatz3t4g1st}
\ee
up to a numerical constant $c_0$ by dimensional analysis. The constant can be fixed from the $T^2g$ and $T^3$ three point amplitude found in the Veneziano amplitude computation, so that $c_0 \propto g_0^2$. At $A=1$ the residue at the pole is parametrised by 
\be\label{eq:ansatz3t4g2nd}
\Res_{s_{12}\to 0} \Amp(1,2,3,4_g) = c_1 \,\xi_4 \cdot k_1 + c'_1 (s_{23} + c''_1) \,\xi_4 \cdot k_3,
\ee
with numerical constants $c_1$ and $c'_1$. Here  the fact that the maximal spin of the exchanged particle is $1$ at this level was used. This either gives a contraction of the polarisation vector into a momentum 'at the other side of the pole', i.e. the $\xi_4 \cdot k_1$ term, or an additional power of momentum. 

Tuning to $s_{23}=-1, s_{13}=-2$ gives by equation \eqref{eq:monod3T1g}
the pole in the $(2,3)$ channel of the right-hand side amplitude which leads to 
\be
c_1 \xi_4 \cdot k_1  + c'_1 (-1+c''_1) \xi_4 \cdot k_3 = - c_0 \,\xi_4 \cdot k_1 
\ee
so that immediately $c''_{1}=1$ follows.  
Tuning $s_{13}=-1, s_{23}=-2$ gives similarly
\be
c_1 \xi_4 \cdot k_1 + c'_1 (-2+1) \xi_4 \cdot k_3 =  c_0\, \xi_4 \cdot k_2.
\ee
Hence there are two equations in two unknowns which can be solved
\be
c'_1 = - c_1 = c_0,
\ee
so that \eqref{eq:ansatz3t4g2nd} becomes
\be\label{eq:ansatz3t4g3rd}
\Res_{s_{12}\to 0} \Amp(1,2,3,4_g) = - c_0 \left(\xi_4 \cdot k_1 - (s_{23} + 1) \,\xi_4 \cdot k_3 \right) .
\ee
Note that this computation has in effect fixed the numerical coefficient of the tachyon-gluon-gluon
coupling in terms of the tachyon-tachyon-tachyon coupling. Generalising to higher values of $A$ is
straightforward since the ansatz in equation \eqref{eq:ansatz3t4g2nd} captures all possible
polarisation structures. At a generic level then the roots appearing in  \eqref{eq:ansatz3t4gpolyn}
can be included as multiplicative factors. To fix the coefficients at level $A$, one tunes to the
two data-points  $s_{23}=-1, s_{13}=-1-A$ as well as  $s_{13}=-1, s_{23}=-1-A$. The result is 
\be \label{3t1g_result}
\Res_{s_{12}\to A-1} \Amp(1,2,3,4_g) = c_A \, (-1)^{A} \left(-\, \xi_4 \cdot k_1 + \frac{1}{A}(s_{23}+1)\, \xi_4 \cdot k_3 \right) \binom{k_{23}}{A-1} ,
\ee
where $c_A$ is a constant that can be different for each $A$.
This completes the calculation of all residues in the $(1,2)$ channel. 

\subsubsection*{The complete four point function through on-shell recursion}
At this stage on-shell recursion can be used to obtain the complete four point amplitude from its residues. As above, a shift in the (2,3) channel will be implemented. Following the same steps this yields
\bea
\Amp(1,2,3,4_g) &= c_A A \left( \xi_4 \cdot k_1 \sum\limits_{A=1}^{\infty} \frac{(-1)^{A-1}}{s_{12}-A+1} \binom{k_{23}}{A-1}
+ \xi_4 \cdot k_3 \sum\limits_{A=0}^{\infty} \frac{(-1)^{A}}{s_{12}-A+1} \binom{k_{23}+1}{A} \right) \\
&= (g'_o)^2 \left(  \xi_4 \cdot k_1 \frac{\Gamma[-s_{12}] \Gamma[-s_{23}-1]}{\Gamma[-s_{12}-s_{23}-1]} +  \xi_4 \cdot k_3 \frac{\Gamma[-s_{12}-1] \Gamma[-s_{23}]}{\Gamma[-s_{12}-s_{23}-1]} \right) .
\eea{eq:TTTG_result}
As a cross-check it can be verified straightforwardly that this colour-ordered amplitude
is invariant under interchange of particles $1 \leftrightarrow 3$ as it must be since
\begin{equation}
\Amp(1234)  = \Amp(4321) = \Amp(3214) .
\end{equation}
Since the particles $1$ and $3$ are tachyons, this amounts simply to an exchange of their momenta. In particular $s_{12} \leftrightarrow s_{23}$. The string coupling constant squared $(g')_o^2$ can be traced to a tachyon factorisation channel where two amplitudes appear which was already computed above: tachyon-tachyon-gluon and $(\text{tachyon})^3$.

\subsubsection{Example: four gluons in the superstring}
Since the monodromy relations hold for all string amplitudes, they are relations between superamplitudes which contain all amplitudes that are related by supersymmetry as components. It is useful for computational purposes to use an on-shell superspace formalism. Here the formalism of \cite{Boels:2012ie} will be used for massless fields which necessarily involves complex chiral spinors. The minimal on-shell superspace in $10$ dimensions constructed through this method therefore has $(2,0)$ supersymmetry. For open strings one has to restrict all momenta to a $D=8$ subspace to be able to employ unrestricted massless on-shell superfields. Note that this is only a (kinematic) restriction above $9$ points. It will mostly be important below that the massless superfields used here are scalar.  

The superamplitudes are given by a kinematic function $\tilde \Amp$ times a momentum conserving
delta function $\delta^8 (K)$ which depends on the kinematic variables $K$ and a fermionic
super-momentum conserving delta function
$\delta^8 (Q)$ which assures that the Ward identities of on-shell supersymmetry are satisfied
\be
\Amp^{D=8} = \delta^8 (K) \delta^8 (Q)  \tilde \Amp (Q,K) .
\ee
For four points, the function $\tilde \Amp(Q,K)$ has no fermionic weight, 
\be
\tilde \Amp(Q,K) = \tilde \Amp(K) , \qquad \textrm{four points} .
\ee
As a function of the momenta $\tilde \Amp(K)$ has roots and poles. The sums over parts of the states at the residues of the poles can be performed using a fermionic integral. As here the interest is in the result of this integral, it actually mostly does not have to be considered. See \cite{Boels:2012if} for an explanation of the massive spinor helicity formalism in higher dimensions. The only thing important for the discussion here is that this makes the computation manifestly on-shell supersymmetric. In field theory, the four point function reads:
\be\label{eq:d8n4ftamp}
\Amp^{D=8, \textrm{YM}} = \delta^8 (K) \delta^8 (Q) \frac{g_{\textrm{ym}}}{k_{12} k_{23}} , \qquad \textrm{field theory} .
\ee

As the delta functions are completely symmetric the functions $\tilde \Amp (K)$ satisfy the same monodromy relations as before. Hence the roots can be derived analogously, with the poles starting at $0$ instead of $-1$\footnote{This is actually not an essential assumption. There is a more complicated version of this derivation which takes an arbitrary starting point for the series of poles, basically introducing an `intercept'. Then, as will be clear from the discussion in Section \ref{sec:no-gost_conditions}, unitarity restricts the starting point to be $0$.}. 
\be\label{eq:superstring4ptmono}
\tilde \Amp(1,2,3,4) = \frac{\sin(\pi k_{13})}{\sin(\pi k_{12})}  \tilde \Amp(1,3,2,4) ,
\ee
leads to $\tilde \Amp(1,2,3,4)=0$ at $k_{12}=-A$ for 
\be
\begin{gathered}
k_{23} \in \mathbb{Z} , \\
0 < k_{13} \quad \wedge \quad 0 < k_{23} \qquad \Leftrightarrow \qquad 0 < k_{23} < A.
\end{gathered}
\label{eq:conditions_superstring_A4=0}
\ee
This gives us the following $A-1$ roots for $A\geq 1$
\be
\Res_{k_{12} \to -A} \tilde \Amp \propto \binom{k_{23}-1}{A-1} .
\label{eq:res_A4_superstring}
\ee
The maximum power of $k_{23}$ can be determined from a BCFW supershift in the $(1,2)$ channel. Compared to the residue of the tree level Yang-Mills amplitude at the $s_{12}$-channel pole, $ (\frac{g_{\textrm{ym}}}{k_{23}})$, this power is $A$. 

This can also be argued on the basis of the known spectrum. The spectrum for the open superstring in $10$ dimensions was worked out in \cite{Hanany:2010da}.
Structurally, the highest spin field in the spectrum at mass level $A$ transforms as the symmetric
traceless $A+1$-tensor of the massive little group $SO(9)$. In the massive superfield formalism
this translates into a $A-1$-tensor massive on-shell superfield.  The fermionic integral in this
case contributes an overall constant \cite{Boels:2012if}. This shows that the obtained polynomials
at the residues contain the complete dependence on kinematic invariants and that the overall
numerical constants are all that is left to be determined. 

These overall constants can, as before, be fixed by unitarity in the cross-channel. That is, first take the residue of \eqref{eq:superstring4ptmono},
\begin{equation}
\textrm{Res}_{s_{12} \rightarrow A} \tilde \Amp(1234) =  \frac{(-1)^{A-1}}{\pi} \left[ \sin(\pi s_{13})  \tilde \Amp(1324) \right]_{s_{12} = A} .
\end{equation}
Then one inserts the ansatz for the left-hand side,
\begin{equation}
c  \binom{k_{23}-1}{A-1} =  \frac{(-1)^{A-1}}{\pi} \left[ \sin(\pi s_{13})  \tilde \Amp(1324) \right]_{s_{12} = A} .
\end{equation}
and tunes $s_{13} = - k_{13} \rightarrow 0$ to obtain
\begin{equation}
c  =  \frac{(-1)^{A}}{A}  ,
\end{equation}
where instead of writing the unitarity expression for the $s_{13}$ pole on the right-hand side the known expression of equation \eqref{eq:d8n4ftamp} was used. Note this last step fixes the residues of the four point superstring amplitude in terms of the field theory limit.  

Assembling the full amplitude through on-shell recursion now follows by repeating
basically the same computation as in the Veneziano amplitude case and simply yields 
\be
\Amp^{D=8} = \delta^8 (K) \delta^8 (Q) \frac{\Gamma[-s_{12}] \Gamma[-s_{23}]}{\Gamma[-s_{12}-s_{23}+1]}.
\ee

\subsubsection{Example: four closed string tachyons}
Closed string amplitudes are defined by the KLT relations. For four points these can be written as
\begin{equation}\label{eq:detrootsclosedstringsII}
M(1234) = \sin(\pi k_{23}) \Amp(1234) \Amp(1324) ,
\end{equation}
with all coupling constants stripped off. In this subsection the direct application of a similar reasoning as above to determine the residues at poles is briefly explored for closed strings.

The closed string amplitude has poles in all channels and is completely symmetric. Consider without loss of generality the residue at the $s_{12}$ channel pole,
\begin{equation}
\textrm{Res}_{s_{12} \rightarrow A-1} M(1234) = \sin (\pi k_{23}) \left( \Amp(1324)\right)_{s_{12} \rightarrow A-1} \left( \textrm{Res}_{s_{12} \rightarrow A-1} \Amp(1234)\right) .
\end{equation}
Now by the following analog of equation \eqref{eq:fixresidue4pt},
\begin{equation}
\left[ \Amp(1324) \right]_{s_{12} = A-1} =(-1)^{A-1} \frac{\pi}{ \sin(\pi k_{13}) }  \textrm{Res}_{s_{12} \rightarrow A-1} \Amp(1234) 
\end{equation}
the residues of the closed string amplitudes simply reduce to a double copy of the residues of the open string amplitude by momentum conservation at the residue,
\begin{equation}
\textrm{Res}_{s_{12} \rightarrow A-1} M(1234) = - \pi \left( \textrm{Res}_{s_{12} \rightarrow A-1} \Amp(1234)\right)^2 .
\end{equation}
By the holomorphic factorisation property of the closed string worldsheet vertex operators this is expected. 

The residues of the open string amplitudes were determined above. This fixes the residue at the pole of the closed string amplitudes. The overall numerical factor is now the product of the two open string coupling constants squared. This can now be defined as the closed string coupling constant.  It should be clear a similar reasoning will go through for tree level closed string amplitudes with arbitrary field content. 

Of course, one can also use monodromy relations to write the KLT relation here as
\begin{equation}\label{eq:detrootsclosedstrings}
M(1234) = \frac{\sin(\pi k_{12}) \sin(\pi k_{23})}{\sin(\pi k_{13})} \Amp(1234) \Amp(1234) .
\end{equation}
Now all poles in the $s_{13}$ channel are explicitly factored into the $\sin$ denominator. This generalises to multiple points: there is always an expression of the closed string amplitudes in terms of a $(N-3)!$ basis of open string amplitudes with three particles fixed in consecutive positions. If these particles are labelled $1,2,3$, then all the poles of the closed string amplitude which involve momentum $k_2$ and multiple momenta not equal to $k_1$ or $k_3$ will be explicit in the denominator. This simply follows since the open string amplitudes in the chosen basis do not have poles in these channels. 

Further and more direct exploration of the closed string sector is left to future work, save for one comment. By Bose symmetry, the complete closed string tachyon amplitude must be completely symmetric. Note that in equation \eqref{eq:detrootsclosedstrings} there are roots of the closed string amplitude manifest in the $s_{12}$ and $s_{23}$ channel while those in the $s_{13}$ channel are contained in the open string amplitude squared, moderated by corresponding poles from the sine function in the denominator. In the first way of writing in equation \eqref{eq:detrootsclosedstringsII} only one series of roots is manifest.

\subsubsection*{Extensions}
The main technical complication in extending the argument given above to four point amplitudes with other external states is
the appearance of more and more polarisation tensors. These may be treated by parametrising
the residues in terms of all possible tensor structures built out of metrics and external
momenta on the three point amplitudes which appear at the residue. Since these tensor
structures are independent, their coefficient polynomials can be fixed as in the example
above from the roots at least to some extent. If the monodromy relations are strong 
enough\footnote{This will be shown below for Koba-Nielsen amplitudes.}, this leaves fixing the overall constants at each mass level. We
strongly suspect that one needs all three point amplitudes up to the level of the highest
level external particle involved in the scattering to fix all coefficients: this ensures
all possible tensor structures appear on the residue.

In the superstring case the same complications start to appear in the massive sector as
long as one considers superfields. Massless vector fields are components of scalar on-shell superfields, which are treated
analogously to tachyons in the bosonic string, at least in the $8$ dimensional formalism.

From the structure of the argument it should be clear that in the four point case one always ends up with sums over $\beta$ function type functions times possibly complicated coefficients. This is of course well known from the worldsheet formalism. 

\subsection{Five and higher point amplitudes}

\subsubsection{Five tachyon amplitude}

At five points the monodromy relations can be solved to give
\begin{equation}\label{eq:monodrat5pt}
\Amp(12345) = \frac{1}{\sin(\pi s_{12})}\left[ \sin(\pi (-s_{12}+ k_{23})) \Amp(13245) + \sin(\pi (-s_{12}+ k_{23} + k_{24})) \Amp(13425)\right] ,
\end{equation}
so the residues of the amplitude in the $s_{12}$ channel are
\be
\Res_{ s_{12} \to A-1} \Amp(12345) = \frac{1}{\pi} \left[ \sin(\pi k_{23}) \Amp(13245) + \sin(\pi (k_{23} + k_{24})) \Amp(13425) \right]_{s_{12}=A-1}.
\ee
This has roots for 
\be
k_{23}, k_{24}  \in \mathbb{Z} ,
\label{k23k24Integers}
\ee
but only if the two amplitudes on the right-hand side do not have a pole at these values, which leads to the conditions
\begin{equation}
\begin{aligned}
k_{23} &\geq 0\;,\\
k_{24} &\geq 0 \;,\\
k_{25} &\geq 0 \quad \Leftrightarrow \quad k_{23} + k_{24} \leq A - 1\;.
\end{aligned}
\label{s12conds}
\end{equation}
The condition for $k_{25}$ is required because $k_{25}$ becomes an integer due to momentum conservation when $k_{12}, k_{23}, k_{24}$ are integers.

The conditions are solved by the polynomials
 \begin{align}\label{eq:basisofsols5pt}
\binom{k_{23}}{A-a} \binom{k_{24}}{a} , \qquad 0 \leq a \leq A .
\end{align}
Each of this terms contains $A$ powers of $k_2$, the maximally allowed number. So multiplying them by further polynomials containing $k_{23}$ or $k_{24}$
is not allowed. 

The polynomials just written down are a basis of the space of polynomials of total order $\leq A$ which vanish under the conditions \eqref{k23k24Integers} and \eqref{s12conds}. Since the main theorem of algebra does not hold for functions of more than one variable proving this requires some work. For this, note that 
\begin{align}
\binom{k_{23}}{B-a} \binom{k_{24}}{a} , \qquad 0 \leq a \leq B , \qquad B \leq A ,
\label{eq:5pt_poly_basis}
\end{align}
is a basis for all polynomials of maximal total degree A labelled by indices $B$ and $a$.
This follows as they are linear combinations of the natural basis monomials $(k_{23})^i (k_{24})^j$ for $i+j \leq A$.
The most generic polynomial of maximal total degree $A$ is therefore a linear combination of this basis.
Now consider the set of roots in equation \eqref{s12conds}. By first setting $k_{23}$ and $k_{24}$ to zero it is
easy to see there can be no constant term. Then, considering the two points $(k_{23},k_{24}) =(0,1)$ and $(1,0)$
one can rule out all linear polynomials. Continuing along these lines one sees that none of the polynomials in 
\eqref{eq:5pt_poly_basis} with $B<A$ has the required roots s.t.\ equation \eqref{eq:basisofsols5pt}
is the basis of all polynomials which satisfy the conditions of equation \eqref{s12conds}.

For the channel $s_{123}=B-1$ the monodromy relation \eqref{eq:monod_im_lvl2} can be used,
\begin{equation}\label{eq:monodrat5pt2}
\Amp(12345) = \frac{1}{\sin(\pi s_{123})}\left[ \sin(\pi (-s_{123}+ k_{34})) \Amp(12435) + \sin(\pi (-s_{123}+ k_{34} + k_{24})) \Amp(14235)\right] ,
\end{equation}
which implies the following conditions for a vanishing residue $\Res_{s_{123} \to B-1} \Amp(12345)$
\begin{equation}
\begin{aligned}
k_{24}, k_{34}  &\in \mathbb{Z} \;,\\
k_{24} &\geq 0\;,\\
k_{34} &\geq 0 \;,\\
k_{14} &\geq 0 \quad \Leftrightarrow \quad k_{24} + k_{34} \leq B - 1\;.
\end{aligned}
\label{s45conds}
\end{equation}
The polynomials solving them are
 \begin{align}
\binom{k_{24}}{a} \binom{k_{34}}{B-a} , \qquad 0 \leq a \leq B .
\end{align}
If  both internal particles are send on-shell, that is the channel
\be
{\mathord{\vcenter{\hbox{\scalebox{0.5}{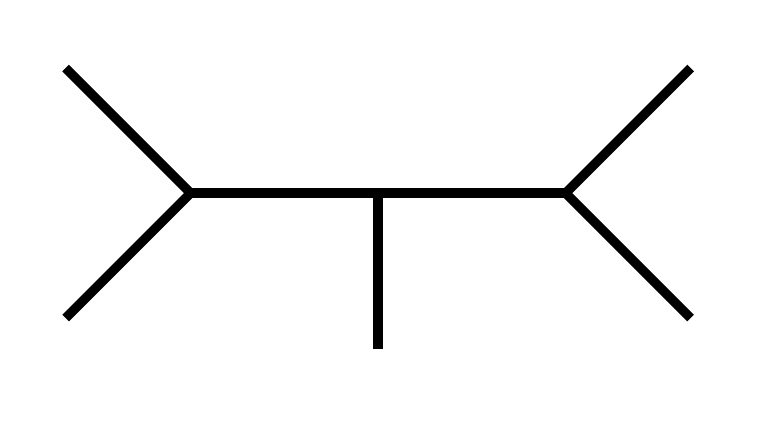}}}}} ,
\ee
is considered, the residues have to vanish when either conditions (\ref{k23k24Integers}, \ref{s12conds}) or \eqref{s45conds} are satisfied.
At the same time, $k_2$ is only allowed to appear to the $A$th power and  $k_4$ to the $B$th power. These conditions follow from considering BCFW shifts of the residue in the $(1,2)$ channel as well as the $(4,5)$ channel. Just as in the four-point case one finds
\begin{align}\label{eq:largeztachyonsatresidue5pt}
\textrm{Res}_{s_{12} \rightarrow A-1} & \Amp(12345)  \sim z^{A} \left(f_1\left(\frac{1}{z}\right) \right) , \\
\textrm{Res}_{s_{45} \rightarrow B-1} & \Amp(12345)  \sim z^{B} \left(f_1\left(\frac{1}{z}\right) \right) .
\end{align}
The only polynomials fulfilling all roots as well as the power counting constraints just derived are
\begin{align}
\binom{k_{23}}{A-a} \binom{k_{24}}{a} \binom{k_{34}}{B-a} , \qquad 0 \leq a \leq \min(A,B) .
\label{eq:ResResA5_poly_basis}
\end{align}

\subsubsection*{Fixing the coefficients}
The coefficient for each of these polynomials can be fixed by using the monodromy relations again or, alternatively, by assuming cyclicity of the amplitude which is shown in Appendix \ref{sec:cyclicity_fixes_coefficients_5point}. As a warm-up for the the $N$-point case discussed below,  the exact linear combination of polynomials \eqref{eq:ResResA5_poly_basis} that is the double residue of $\Amp(12345)$ will be determined.
Just as in the four point case, the overall factors will follow by considering the right-hand side of equation \eqref{eq:monodrat5pt} at an integer-valued kinematic point where it does \emph{not} vanish. 
It is convenient to take this point to be  
\begin{equation}
k_{23} + k_{24} = A , 
\end{equation}
with $k_{23}$ and $k_{24}$ non-negative integers. The polynomials \eqref{eq:ResResA5_poly_basis}  are special at this point. To see this, assume w.l.o.g.\ that $A \leq B$ and consider the expression
\begin{equation}
\binom{A-k_{24}}{A-a} \binom{k_{24}}{a} \binom{k_{34}}{B-a}, \qquad 0 \leq a \leq A .
\end{equation}
Now, the second binomial coefficient vanishes at these integer values when $k_{24} < a$, while the first vanishes when $A-k_{24} < A-a$.
Hence at this particular kinematic point the only one of these polynomials that is non-zero is the one with $a = k_{24}$. By choosing different integers for $k_{24}$ the coefficients of all
the polynomials can now be calculated. This can be done by calculating the right-hand side of equation \eqref{eq:monodrat5pt} at this particular kinematic point.
The term containing $\Amp(13245)$ vanishes while the amplitude $A(13425)$ develops a tachyonic pole in the $(25)$ channel which cancels against the root from the sine function that multiplies it
\begin{multline}
\lim_{k_{23} + k_{24} \rightarrow A} \lim_{k_{24}\rightarrow a} \left[ \frac{1}{\pi} \sin(\pi (k_{23} + k_{24})) \Amp(1,3,4,2,5) \right]_{s_{12}=A-1} = \\ -(-1)^{A}  \left[\, \Amp(1,3,4,P) \, \Amp(-P,2,5) \, \right]_{\small \left\{\begin{array}{c} s_{12}=A-1 \\ k_{23} + k_{24} = A \\  k_{24}=a\end{array}\right\}} .
\end{multline}
Note the amplitudes in this equation all involve tachyons only and the open string coupling constant has been suppressed. 
The four point amplitude is easy to evaluate on a further special kinematic point.  Now use monodromy relation \eqref{eq:monod_im_lvl1}
again to expose the residue in $s_{13} = s_{123}-a$
\be
\Res_{s_{123}\to B-1} \Amp (1,3,4,P) = \frac{1}{\pi} \left[ \sin (\pi k_{34}) \Amp(1,4,3,P) \right]_{s_{123} = B-1} .
\ee
Setting $k_{34}=B-a \in \mathbb{Z}$ will hit a root of the sine
and a tachyon pole in the amplitude $\Amp(1,4,3,P)$ because at this value of $k_{34}$ the equation $s_{3P}=s_{235}=-1$ holds. This lead to
\bea
& \lim_{k_{34} \rightarrow B-a} \left[ \frac{1}{\pi} \sin(\pi k_{34}) \Amp(1,4,3,P) \right]_{\small \left\{\begin{array}{c} s_{12}=A-1 \\ s_{123}=B-1 \\k_{23} + k_{24} = A \\  k_{24}=a\end{array}\right\}} \\
={}& -(-1)^{B-a}  \left[\, \Amp(1,4,Q) \, \Amp(-Q,3,P) \, \right]_{\small \left\{\begin{array}{c} s_{12}=A-1 \\ s_{123}=B-1 \\ k_{23} + k_{24} = A \\  k_{24} =a \\ k_{34} + k_{24} = B\end{array}\right\}} .
\eea{eq:5pt_coef2}
Plugging everything back into \eqref{eq:monodrat5pt} the final result reads
\bea
\Res_{s_{12}\to A-1} \Res_{s_{123}\to B-1} \Amp(12345) =
\sum\limits_{a=0}^{\infty} \binom{k_{2 3}}{A-a} \binom{k_{2 4}}{a} \binom{k_{34}}{B-a} (-1)^{A+B-a} .
\eea{eq:ResResA5}

\subsubsection*{The complete five point function through on-shell recursion}
In this example it will now be shown explicitly how the double residues can be combined with BCFW on-shell recursion to obtain the full amplitude.
First perform a BCFW shift on particles 1 and 5 by a vector $q_{15}$ scaled by a complex parameter $z_{15}$
\be
\hat k_1 = k_1 + z_{15} q_{15} , \qquad \hat k_5 = k_5 - z_{15} q_{15} ,
\ee
where
\be
k_1 \cdot q_{15} = k_5 \cdot q_{15} = q_{15}^2 = 0.
\ee
Using BCFW on-shell recursion, 
\bea
\Amp(12345)={}&
- \sum\limits_{A=0}^{\infty} \sum\limits_{\alpha} \sum\limits_{\text{polarisations}} \frac{\Amp(\hat 1, 2, \hat M^{A,\alpha}) \Amp(\hat M^{A,\alpha},3,4,\hat 5)}{s_{12}-A+1} \\
&- \sum\limits_{B=0}^{\infty} \sum\limits_{\beta} \sum\limits_{\text{polarisations}} \frac{\Amp(\hat 1, 2, 3, \hat M^{B,\beta}) \Amp(\hat M^{B,\beta},4,\hat 5)}{s_{123}-B+1} ,
\eea{eq:BCFW-5pt-1}
is obtained. For details about the sums over irreps $\alpha, \beta$ and polarisations of the intermediate particles see Section \ref{sec:unit}.
Now implement another shift for each of the four point amplitudes, namely for the first term
\be
\tilde k_3 = k_3 + z_{34} q_{34} , \qquad \tilde k_4 = k_4 - z_{34} q_{34} ,
\ee
and for the second term
\be
\bar k_2 = k_2 + z_{23} q_{23} , \qquad \bar k_3 = k_3 - z_{23} q_{23} .
\ee
Using this 
\bea
\Amp(12345) ={}&
\sum\limits_{A,B=0}^{\infty} \sum\limits_{\alpha,\beta} \sum\limits_{\text{polarisations}} \frac{\Amp(\hat 1, 2, \hat M^{A,\alpha}) \Amp(\hat M^{A,\alpha},\tilde 3,\tilde M^{B,\beta}) \Amp(\tilde M^{B,\beta}, \tilde 4,\hat 5)}{(s_{12}-A+1)(s_{\hat{1}23}-B+1)} \\
&+ \sum\limits_{A,B=0}^{\infty} \sum\limits_{\alpha,\beta} \sum\limits_{\text{polarisations}} \frac{\Amp(\hat 1, \bar 2, \bar M^{A,\alpha}) \Amp(\bar M^{A,\alpha}, \bar 3, \hat M^{B,\beta}) \Amp(\hat M^{B,\beta},4,\hat 5)}{(s_{\hat{1}2}-A+1)(s_{123}-B+1)} ,
\eea{eq:BCFW-5pt-2}
is obtained. In each term the BCFW shifts are tuned in such a way that in the first line
\be
s_{\hat 1 2} = A-1 , \qquad s_{\hat 1 2 \tilde 3} = B-1 ,
\ee
and in the second line
\be
s_{\hat 1 \bar 2} = A-1 , \qquad s_{\hat 1 \bar 2 \bar 3} = B-1 .
\ee
The first and second line in \eqref{eq:BCFW-5pt-2} are very similar up to a difference
in the BCFW shifts and the rather subtle difference in denominators. Practically this
means that taking first a $s_{12} \rightarrow A'-1$ and then a $s_{123} \rightarrow B'-1$
limit of the full result selects the first term, while doing this in the opposite order
selects the second. This follows as the second expression for instance generically does
not have a pole at $s_{12}=a$ for \emph{any} integer $a$ unless $s_{123}=B-1$ holds. 

The residues appearing in both terms were derived from the monodromy relations above in equation \eqref{eq:ResResA5}. These can be plugged in
\bea
\Amp(12345) ={}&
 \sum\limits_{A,B=0}^{\infty} \sum\limits_{a=0}^{\infty} \binom{k_{2 \tilde 3}}{A-a} \binom{k_{2 \tilde 4}}{a} \binom{k_{34}}{B-a} \frac{ (-1)^{A+B-a}}{(s_{12}-A+1)(s_{\hat{1}23}-B+1)} \\
&+ \sum\limits_{A,B=0}^{\infty} \sum\limits_{a=0}^{\infty} \binom{k_{23}}{A-a} \binom{k_{\bar 24}}{a} \binom{k_{\bar 34}}{B-a} \frac{ (-1)^{A+B-a}}{(s_{\hat{1}2}-A+1)(s_{123}-B+1)} .
\eea{eq:BCFW-5pt-3}
Note that the secondary BCFW shifts can be chosen\footnote{Choosing BCFW shift vectors like this should always be done with care, the obtained poles must always be at finite values of the shift parameters. For these particular shifts this is the case.} such that $q_{34} \cdot k_2 = q_{23} \cdot k_4 = 0$. In this case the dependence on these shifts trivially drops out of the numerator.
This is significant as a form of internal recursion relations for open string tachyon amplitudes were already proposed more than 40 years ago by Hopkinson and Plahte \cite{Hopkinson:1969er}.
Here the full amplitude is just the maximal residue summed over the mass levels. These results seem to suggest much simpler formulae are possible. We leave this for future work.

\subsubsection{Koba-Nielsen amplitude}

Equation \eqref{eq:monod_im_lvl_s} can be used to derive the residue of the $N$ tachyon
amplitude in $s_{1\ldots l}$. This amplitude will be referred to as the Koba-Nielsen amplitude. First note that none of the amplitudes on the right-hand side has a pole in 
$s_{1\ldots l}$, where $2 \leq l \leq N-2$. Furthermore, all sines vanish at a pole at  $s_{1\ldots l} = A_l-1$ under the condition
\be
k_{ij} \in \mathbb{Z} \qquad \forall \ i \in \{ 2, \ldots, l \} , j \in \{ l+1, \ldots, N-1 \} .
\ee
as in the previous examples, these momenta must be in the range where the amplitudes on the right-hand side do not have poles 
\be
k_{ij} \geq 0 \qquad \forall \ i \in \{ 2, \ldots, l \} , j \in \{ l+1, \ldots, N-1 \} .
\ee
There is one further pole in one of the amplitudes that has to be taken into account, namely the one in $s_{2 \ldots l,N}$,
because this Mandelstam variable becomes integer at the considered configuration
\be
s_{2 \ldots l,N} = \sum\limits_{\substack{1<i\leq l\\l < j < N}}  k_{ij} - s_{1\ldots l} + \alpha' (m_1^2 + m_N^2) .
\label{eq:s_2lN}
\ee
Avoiding the pole leads to the condition
\be
s_{2 \ldots l,N} \leq -2 \quad \Leftrightarrow \quad \sum\limits_{\substack{1<i\leq l\\l < j < N}} k_{ij} \leq A_l-1 .
\label{eq:s_2lN_cond}
\ee
The combined conditions are naturally solved by the polynomials
\be
\prod\limits_{\substack{1< i \leq l\\l < j < N}}
\binom{k_{ij}}{a_{ij}} , \qquad \text{where}\ a_{ij} \in \mathbb{N}_0 \wedge \sum\limits_{\substack{1<i \leq l\\l<j<N}} a_{ij} = A_l .
\label{eq:A_N_1_residue_basis}
\ee
To obtain the multiple residue where all the internal particles in the multiperipheral channel are on-shell $s_{1\ldots l}=A_l-1$ $\forall l \in \{2,\ldots,N-2\}$
 take the polynomials that solve the above conditions for all those $l$
\be
\prod\limits_{\substack{i,j\\1<i<j<N}} \binom{k_{ij}}{a_{ij}} , \qquad \text{where}\ a_{ij} \in \mathbb{N}_0 \wedge \sum\limits_{\substack{1<i \leq l\\l<j<N}} a_{ij} = A_l \quad \forall l.
\label{eq:A_N_max_residue_basis}
\ee
The multiperipheral channel is visualised by the diagram
\be
{\mathord{\vcenter{\hbox{\scalebox{0.5}{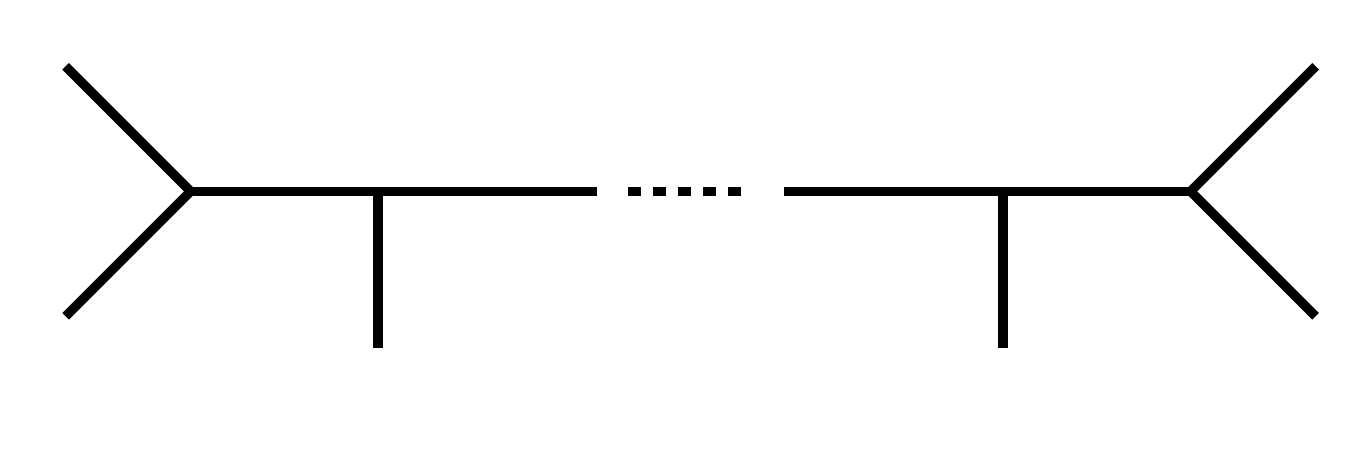}}}}} .
\ee
Although it is possible to consider other channels, for our purposes the multiperipheral channel is enough: this channel has enough information to determine the full amplitude through on-shell recursion.

\subsubsection*{Completeness of the basis}

The bases of polynomials \eqref{eq:A_N_1_residue_basis} and \eqref{eq:A_N_max_residue_basis} are
complete. This follows as the spin is limited by the mass level and the spin determines how many indices can be contracted across an internal line that 
is put on-shell (this will be discussed in much more detail in Section \ref{sec:unit}).

For amplitudes in the multiperipheral channel as discussed above the limit on spin by level implies that the residue at $s_{1\ldots l} = A_l-1$ is proportional to
a polynomial of degree $A_l$ in Lorentz invariants which involve a contraction across the pole under consideration. For tachyon amplitudes this
statement can be written as 
\be
\Res_{s_{1\ldots l}\to A_l-1} \Amp_N \propto \text{Pol}[k_{ij} | 1\leq i \leq l , l < j \leq N] \ \text{of degree} \ A_l .
\label{eq:poly_order}
\ee
For other external particles the same statement holds true, but now the Lorentz invariants can also be constructed from polarisations
as for example in \eqref{3t1g_result}. The polynomials in \eqref{eq:A_N_1_residue_basis} saturate condition \eqref{eq:poly_order} for one
choice of residue $l$ while the polynomials in \eqref{eq:A_N_max_residue_basis} saturate the condition for each $l$ individually.
Hence they constitute bases of polynomials fulfilling the requirements for roots and degree.

The same result can also be derived by utilising a modified BCFW-type shift.
Note that two-particle shifts were enough to fix the polynomials up to five
external particles. This is because any pole in the multiperipheral
channel for a 5 particle amplitude splits the external particles into at
least one set with two particles.  Above $5$ points however one also has
a multiperipheral pole which splits the external lines into two sets, both
of which contain more than two particles. Consider such a pole with particles
$1$ through $l$ in the left-hand side set. Consider the shift
\begin{equation}
k_1 \rightarrow k_1 - (l-1) q z , \quad k_2 \rightarrow k_2 + q z , \quad \ldots , \quad k_l \rightarrow k_l+ q z ,
\end{equation}
for a non-trivial vector $q$ for which $q^2 = q \cdot k_1 = \ldots = q\cdot k_l =0$, but for which also
$q \cdot k_{l+1} \neq 0$. This shift always exists for up to $27$ particle kinematics in the bosonic
string, above it requires an analytic continuation in the dimension\footnote{This argument will only
be used to estimate the maximal degree of a polynomial, so this continuation will not have drastic
consequences at string tree level. Moreover, in the analysis of Section \ref{sec:unit} it will be
manifest that the target space dimension only affects unitarity in sub-leading coefficients.}. The
large $z$ behaviour of a string scattering amplitude can be argued for using a saddle-point-type
argument just as in \cite{Cheung:2010vn} and \cite{Boels:2010bv} which leads to 
\begin{align}\label{eq:largeztachyonsatresidueNpt}
\lim_{z\rightarrow \infty }\textrm{Res}_{s_{1\ldots l} \rightarrow A_l -1} & \Amp_N \sim z^{A_l} \left(f_1\left(\frac{1}{z}\right) \right) ,
\end{align}
under this shift which is equivalent to the statement above.

\subsubsection*{Fixing the coefficients}
The residues can contain only polynomials from the basis \eqref{eq:A_N_max_residue_basis} which are labeled by the mass levels $\{A_l\}$ and further parameters $\{a_{ij}\}$. All that is left to do is to fix the coefficient $h_{\{A_l\}, \{a_{ij}\}}$ for each basis element. For this, start with the ansatz
\bea
&\left( \prod\limits_{l=2}^{N-2} \Res_{s_{1\ldots l}\to A_l-1} \right) \Amp_N 
= \sum\limits_{a_{23},\ldots,a_{N-2,N-1}=0}^\infty h_{\{A_l\}, \{a_{ij}\}}
\prod\limits_{\substack{i,j\\1<i<j<N}} \binom{k_{ij}}{a_{ij}} \prod\limits_{l=2}^{N-2} \delta_{A_l,\sum\limits_{\substack{ \scriptscriptstyle 1<u \leq l\\ \scriptscriptstyle l<v<N}} a_{uv}} .
\eea{eq:A_N_max_residue_ansatz}
To fix the coefficients consider certain kinematic limits where the ansatz reduces to a single coefficient $h_{\{A_l\}, \{a_{ij}\}}$. These limits are reached when the Mandelstams $s_{2\ldots l,N}$ that were considered in \eqref{eq:s_2lN_cond} are set to $-1$ which implies $\sum\limits_{\substack{1<i\leq l\\l < j < N}} k_{ij} = A_l$. With this constraint the ansatz becomes
\bea
&\left( \prod\limits_{l=2}^{N-2} \Res_{s_{1\ldots l}\to A_l-1} \right) \Amp_N 
= \hspace{-.7em} \sum\limits_{a_{24},\ldots,a_{N-3,N-1}=0}^\infty  \hspace{-.7em}
h_{\{A_l\}, \{a_{ij}\}}
\hspace{-.5em} \prod\limits_{\substack{i,j\\1<i<j-1<N-1}} \hspace{-.5em} 
\binom{k_{ij}}{a_{ij}} \prod\limits_{l=2}^{N-2} 
\begin{pmatrix}
{A_l-\sum\limits_{\substack{ \scriptscriptstyle 1<u \leq l\\ \scriptscriptstyle l<v<N\\ \scriptscriptstyle v-u \geq 2}} k_{uv}}\\
{A_l-\sum\limits_{\substack{ \scriptscriptstyle 1<u \leq l\\ \scriptscriptstyle l<v<N\\ \scriptscriptstyle v-u \geq 2}} a_{uv}}
\end{pmatrix}  .
\eea{eq:A_N_max_residue_ansatz2}
The next step is to set all the remaining $k_{ij}$ to non-negative integer values. For every term in the sum the first product of binomial coefficients vanishes if $a_{ij}>k_{ij}$ for any values of $i,j$. A binomial coefficients in the second product vanishes if $\sum\limits a_{uv} < \sum\limits k_{uv}$ for some summation range as given in \eqref{eq:A_N_max_residue_ansatz2}. Every relevant pair $u,v$ appears in the sum of such a condition at least once. Together these two observations imply that the only term that does not vanish is the one for $a_{ij}=k_{ij} \ \forall i,j$. The coefficient $h_{\{A_l\}, \{a_{ij}\}}$ is extracted from the relation by tuning the $\{k_{ij}\}$ to the desired $\{a_{ij}\}$
\bea
&\left. \left( \prod\limits_{l=2}^{N-2} \Res_{s_{1\ldots l}\to A_l-1} \right) \Amp_N \right|_{\scriptsize\begin{aligned}&\{s_{2\ldots l, N}=-1\}_{1<l<N-1} \\ &\{k_{ij}\}_{1<i<j-1<N-1} \subset {\mathbb N}_0\end{aligned}}
= h_{\{A_l\}, \{k_{ij}\}} .
\eea{eq:A_N_max_residue_ansatz3}

To determine the number on the left-hand side (which must be a number because all momentum invariants are fixed) the monodromy relations can be employed again. 
Only  the first relation, equation \eqref{eq:monod_im_lvl1}, is needed. At the $s_{12}$ residue only the $s_{2N}$ pole is hit in the last amplitude and all other
terms vanish due to the sines, so
\be
\Res_{s_{12}\to A_2-1} \Amp_N(1,2,\ldots ,N) = \frac{1}{\pi} \sin\left(\pi \sum\limits_{i=3}^{N-1}k_{2i} \right) \Amp_N(1,3,4,\ldots, N-1,2,N) ,
\ee
follows. The remaining amplitude on the right-hand side factorises in the tachyon channel since $s_{2N}=-1$. This leaves a $N-1$ tachyon amplitude where one external leg has the momentum $k_2 + k_N$. The argument of the sine function equals $\pi A_2$ and determines the sign
\be
\Res_{s_{12}\to A_2-1} \Amp_N(1,2,\ldots ,N) = - (-1)^{A_2} \Amp_{N-1}(1,3,4,\ldots, N-1,(2+N)).
\ee
The same monodromy relation can be used again to move leg $3$ to the right. This time the sine in the denominator has the argument $\pi s_{13}$ but this can also be related to the $s_{123}$ channel since $k_{12}$ and $k_{23}$ are integer
\be
\Res_{s_{123}\to A_3-1} \Amp_{N-1}(1,3\ldots N-1,(2+N)) = \frac{1}{\pi} \sin\left(\pi \sum\limits_{i=4}^{N-1}k_{3i} \right) \Amp_{N-1}(1,4\ldots N-1,3,(2+N)).
\ee
Using $\sum\limits_{i=4}^{N-1}k_{3i} = A_3 - \sum\limits_{i=4}^{N-1}k_{2i}$ and the factorisation in the $s_{23N}$ tachyon channel
\be
\Res_{s_{123}\to A_3-1} \Amp_{N-1}(1,3\ldots N-1,(2+N)) = - (-1)^{A_3 - \sum\limits_{i=4}^{N-1}k_{2i}}  \Amp_{N-2}(1,4\ldots N-1,(3+2+N))
\ee
is obtained. This procedure can be repeated until one arrives at the $3$-tachyon amplitude which is $1$. In general each step contributes a factor 
\be
- (-1)^{A_l - \sum\limits_{\substack{\scriptscriptstyle 1<i<l\\ \scriptscriptstyle l < j < N}} k_{ij}}
\ee
so that
\be
h_{\{A_l\}, \{a_{ij}\}} = (-1)^{N-3} (-1)^{\sum\limits_{l=2}^{N-2} \left(A_l - \sum\limits_{\substack{\scriptscriptstyle 1<i< l\\ \scriptscriptstyle l < j < N}} a_{ij} \right)} 
\ee
follows. This agrees with the known result from the worldsheet computation, equation \eqref{eq:A_N_result}. Note that this result is again basically a simple sign, an indication that the polynomial basis chosen is very natural.

%%%%%%%%%%%%%%%%%%%%%%%%%%%%%%%%%%%%%%%%%%%%%%%%%%%%%%%%%%
%%%%%%%%%%%%%%%%%%%%%%%%%%%%%%%%%%%%%%%%%%%%%%%%%%%%%%%%%%

\section{Unitarity in the target space}\label{sec:unit}

For a theory to be physical the S-matrix must be unitary. The non-trivial part of the S-matrix is captured by the $T$ matrix: $S=1 + \ii T$. The demand for the S-matrix to be unitary leads to the in principle exact equation
\begin{equation}
- \ii \left(T - T^{\dagger}\right) = T^\dagger T .
\end{equation}
At tree level (for real momenta) the only source of imaginary parts for the left-hand side is the `$+i\epsilon$' in every propagator. For an imaginary part to arise, the momentum flowing through a propagator must go on-shell. For the right-hand side, insert $1$ as a sum over all states of the theory between $T$ and $T^{\dagger}$. At tree level, only single particle on-shell states can contribute. This is the source of equation \eqref{eq:unitarity_recursion} that relates residues of amplitudes to an expression involving lower point amplitudes
summed over the spectrum of the theory.  For three point amplitudes the same reasoning leads to the constraint that the coupling constant of the three point amplitude must be real.

This can also be seen as follows: if the $3$-point amplitudes are defined as the coupling times a real function of polarisations and momenta,
the corresponding terms in the interacting Hamiltonian contain only real fields, derivatives $\ii \partial^\mu$ and the coupling.
The amplitude of two tachyons and one tensor particle \eqref{eq:A_TTM} for example comes (up to a possible real symmetry factor) from the term
\bea
c_{A,\alpha} \left( \frac{\alpha'}{2} \right)^{|\alpha| /2} \Phi^{A,\alpha}_{\mu_1 \ldots \mu_{|\alpha|}}     \phi \, \ii^{|\alpha|} \partial^{\mu_1} \ldots \partial^{\mu_{|\alpha|}} \phi ,
\eea{eq:Hamiltonian_TTM}
in the Hamiltonian where $\Phi^{A,\alpha}$ is an irreducible tensor field, $\phi$ is the
tachyon field and $c_{A,\alpha}$ is the coupling constant that appears in the 3-point amplitude.
In this setup real couplings imply a hermitian Hamiltonian and thus a unitary S-matrix.
To study unitarity in string theory, one therefore has to inspect all three point amplitudes.
These can be obtained, at least in principle, by factorising higher point amplitudes on poles.
The sums over the spectrum which appear in the residue have so far mostly been avoided by using monodromy relations.

In this section it will be explained how to honestly do the sum over the spectrum. Apart from unitarity, the computation is interesting in its own right.
While the spectrum of most field theories contains only scalars, spinors and vector particles here arbitrary irreducible tensor representations of $SO(D-1)$ appear.
Since it is known from the previous section what the result of summing over the spectrum must be, say for the residue of a tachyon amplitude,
comparing the two resulting expressions also allows us to calculate lower point amplitudes with arbitrary external states. More specifically, $3$-point amplitudes with massive legs will be studied.

The $3$-point amplitudes consist in principle of simple building blocks, see e.g \cite{Boels:2012if}. However even for two tachyons and one arbitrary massive tensor state the general formula for their constant coefficients is quite messy, as is shown in Appendix \ref{sec:one_massive_coeffs}. The relation between $3$-point amplitudes with tachyons and one or two tensor particles and the well-known $N$-tachyon amplitudes is explored in various other ways in Section \ref{sec:3point_vs_tachyons}.

A general string theory $3$-point amplitude of states $i=1,2,3$ with polarisation tensors $\xi^i$ and momenta $k_i$ can be written as 
\be
\Amp_3 = \xi^1_{\mu_1 \mu_2 \ldots}  \xi^2_{\nu_1 \nu_2 \ldots} \xi^3_{\rho_1 \rho_2 \ldots} f^{\mu_1 \mu_2 \ldots,\nu_1 \nu_2 \ldots,\rho_1 \rho_2 \ldots}(k_1,k_2,k_3) ,
\label{eq:3point_general}
\ee
where $f$ is a correlation function that is usually calculated from the worldsheet string theory with DDF operators \cite{DelGiudice:1971fp}.
It is shown in examples below how this general form is related to the known binomials that arise in the 
residues of higher-point amplitudes and how this relation
allows us to compute the combinatorial coefficients that arise in the function $f$ in \eqref{eq:3point_general}.
For the examples studied it is found that the obtained $3$-point couplings are real if the no-ghost theorem conditions hold. 

\subsection{Summing over the string spectrum}
\label{sec:summing_over_the_string_spectrum}

It will be explored in this section how the sum over all polarisation states can be performed
when the polarisations are tensors. Although the polarisations themselves can be arbitrary,
the sum over all polarisations is governed by completeness relations and all polarisations
are ultimately replaced by projectors. Apart from the completeness relations checking unitarity
requires that one takes into account that elementary particle states are irreducible
representations (irreps) of the little group. In the case of massive particles, these are irreps of
$SO(D-1)$. For later convenience define
\begin{equation}
d \equiv D-1 \, .
\end{equation}
The amount of index contractions that have to be done make it favourable to use the birdtrack notation of \cite{Cvitanovic:2008zz}
to visualise the calculations.

\subsubsection*{Completeness relation for massive tensor states}

The polarisation $\xi$ of a vector particle is a vector and for a massive particle of momentum $k$, it must satisfy $k^\mu \xi_\mu = 0$.
An orthonormal basis of polarisation vectors $\xi_\mu^I, I=1, \ldots, d$ is chosen, where
$\xi_\mu^I$ is in the fundamental representation of $SO(d)$ with respect to the $I$ index.
Summing over the basis of polarisations
yields a completeness relation, which can be considered a projection in the spacetime indices, projecting out the direction of the particle momentum
\begin{equation}
\sum\limits_{I=1}^{d} \xi_\mu^I \xi_\nu^I = \eta_{\mu\nu} - \frac{k_\mu k_\nu}{k^2} \equiv \textbf{P}_{\mu, \nu}^{\perp  k} .
\label{eq:completeness-relation}
\end{equation}

Let us now generalise this to states where the polarisation is in an arbitrary tensor
representation of the little group. The polarisation tensor is orthogonal to the
momentum of the state in all indices, i.e.\ satisfies
$k^{\mu_1} \xi_{\mu_1 \mu_2 \ldots} = k^{\mu_2} \xi_{\mu_1 \mu_2 \ldots} = \ldots = 0$.
A basis for these polarisation tensors can be constructed by taking tensor products of the vector polarisation $\xi_\mu^I$ 
\begin{equation}
\xi_{\mu_1 \mu_2 \ldots} = \xi_{\mu_1}^{I_1}  \xi_{\mu_2}^{I_2} \ldots
\end{equation}
Note that the right-hand side is in general in a representation of $SO(d)$,
but not in an irrep. Using the completeness relation \eqref{eq:completeness-relation}
the sum over polarisations consists of a projector for each index
\begin{equation}
\sum\limits_{I_1,I_2,\ldots=1}^{d} \xi_{\mu_1}^{I_1}  \xi_{\mu_2}^{I_2} \ldots \xi_{\nu_1}^{I_1}  \xi_{\nu_2}^{I_2} \ldots
= \left(\eta_{\mu_1 \nu_1} - \frac{k_{\mu_1} k_{\nu_1}}{k^2}\right) \left(\eta_{\mu_2 \nu_2} - \frac{k_{\mu_2} k_{\nu_2}}{k^2}\right) \ldots 
\equiv  \textbf{P}_{\mu_1, \nu_1}^{\perp  k} \textbf{P}_{\mu_2, \nu_2}^{\perp  k} \ldots .
\label{eq:defPperp}
\end{equation}
Here it seems arbitrary that $\mu_1$ is contracted with $\nu_1$ etc., but this will be taken care of in the next step by introducing the projector to (anti-)symmetrised irreps.

\subsubsection*{Completeness relation for massless vector states}
Since the polarisations of massless states are representations of $SO(D-2)$, their completeness relation is different. The only massless particles in the spectrum of bosonic string theory are vector particles. Let $k$ again be the momentum of the particle and $q$ be another lightlike momentum that is orthogonal to the polarisation
\be
k^\mu \xi_\mu = q^\mu \xi_\mu = k^2 = q^2 = 0 .
\ee
The vector $q$ is a choice of light cone gauge, so that $q\cdot k \neq 0$ must hold. Then there is an orthonormal basis of polarisations $\xi_\mu^i, i=1, \ldots, D-2$ satisfying the completeness relation
\begin{equation}
\sum\limits_{i=1}^{D-2} \xi_\mu^i \xi_\nu^i = \eta_{\mu\nu} - \frac{k_\mu q_\nu + q_\mu k_\nu}{k \cdot q} .
\label{eq:defPmassless}
\end{equation}

\subsubsection{Projecting to irreducible representations}

Elementary particles correspond to irreps of the little group. So far, the sum \eqref{eq:defPperp} includes arbitrary tensors which are null with respect to the momentum $k$. 
Irreps of $SO(d)$ can be obtained from arbitrary tensors by (anti-)symmetrising in the indices and then further decomposing into traceless representations and the remaining traces.
For example, it is a well known fact that $d \times d$ matrices are decomposed into $SO(d)$ irreps by separating the symmetric and antisymmetric parts and then splitting the trace from the symmetric representation.
There are projectors that project onto these three irreps, and they sum up to unity
\bea
\delta_{I_1 J_1}  \delta_{I_2 J_2} =& \left\{ \frac{1}{2} \left( \delta_{I_1 J_1} \delta_{I_2 J_2} + \delta_{I_1 J_2} \delta_{I_2 J_1} \right) - \frac{1}{d} \delta_{I_1 I_2} \delta_{J_1 J_2}\right\} \\
&+ \frac{1}{d} \delta_{I_1 I_2} \delta_{J_1 J_2} + \frac{1}{2} \left( \delta_{I_1 J_1} \delta_{I_2 J_2} - \delta_{I_1 J_2} \delta_{I_2 J_1} \right) .
\eea{eq:matrix_projectors_ex}
In birdtrack notation, the same equation reads
\be
\bt{bt/2lines} = \left\{ \bt{bt/2lines_sym} - \frac{1}{d} \bt{bt/2lines_trace} \right\} + \frac{1}{d} \bt{bt/2lines_trace} + \bt{bt/2lines_asym} ,
\label{eq:birdtrack_matrix_projectors}
\ee
where the following symbols for (anti-)symmetrisation were introduced
\bea
\bt{bt/def_sym} &= \frac{1}{n!} \left\{ \bt{bt/def_0twists} + \bt{bt/def_1twists} + \bt{bt/def_2twists} + \ldots \right\} ,\\
\bt{bt/def_asym} &= \frac{1}{n!} \left\{ \bt{bt/def_0twists} - \bt{bt/def_1twists} + \bt{bt/def_2twists} - \ldots \right\} .
\eea{eq:def_bt_sym_asym}
The projection onto the representation $\alpha$ is denoted by $\textbf{P}_\alpha$. These
projectors to $SO(d)$ irreps are introduced into the sum over polarisations in
\eqref{eq:defPperp} to extract just the contribution of a given irrep when performing
on-shell recursion.
Given that the $\textbf{P}_\alpha$ are just a bunch of Kronecker deltas, each contracting
either two indices belonging to different or to the same polarisation tensor, let us perform 
the sum \eqref{eq:defPperp} with an arbitrary projection inserted
\begin{equation}
\begin{aligned}
& \xi_{\mu_1}^{I_1} \xi_{\mu_2}^{I_2} \ldots   \left( \delta_{I_1 J_1} \delta_{I_2 J_2} \ldots + \delta_{I_1 I_2} \delta_{J_1 J_2}\ldots + \ldots \right)  \xi_{\nu_1}^{J_1} \xi_{\nu_2}^{J_2}  \ldots\\
={} & \textbf{P}_{\mu_1, \nu_1}^{\perp  k} \textbf{P}_{\mu_2, \nu_2}^{\perp  k} \ldots + \textbf{P}_{\mu_1, \mu_2}^{\perp  k} \textbf{P}_{\nu_1, \nu_2}^{\perp  k} \ldots + \ldots \\
={} & \textbf{P}_{\mu_1, \rho_1}^{\perp  k} \textbf{P}_{\mu_2, \rho_2}^{\perp  k} \ldots \left(  \eta^{\rho_1 \sigma_1}  \eta^{\rho_2 \sigma_2} \ldots + \eta^{\rho_1 \rho_2} \eta^{\sigma_1 \sigma_2} \ldots + \ldots   \right) \textbf{P}_{\sigma_1, \nu_1}^{\perp  k} \textbf{P}_{\sigma_2, \nu_2}^{\perp  k} \ldots .
\end{aligned}
\end{equation}
In the last step is was used that the $\textbf{P}^{\perp  k}$ are idempotent.
This shows that this operation can be written diagrammatically as
\be
{\mathord{\vcenter{\hbox{\scalebox{0.3}{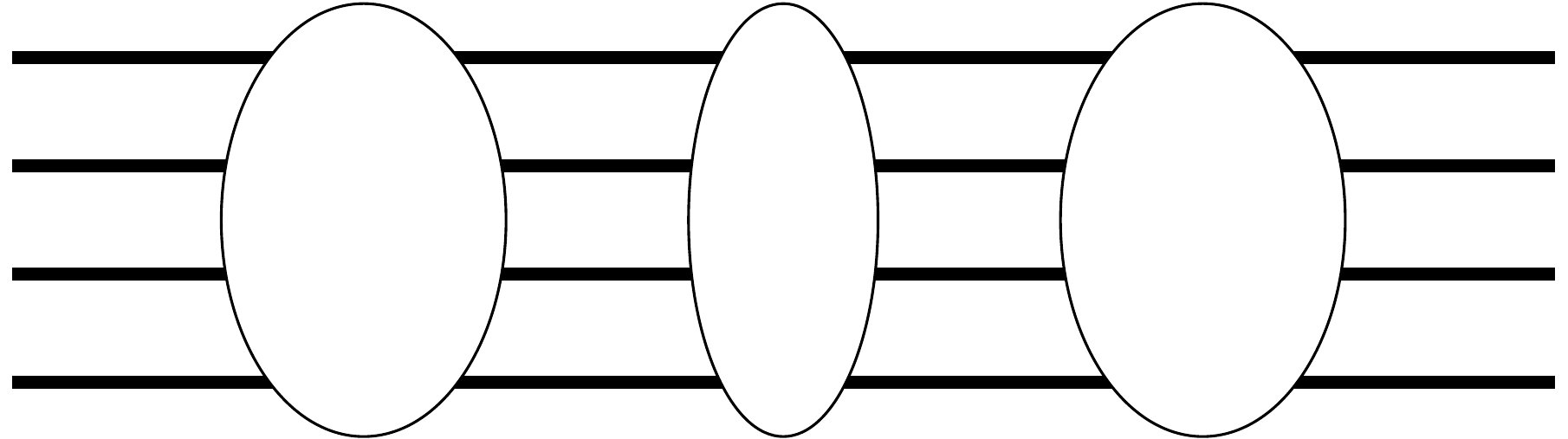}}}}} ,
\label{eq:projector}
\ee
where $\textbf{P}_\alpha$ is the projection onto the $SO(d)$ representation $\alpha$
with each $\delta_{IJ}$ replaced by an $\eta^{\mu \nu}$ and every line is a contraction of spacetime indices.
This is now the recipe to perform the sum over polarisations in string theory:
Sum over all irreps and replace the polarisation tensors of both amplitudes by the combination of projectors \eqref{eq:projector}.
Which irreps appear at which mass level can be elegantly computed using a generating function that was found in \cite{Hanany:2010da}. To use this effectively, a few more facts about the irreps of $SO(n)$ will be needed.

\subsubsection{The covariant string spectrum}

The  index symmetries that can appear in an irreducible $k$-tensor representation of $SO(n)$ are labelled by $k$-box Young tableaux with not more than $\lfloor \frac{n}{2} \rfloor$ rows.
Horizontally aligned boxes correspond to symmetrised indices and vertically aligned boxes correspond to antisymmetrisation.
The limitation to $\lfloor \frac{n}{2} \rfloor$ rows stems from the fact that there is no antisymmetric tensor with more than $n$ indices if the indices run from 1 to $n$,
and that any $m$-box column with $m>\lfloor \frac{n}{2} \rfloor$ can be transformed into a  $(n-m)$-box column by contraction with the fully antisymmetric Levi-Civita tensor.
A Young diagram can be described by its Dynkin label $\alpha = [\alpha_1,\alpha_2,\ldots,\alpha_{\lfloor \frac{n}{2} \rfloor}]_n$, which lists the numbers $\alpha_i$ of columns with $i$ boxes. 
An (anti-) symmetrised tensor can be decomposed into $SO(n)$ irreps by separating a
traceless irrep and further irreps containing traces. The traceless
irreps will be labelled by the Dynkin label or Young diagram.
The number of indices on a tensor in representation  $\alpha$ will be denoted by $|\alpha|$. All bosonic string states are traceless irreps, for which
\be
|\alpha| = \sum\limits_{i} i \alpha_i ,
\ee
holds. All this, including the construction of the projectors $\textbf{P}_\alpha$ using birdtrack notation, is described in detail in \cite{Cvitanovic:2008zz}.

To get back to the previous example the following Young diagrams will be needed
\be
\textbf{P}_{\bts{bt/young_2}} = \left\{ \bt{bt/2lines_sym} - \frac{1}{d} \bt{bt/2lines_trace} \right\}, \qquad
\textbf{P}_{\bts{bt/young_01}} = \bt{bt/2lines_asym}.
\ee
These correspond to the traceless projectors in \eqref{eq:birdtrack_matrix_projectors}.

After having introduced the concept of labelling $SO(n)$ irreps by Dynkin labels, the bosonic string spectrum up to mass level $A = 5$, as given in \cite{Hanany:2010da} reads
\bea
Z_{Bosonic} ={}& \frac{1}{q} + [1,0,\ldots,0]_{24}+ [2,0,\ldots,0]_{25}q \\
&+ ([3,0,\ldots,0]_{25}+[0,1,0,\ldots,0]_{25})q^2 \\
&{}+([4,0,\ldots,0]_{25}+[2,0,\ldots,0]_{25}+[1,1,0,\ldots,0]_{25}+1)q^3 \\ 
&{}+([5,0,\ldots,0]_{25}+[3,0,\ldots,0]_{25}+[2,1,0,\ldots,0]_{25} \\
&{}+[1,1,0,\ldots,0]_{25}+[1,0,\ldots,0]_{25}+[0,1,0,\ldots,0]_{25})q^4 + O(q^5).
\eea{partition_function}
The exponent of $q$ indicates the mass $\alpha'm^2 = A-1$ of a state.

\subsection{From Koba-Nielsen to arbitrary $3$-point amplitudes and back}
\label{sec:3point_vs_tachyons}

In this section  $3$-point amplitudes will be glued together to compare the result with the known residues of higher point tachyon amplitudes.
Fortunately, the three point amplitudes with one or two massive legs can be predicted easily up to a few constants by physical considerations, without the need to do any string theory computation.
One can then check that the ansatz matches the known result and unambiguously compute the missing coefficients. Various consistency checks will be performed: The same coefficients must appear when the same $3$-point amplitude is part of a different (higher point) tachyon amplitude and the coefficients must be real, which is required for unitarity.

\subsubsection{One tensor, two tachyons}
\label{sec:unit_ttm}
Following \cite{Boels:2012if}, consider the amplitude of two tachyons with momenta $k_1, k_2$ and a massive particle with momentum $k_1+k_2$.
It is clear that $k_1 - k_2$ is the only possible term that can be contracted to the polarisation of the massive particle,
or equivalently, survives the projection by $P^{\perp}_{k_1+k_2}$ (which can in this case also be \eqref{eq:defPmassless})
\be
{\mathord{\vcenter{\hbox{\scalebox{0.3}{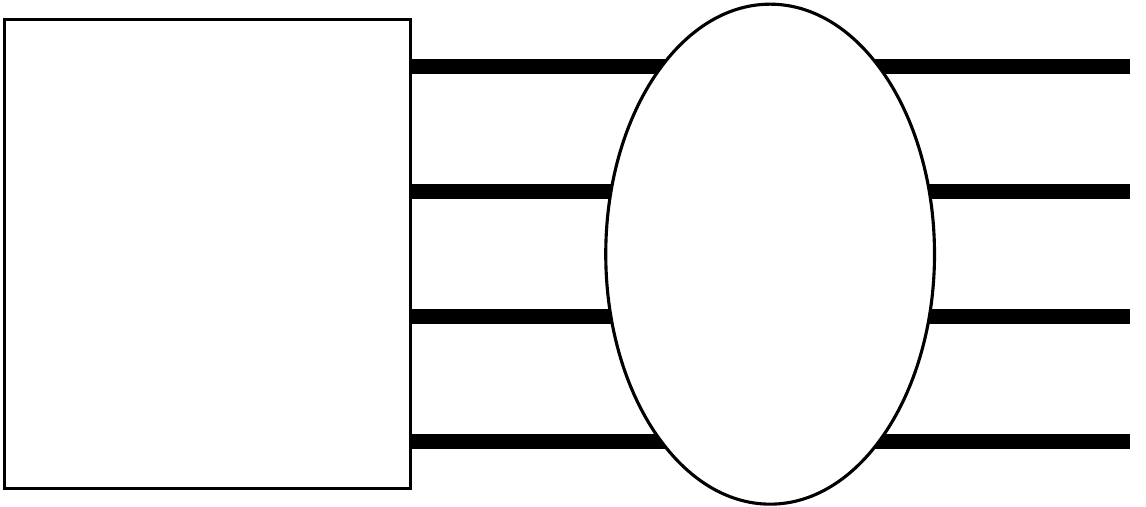}}}}} = {\mathord{\vcenter{\hbox{\scalebox{0.3}{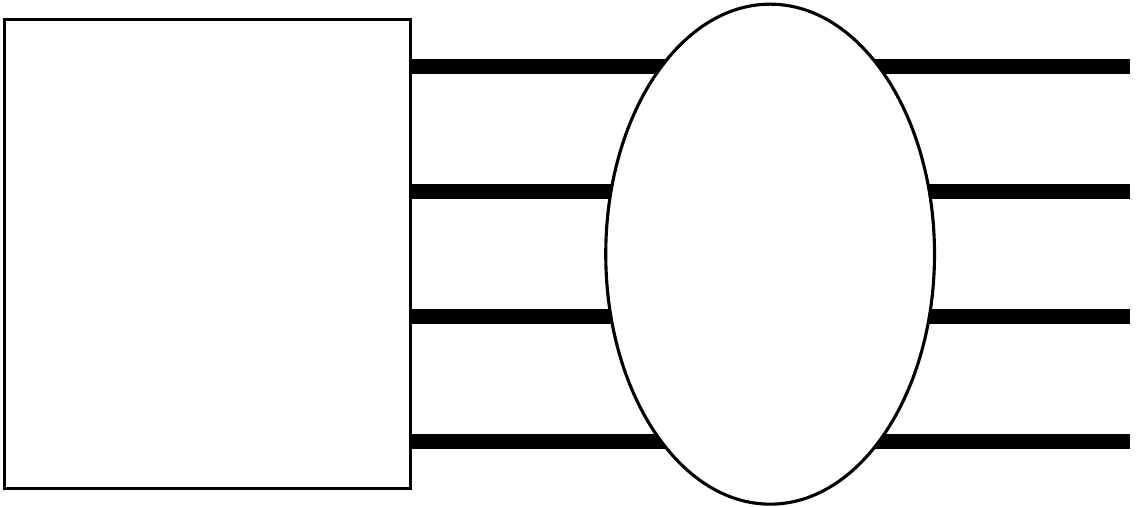}}}}} = {\mathord{\vcenter{\hbox{\scalebox{0.3}{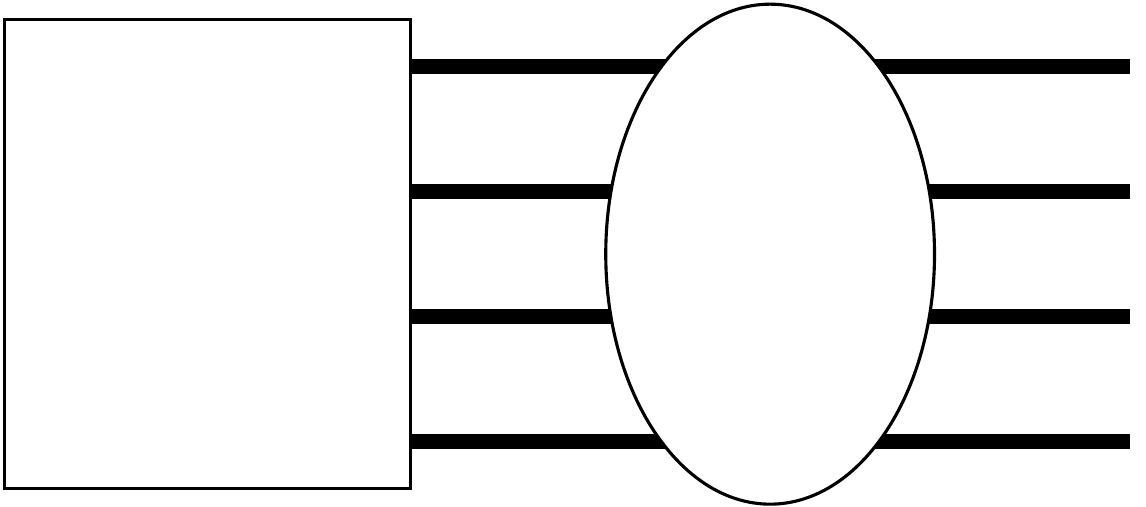}}}}} = {\mathord{\vcenter{\hbox{\scalebox{0.3}{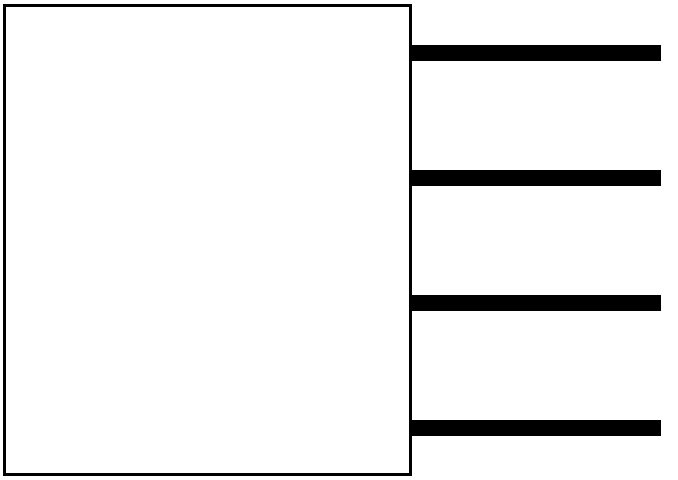}}}}}.
\label{TTM_Pperp}
\ee
The amplitudes have to be this term contracted into the polarisation, times a coefficient $c_{A, \alpha}$,
which depends on the mass level $A$ and irrep $\alpha$ of the massive particle. A circle is drawn to denote
the amplitude including this coefficient, but without the polarisation tensor
\be
 {\mathord{\vcenter{\hbox{\scalebox{0.3}{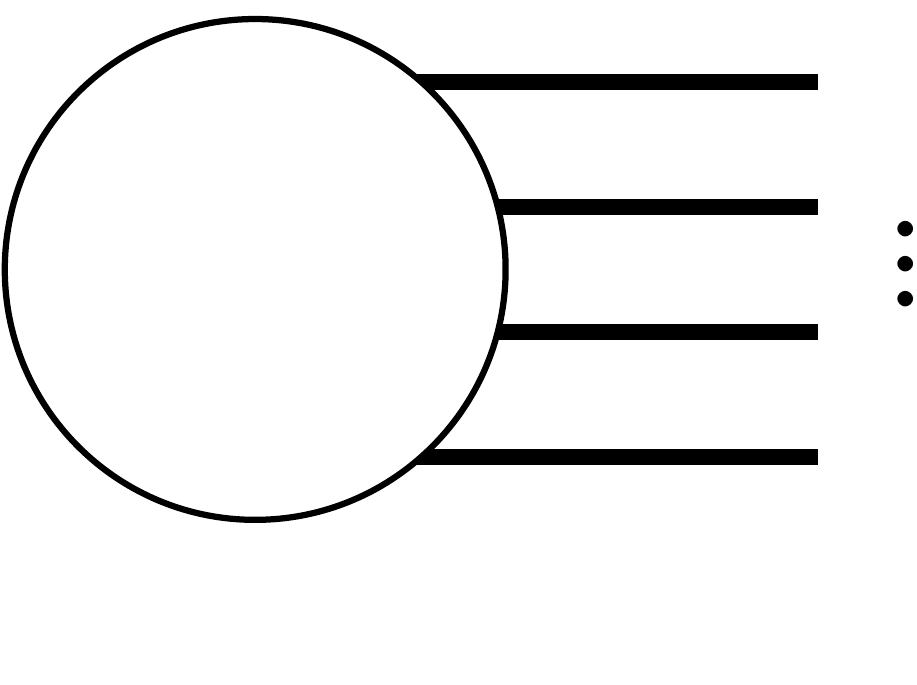}}}}} \quad  = c_{A,\alpha}  \prod\limits_{a=1}^{|\alpha|} \sqrt{\frac{\alpha'}{2}} (k_1-k_2)^{\mu_a} .
\label{eq:def_ttm_blob}
\ee
Note that here and in the following four lines in a birdtrack diagram are meant to represent an arbitrary number
(in this case $|\alpha|$) of lines.
The corresponding amplitude is obtained by contracting with a polarisation tensor $\xi^{\alpha}$ from the respective irrep
\be
 \Amp(T_1,T_2,M^{A,\alpha}) = {\mathord{\vcenter{\hbox{\scalebox{0.3}{\input{bt/A_TTM_diagram.pdf_t}}}}}} \quad \xi^{\alpha}_{\mu_1 \ldots \mu_{|\alpha|}}.
\label{eq:A_TTM}
\ee
Since all indices are contracted to the same expression $(k_1 - k_2)$, only fully symmetric representations couple to two tachyons
\be
 {\mathord{\vcenter{\hbox{\scalebox{0.3}{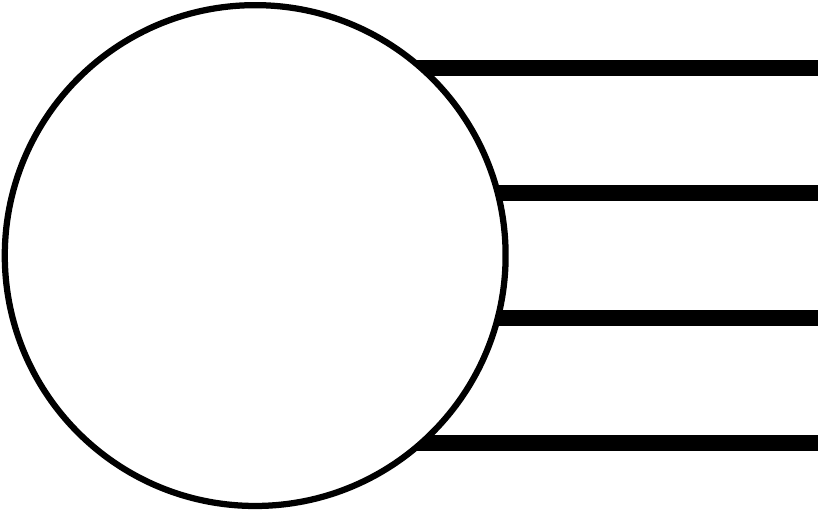}}}}} = {\mathord{\vcenter{\hbox{\scalebox{0.3}{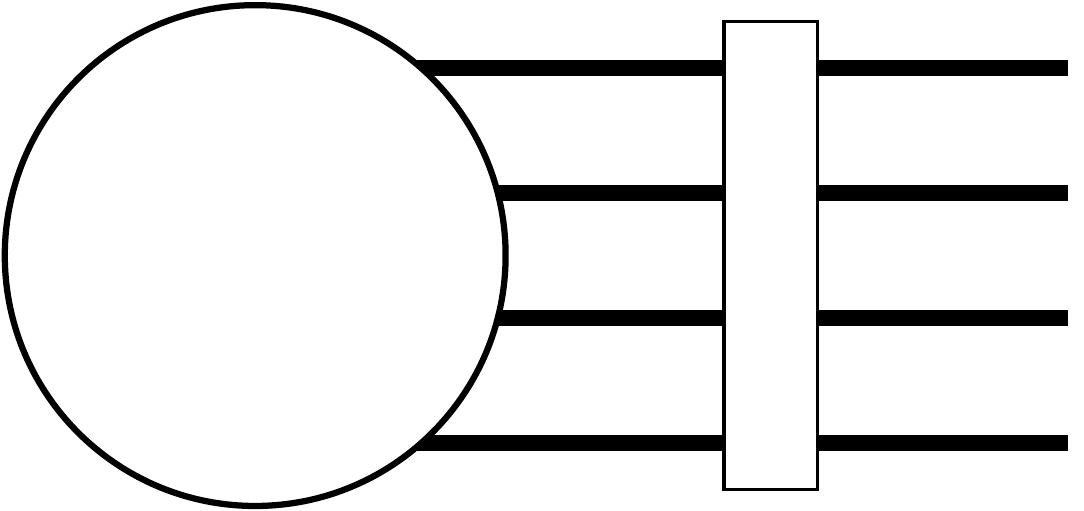}}}}}.
\label{TTM_sym}
\ee
Note this is only a subset of the full spectrum of string theory, cf. equation \eqref{partition_function}.
The residues of the $4$-point tachyon amplitude can be calculated from the $3$-point amplitudes with one massive leg. They are
\begin{equation}
-\Res_{s_{12}\to A-1} \Amp_4 = \sum\limits_{\alpha} {\mathord{\vcenter{\hbox{\scalebox{0.3}{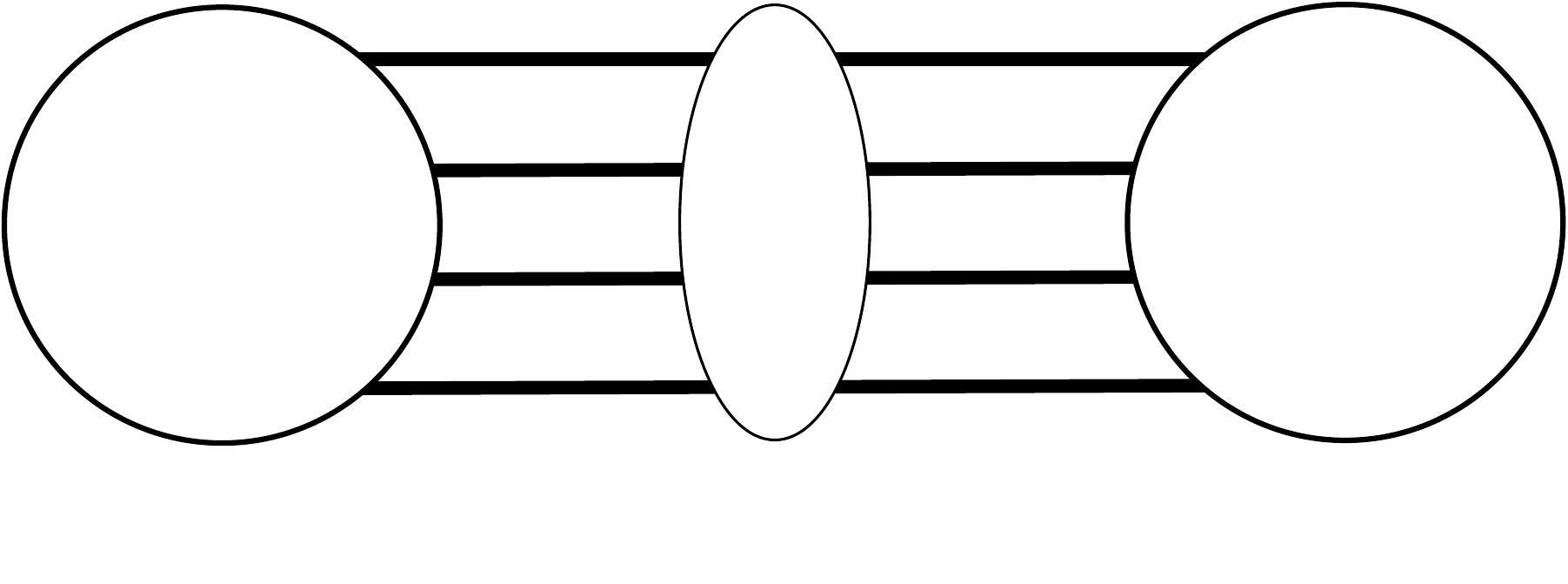}}}}}.
\label{A4fact}
\end{equation}
The projectors $P^{\perp}_{k_1+k_2}$ were left out, because they are annihilated by the $3$-point amplitudes \eqref{TTM_Pperp}, and only fully symmetric representations appear in the sum because of \eqref{TTM_sym}. 
Whenever such contractions are performed, momentum conservation and the on-shell conditions have to be used to write the result in terms of the remaining $\frac{(N-2)(N-3)}{2}$ independent kinematic variables $k_{ij}$ with $1<i<j<N$. This procedure is briefly described in Appendix \ref{sec:on-shell_space}.

As recalled above, one way to prove unitarity of the Veneziano amplitude is to show that all the $c_{A,\alpha}$ are real
and more generally, real coefficients in all $3$-point amplitudes imply unitarity of the complete S-matrix. 
It is not obvious from \eqref{A4fact} that the $c_{A,\alpha}$ are real since the coefficients always appear in this formula as squares $c_{A,\alpha}^2$.
By matching the right-hand side of \eqref{A4fact} to the known residue of the $4$-point tachyon amplitude, the coefficients $c_{A,\alpha}$ can be calculated. While straightforward for low levels, this computation gets harder for the general case. This computation can be found in Appendix \ref{sec:one_massive_coeffs} with the result
\begin{align}
c_{A,|\alpha|}^2 = \begin{cases}
{\displaystyle \sum\limits_{l=0}^{\frac{A-|\alpha|}{2}} V_{\frac{A-|\alpha|}{2}-l,A} \left(\frac{A+3}{4}  \right)^{2l} \frac{(|\alpha|+1)^{(2l)}}{l!(\frac{d}{2}+|\alpha|)^{(l)} } } \qquad &A-|\alpha|\ \text{even} , \\[1em]
0 \qquad &A-|\alpha|\ \text{odd} ,
\end{cases}
\label{eq:TTM_couplingsII}
\end{align}
where $V_{k, A\, \text{even}}$ and $V_{k, A\, \text{odd}}$ are essentially the central factorial numbers $t(A,k)$ and $t_2(A,k)$, (sequences A008955 and A008956 in the Online Encyclopedia of Integer Sequences, \cite{int:seq}), 
\begin{align}
V_{k, A\, \text{even}} &=\frac{(-1)^k}{A! 4^k} t_2\left(\frac{A}{2}, k\right) , &0 \leq k \leq \lfloor \frac{A}{2} \rfloor ,\\
V_{k, A\, \text{odd}} &= \frac{(-1)^k}{A!} t\left(\frac{A-1}{2}, k\right) , &0 \leq k \leq \lfloor \frac{A}{2} \rfloor .
\end{align}
The zero in equation \eqref{eq:TTM_couplingsII} follows from the monodromy relation \eqref{eq:monodfund} for three point amplitudes,
which makes them symmetric or anti-symmetric under interchange of the two tachyon legs depending on whether $A$ is even or odd.
As \eqref{eq:def_ttm_blob} shows this is only the case for $|\alpha|$ even/odd.
 In order to prove tree-level unitarity of the Veneziano amplitude, the question is 
\begin{equation}
c_{A,|\alpha|}^2 \stackrel{{\bf \large ?}}
     {\geq}  0, \qquad \forall A, |\alpha|. \qquad \textrm{to be shown} 
\end{equation}
Despite having formula \eqref{eq:TTM_couplingsII}, this is not straightforward since the $V$ contains an alternating sign and the central factorial numbers complicate the issue.
Explicit checks for $d=25$ and all states up to $A = 400$ show that the squared couplings are positive.

\subsubsection{The no-ghost theorem conditions}
\label{sec:no-gost_conditions}

Before continuing to two massive legs, it will be shown that the techniques which led to \eqref{A4fact} can be used to (re)derive the no-ghost theorem conditions. For this, consider the Veneziano amplitude for arbitrary `intercept' $\alpha_0$,
\begin{equation}
\Amp_4 = \frac{\Gamma(-s_{12} - \alpha_0)\Gamma(-s_{23} - \alpha_0)}{\Gamma(-s_{12} -s_{23} - 2\alpha_0)} ,
\label{veneziano}
\end{equation}
and arbitrary dimension $D$. Above $\alpha_0$ was always assumed to be 1. The intercept $\alpha_0$ appears in the residues
\begin{align}
\lim\limits_{s_{12} \rightarrow A - \alpha_0}{\Amp_4} &= \frac{1}{A!} \frac{-1}{s_{12}-A+\alpha_0}
\prod\limits_{i=1}^A(s_{23} +  \alpha_0 + i), \quad A \in {\mathbb N}_0 . 
\label{venezianoResidues}
\end{align}
In the case $A=0$ a scalar particle with minimal mass $\alpha' m^2 = - \alpha_0$ is exchanged. It will be assumed this is the same as the external particle. Therefore the external tachyons also have this mass.

The left-hand side of equation \eqref{A4fact} becomes with $A=1$
\be
-\Res_{s_{12}\to 1-\alpha_0} \Amp_4 = s_{23} + \alpha_0 + 1 .
\label{relation_intercept_veneziano}
\ee
For the right-hand side of equation \eqref{A4fact} a vector and a scalar particle have to be considered in this case
\bea
& \lim\limits_{s_{12} \to 1-\alpha_0} {\mathord{\vcenter{\hbox{\scalebox{0.3}{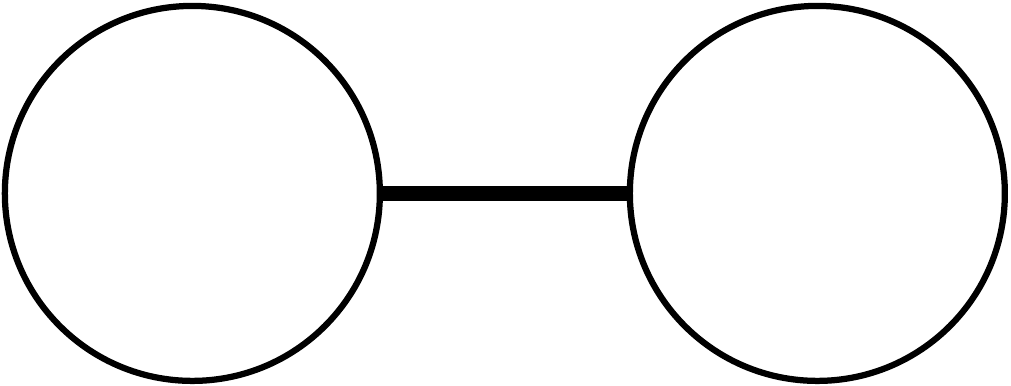}}}}} + c_{1,\bullet}^2 \\
={}& \lim\limits_{s_{12} \to 1-\alpha_0} c_{1,\bts{bt/young_1}}^2 \frac{\alpha'}{2} (k_1 - k_2)\cdot(k_3 - k_4) +  c_{1,\bullet}^2 \\
={}& c_{1,\bts{bt/young_1}}^2 \left( s_{23} +\frac{3}{2} \alpha_0 + \frac{1}{2} \right)+  c_{1,\bullet}^2 .
\eea{relation_intercept_onshell}
Matching up \eqref{relation_intercept_veneziano} and \eqref{relation_intercept_onshell}, it is seen that the overall coefficient is $c_{1,\bts{bt/young_1}}^2 = 1$ and that the intercept is fixed at
\be
\alpha_0 = 1 - 2 c_{1,\bullet}^2 .
\ee
Unitarity requires all couplings $c$ to be real
which implies one of the conditions of the no-ghost theorem
\be
\alpha_0 \leq 1 .
\ee
For the monodromy relations as written in equation \eqref{eq:monod1} to hold $\alpha_0=1$ is required. For more generic $\alpha_0$ the monodromy relations could be modified, see \cite{Plahte:1970wy}.

For the next mass level start by reading off the residue of the Veneziano amplitude \eqref{veneziano}
\be
-\Res_{s_{12}\to 1} \Amp_4 = \frac{1}{2} \left( s_{23}^2 + 5 s_{23} + 6 \right) .
\label{relation_D26_veneziano}
\ee
Note that this result was also obtained above using monodromy relations. The right-hand side of \eqref{A4fact} yields
\bea
& \lim\limits_{s_{12} \to 1} 
\left\{ {\mathord{\vcenter{\hbox{\scalebox{0.3}{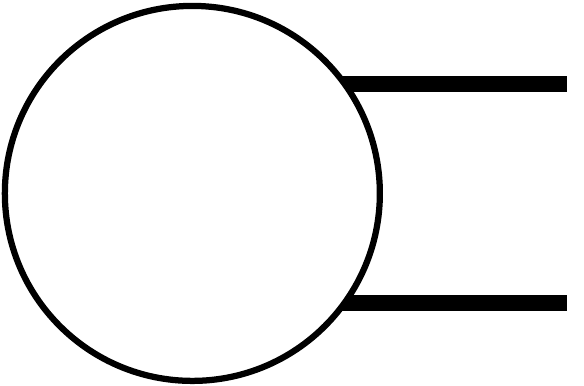}}}}} \left(\bt{bt/2lines_sym} - \frac{1}{D-1} \bt{bt/2lines_trace} \right)  {\mathord{\vcenter{\hbox{\scalebox{0.3}{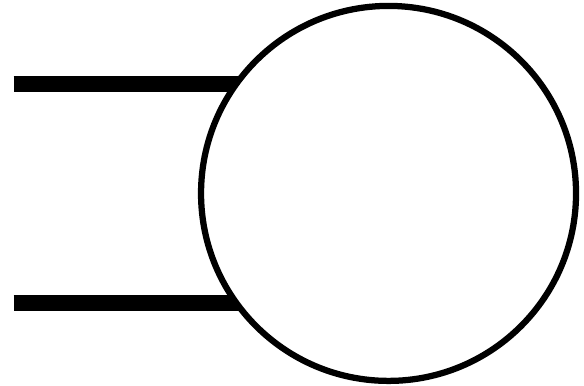}}}}} + {\mathord{\vcenter{\hbox{\scalebox{0.3}{\input{bt/ex21_1234.pdf_t}}}}}} + c_{2,\bullet}^2 \right\} \\
={}& \lim\limits_{s_{12} \to 1} \Biggr\{
 c_{2,\bts{bt/young_2}}^2 \left(\frac{\alpha'}{2}\right)^2 
\left( \left\{ (k_1 - k_2)\cdot(k_3 - k_4)\right\}^2 - \frac{1}{D-1} (k_1-k_2)^2 (k_3-k_4)^2 \right)  \\
& \qquad \quad + c_{1,\bts{bt/young_1}}^2 \frac{\alpha'}{2} (k_1 - k_2)\cdot(k_3 - k_4) +  c_{1,\bullet}^2 \Biggr\} \\
={}& c_{2,\bts{bt/young_2}}^2 \left(s_{23}^2 + 5 s_{23} + \frac{25}{4} - \frac{25}{4} \frac{1}{D-1}  \right)+ c_{2,\bts{bt/young_1}}^2 \left(s_{23} +\frac{5}{2} \right) + c_{2,\bullet}^2 .
\eea{relation_D26_onshell}
This time \eqref{relation_D26_veneziano} and \eqref{relation_D26_onshell} agree for
\be
c_{2,\bts{bt/young_2}}^2 = \frac{1}{2}, \qquad c_{2,\bts{bt/young_1}}^2 = 0, \qquad c_{2,\bullet}^2 = \frac{1}{2} \left( 6 - \frac{25}{4} + \frac{25}{4} \frac{1}{D-1} \right) = \frac{26-D}{8(D-1)} .
\ee
Only for $D=26$ the symmetric traceless $2$-tensor is the only particle appearing, as stated in \eqref{partition_function}.
For $D < 26$ unitarity requires a scalar particle at this mass level\footnote{known as a Brower state \cite{Brower:1972wj} in non-critical string theory }, while for $D > 26$ the required coupling $c_{2,\bullet}$ becomes imaginary, which conflicts with unitarity of the S-matrix. So by unitarity there is an upper bound for $D$ that agrees with the result known from the no-go theorem
\be
D \leq 26 .
\ee
This sub-subsection contains a direct derivation of the dimension and intercept bounds of the no-ghost theorem from the Veneziano amplitude using nothing but locality, unitarity and Poincar\'e invariance. Closest to this in the literature as far as we are aware comes a derivation in \cite{Frampton:1972st} which does still use some worldsheet input about the spectrum.

\subsubsection{Two tensors, one tachyon}
Now consider the amplitude $\Amp(M^{A,\alpha},T,M^{B,\beta})$ of one tachyon and two massive particles on mass levels $A,B$ with polarisations $\xi^\alpha , \xi^\beta$ in the irreducible representations $\alpha, \beta$.
Here $k_A \cdot \xi^\alpha = 0$ holds as well as  $k_B \cdot \xi^\alpha = (-k_A - k_T) \cdot \xi^\alpha = - k_T \cdot \xi^\alpha$ due to momentum conservation. An analogous result holds for the other polarisation. So it is enough to consider 
each polarisation to only be contracted with the tachyon momentum $k_T$ or with the other polarisation.
The index $q$ labels the number of contractions of the two polarisation tensors with each other in a given term. For the couplings introduce an unknown coefficient $c_{A,B,\alpha,\beta,q}$. Since the calculation of the
coefficients was already cumbersome in the case of $3$-point amplitudes with one massive leg, a general computation of the coefficients $c_{A,B,\alpha,\beta,q}$ will not be attempted here. Define
\be
 {\mathord{\vcenter{\hbox{\scalebox{0.3}{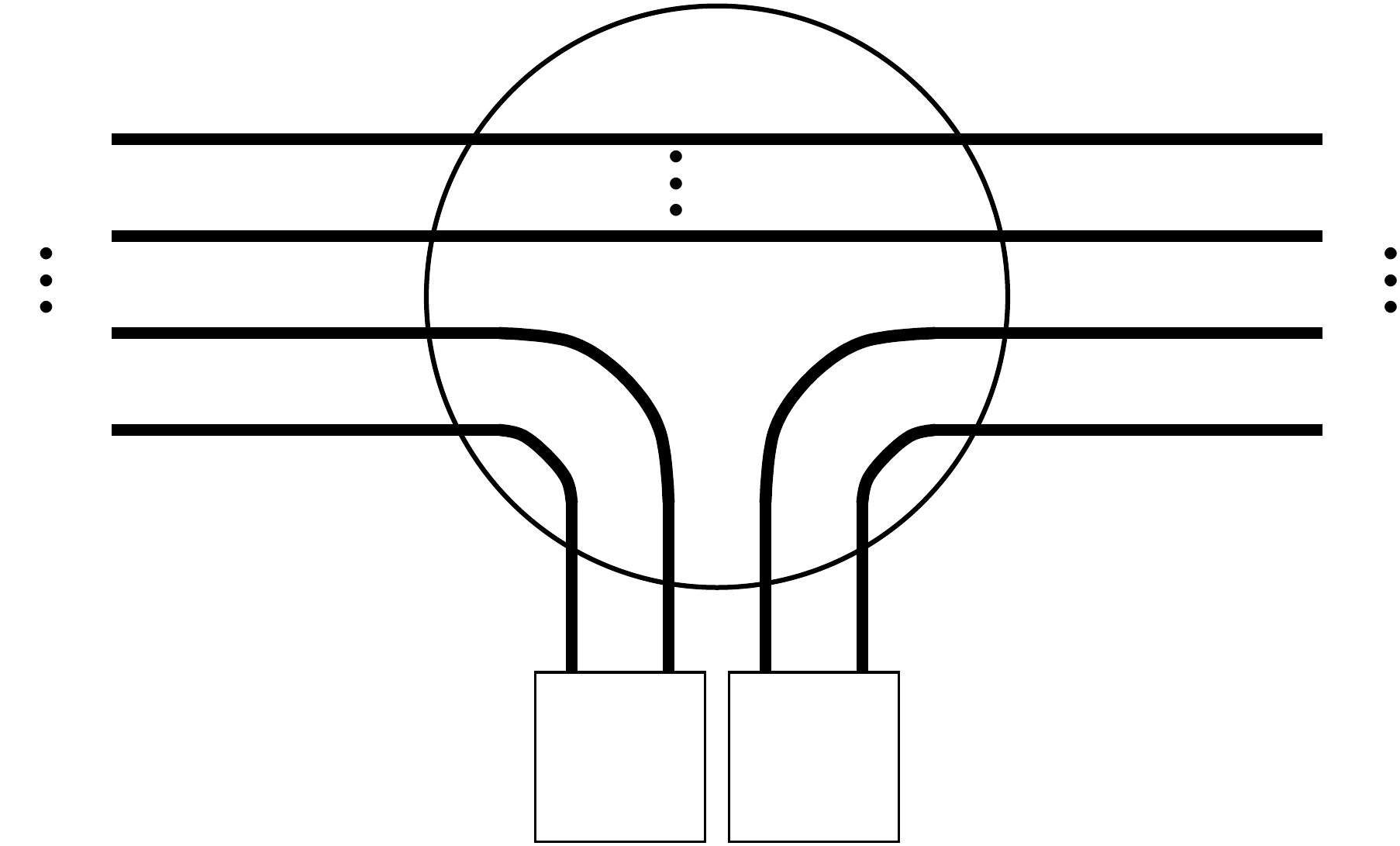}}}}} \quad  = c_{A,B,\alpha,\beta,q} \prod\limits_{a=1}^q \eta^{\mu_a \nu_a} \prod\limits_{b=q+1}^{|\alpha|} k^{\mu_b}  \prod\limits_{c=q+1}^{|\beta|} - k^{\nu_c}.
\ee
The amplitude is then
\be
 \Amp(M^{A,\alpha},T,M^{B,\beta}) = \sum\limits_{q = 0}^{\min(|\alpha|,|\beta|)} \xi^{\alpha}_{\mu_1 \ldots \mu_{|\alpha|}} \quad {\mathord{\vcenter{\hbox{\scalebox{0.3}{\input{bt/A_TMM_diagram.pdf_t}}}}}} \quad \xi^{\beta}_{\nu_1 \ldots \nu_{|\beta|}}.
\ee
This amplitude appears first in the residue of the $5$-point tachyon amplitude
\bea
&\Res_{s_{12} \to A-1} \Res_{s_{45}\to B-1} \Amp_5 \\
={}& \sum\limits_{\alpha,\beta}    \sum\limits_{q = 0}^{\min(|\alpha|,|\beta|)} {\mathord{\vcenter{\hbox{\scalebox{0.3}{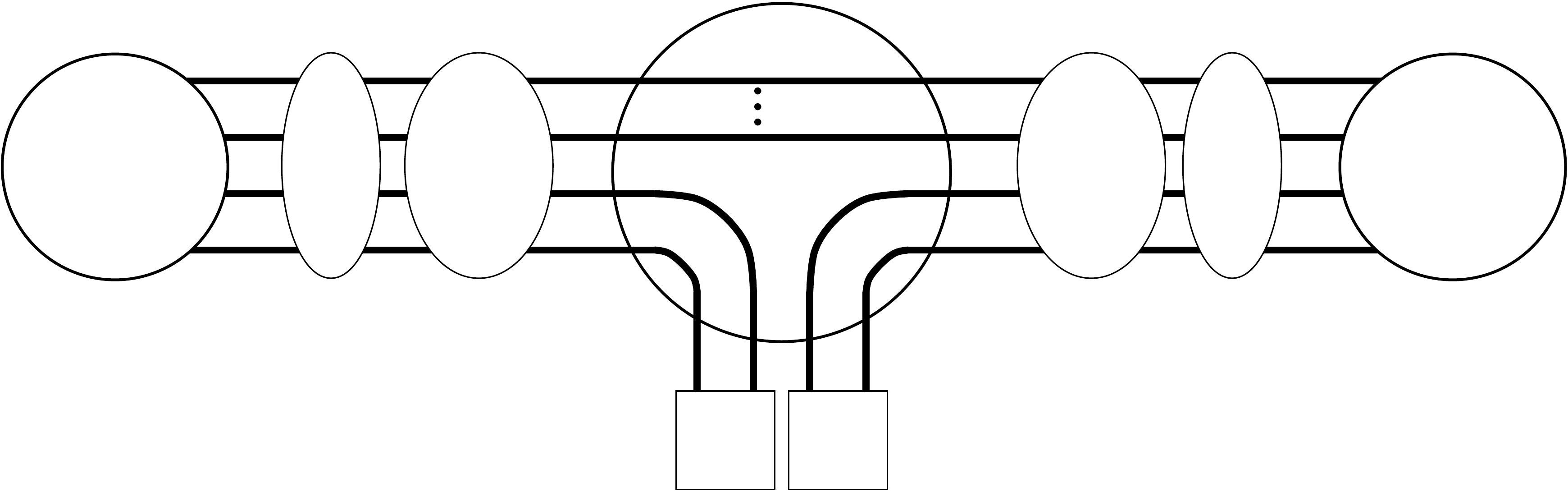}}}}}.
\eea{eq:5point_contraction}
Since both other $3$-point amplitudes that appear have only one massive leg, again only symmetric representations appear.
This suggests that in terms of unitarity the $5$-point amplitude does not add anything to the story
of the $4$-point amplitude. And indeed, assuming $c_{A,\alpha} \in \mathbb{R}$, everything in \eqref{eq:5point_contraction}
is real and the right-hand side linear in the $c_{A,B,\alpha,\beta,q}$. This implies $c_{A,B,\alpha,\beta,q} \in \mathbb{R}$ and thus unitarity of the $5$-point amplitude follows trivially from
unitarity of the $4$-point amplitude.

A nice consistency check is the case $A=B=2$. It is known from \eqref{partition_function} that at this mass level only one irrep appears if $D=26$, the symmetric traceless matrices.
This example can neatly be written explicitly in birdtracks, which ensures the notation is clear.
\bea
&\Res_{s_{12} \to 1} \Res_{s_{45}\to 1} \Amp_5 \\
={}& {\mathord{\vcenter{\hbox{\scalebox{0.3}{\input{bt/ex22_A12.pdf_t}}}}}} \left(\bt{bt/2lines_sym} - \frac{1}{d} \bt{bt/2lines_trace} \right) \\
& \cdot \left(\bt{bt/2lines} +\frac{1}{2} \raisebox{-3pt}{${\mathord{\vcenter{\hbox{\scalebox{0.3}{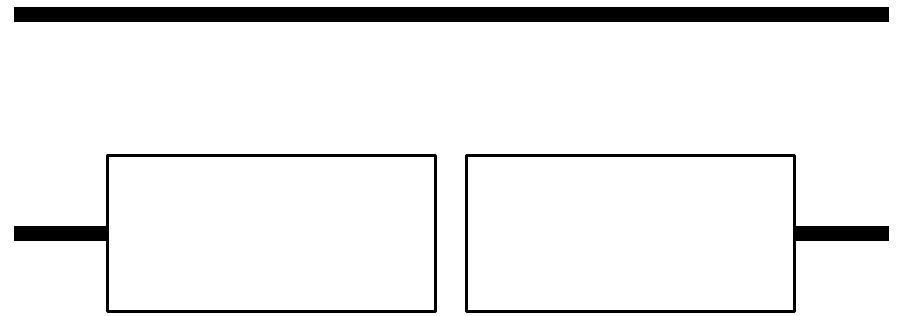}}}}}$} +\frac{1}{2} \raisebox{3pt}{${\mathord{\vcenter{\hbox{\scalebox{0.3}{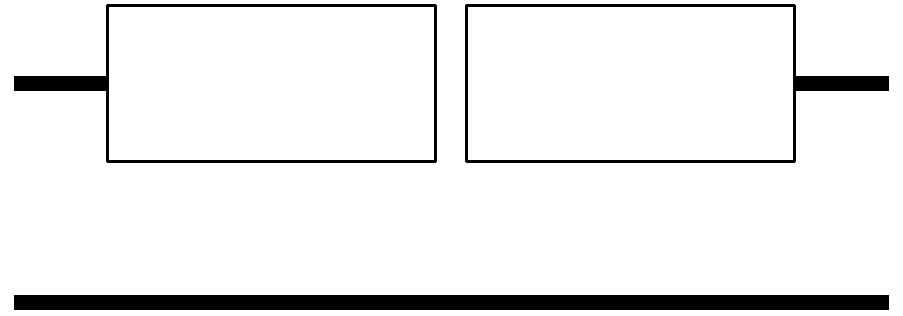}}}}}$} +\frac{1}{4} {\mathord{\vcenter{\hbox{\scalebox{0.3}{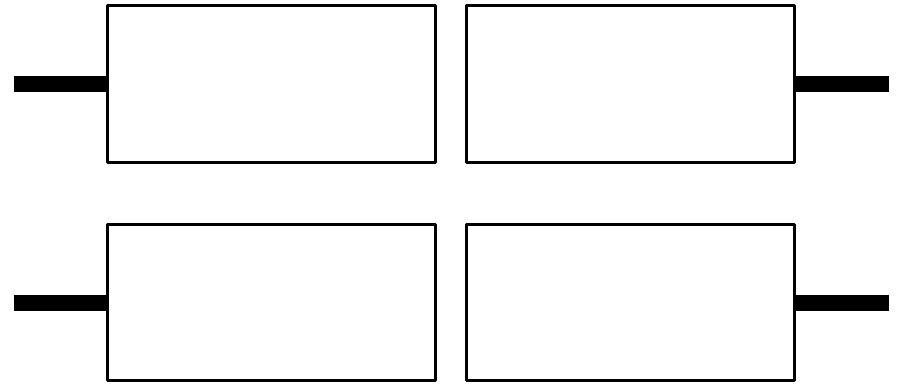}}}}}\right)\\
& \raisebox{-16pt}{$\cdot \left({\mathord{\vcenter{\hbox{\scalebox{0.3}{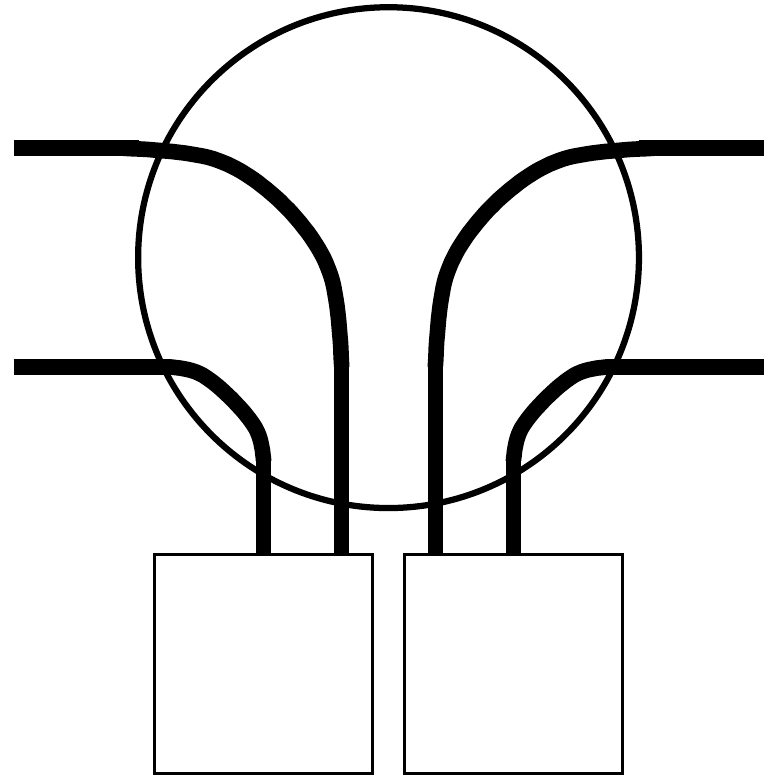}}}}} + {\mathord{\vcenter{\hbox{\scalebox{0.3}{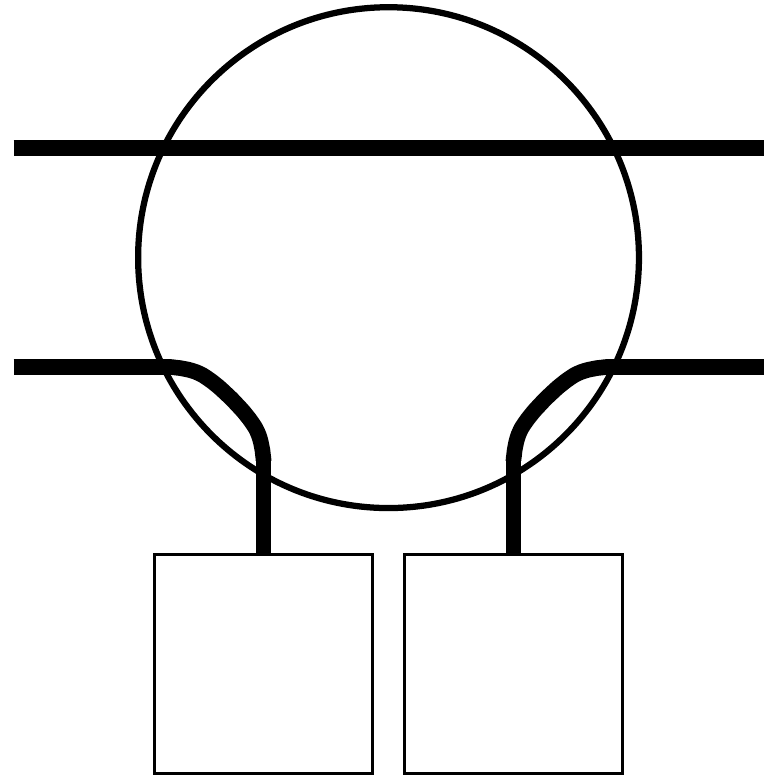}}}}} + \raisebox{11pt}{${\mathord{\vcenter{\hbox{\scalebox{0.3}{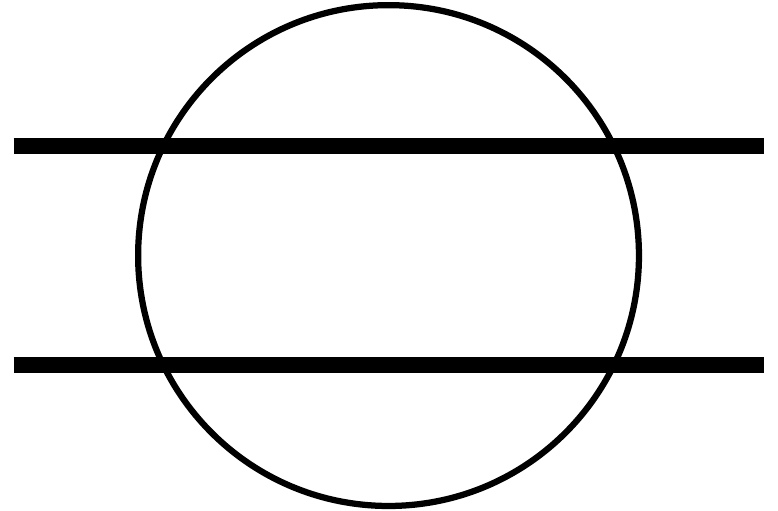}}}}}$} \right)$}\\
& \cdot \left(\bt{bt/2lines} +\frac{1}{2} \raisebox{-3pt}{${\mathord{\vcenter{\hbox{\scalebox{0.3}{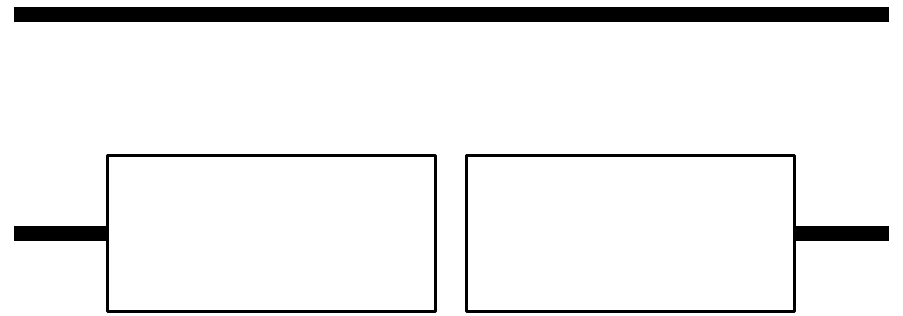}}}}}$} +\frac{1}{2} \raisebox{3pt}{${\mathord{\vcenter{\hbox{\scalebox{0.3}{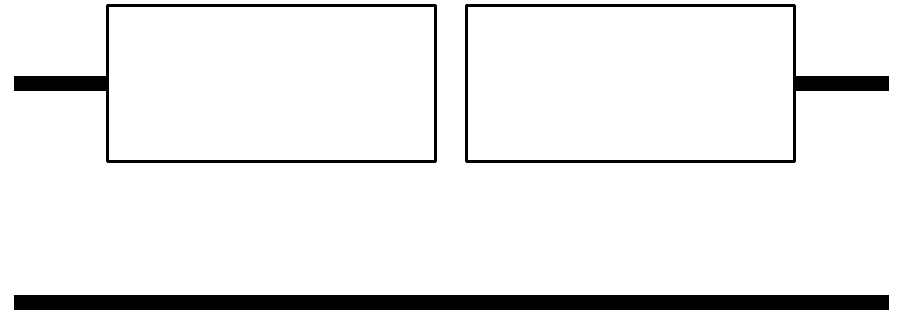}}}}}$} +\frac{1}{4} {\mathord{\vcenter{\hbox{\scalebox{0.3}{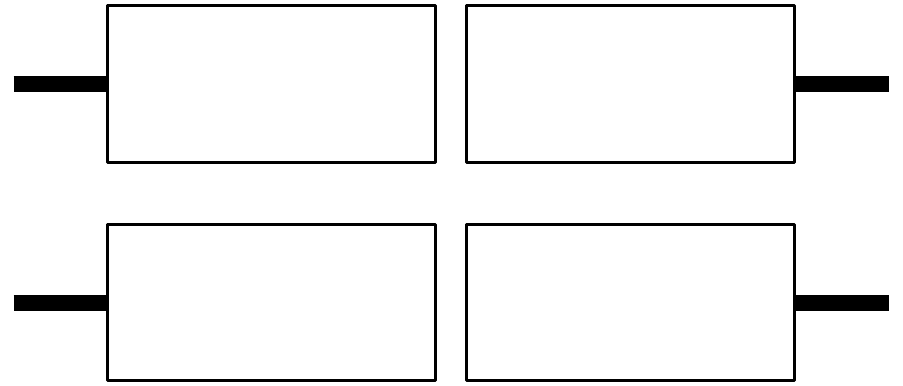}}}}}\right)\\
& \cdot \left(\bt{bt/2lines_sym} - \frac{1}{d} \bt{bt/2lines_trace} \right) {\mathord{\vcenter{\hbox{\scalebox{0.3}{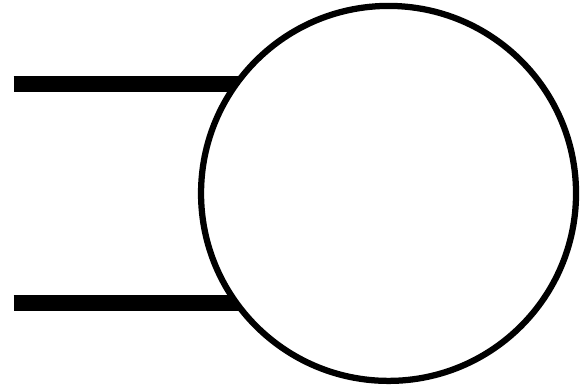}}}}} \;.
\eea{eq:ex22}
The brackets of the projector \eqref{eq:defPperp} were expanded. This has to be compared to the known result
\be
\sum\limits_{a=0}^{2} (-1)^a \binom{k_{24}}{a} \binom{k_{23}}{2-a} \binom{k_{34}}{2-a} = \frac{1}{2} k_{24}^2 - k_{24} k_{23} k_{34} + \frac{1}{4} k_{23}^2 k_{34}^2 + \ldots \;,
\label{eq:A5result}
\ee
where only the highest order terms were written on the right-hand side.
Since $c_{2,\bts{bt/young_2}}$ can be computed from the residue of the $4$-point amplitude \eqref{A4fact}, the factorised expression \eqref{eq:ex22} is determined up to the three coefficients
 $c_{2,2,\bts{bt/young_2},\bts{bt/young_2},0}$, $c_{2,2,\bts{bt/young_2},\bts{bt/young_2},1}$ and $c_{2,2,\bts{bt/young_2},\bts{bt/young_2},2}$  that appear in the central bracket.
The term with $q=0$ is the only one in \eqref{eq:ex22} containing $k_{23}^2 k_{34}^2$ and similarly $k_{24} k_{23} k_{34}$ appears only in the $q=1$ and $k_{24}^2$ only in the $q=2$ term.
This fixes the coefficients unambiguously. With $c_{2,2,\bts{bt/young_2},\bts{bt/young_2},2} = c_{2,2,\bts{bt/young_2},\bts{bt/young_2},1} = c_{2,2,\bts{bt/young_2},\bts{bt/young_2},0} = \frac{1}{2}$, \eqref{eq:ex22} equals the known result \eqref{eq:A5result}.

We checked\footnote{
For computational reasons this check was performed for the higher mass levels without subtracting traces.
This amounts to taking reducible representations and is thus not a suitable approach for checking unitarity.
However, the result of this particular consistency check carries over to irreducible representations.
}
 for the mass levels up to $A=B=C=3$ that the coefficients $c_{A,B,\alpha,\beta,q}$ that can be obtained from the 5 point amplitude by 
matching the polynomials in kinematic invariants $k_{ij}$ order by order to the known result \eqref{eq:ResResA5} give the correct contribution to the factorised six point amplitude, which is related to $3$-point amplitudes by
\bea
&-\Res_{s_{12} \to A-1} \Res_{s_{123}\to B-1} \Res_{s_{56}\to C-1}\Amp_6 \\
={}& \sum\limits_{\alpha,\beta,\gamma}  \sum\limits_{q = 0}^{\min(|\alpha|,|\beta|)}  \sum\limits_{r = 0}^{\min(|\beta|,|\gamma|)}{\mathord{\vcenter{\hbox{\scalebox{0.3}{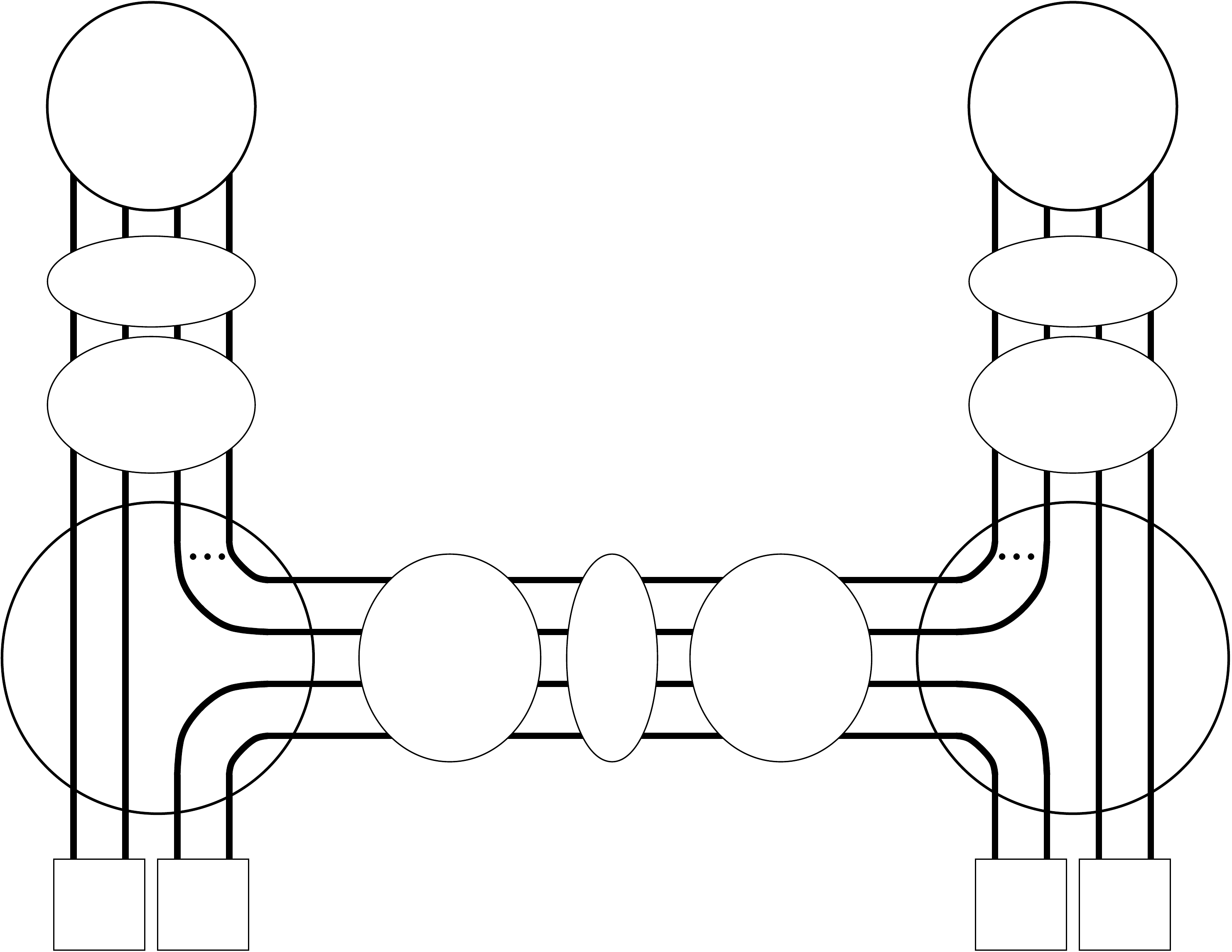}}}}}.
\eea{eq:6point_contraction}
In the six-point amplitude antisymmetric representations appear in the central factorisation channel starting at $B=3$. Since these states do not appear in $5$-point amplitudes,
 these residues of the six point amplitude cannot be fully reconstructed from data obtained from the five point amplitude. Furthermore, the six point amplitude 
does not provide all the data to construct higher point amplitudes, because it does not involve enough momenta to contain all highly antisymmetric representations itself.
However, since the representations of ${SO(2n+1)}$ are limited to Young diagrams with not more than $n$ rows, there is no representation antisymmetric in more than $n$ indices and there must exist
a tachyon amplitude involving all irreps of $SO(25)$ that appear in the spectrum. 

The techniques introduced in this section make summing over physical tensor states in principle straightforward for amplitudes with traceless symmetric irreps: a general algorithm for computing couplings of two massive particles and one tachyon
or even three massive particles could be in principle implemented in computer algebra. 
However, brute force calculations quickly become unfeasible due to the complexity of
the contraction and (anti)symmetrisation of multiple rank $A$ tensors.

For further explicit calculations an ingredient that needs some further thought are the $SO(n)$ irrep projectors $\mathbf{P}_{\alpha}$. In this paper only  the projector to traceless symmetric irreps \eqref{eq:tracefree_projector} was used. All projectors for $3$-index tensors are given in \cite{Cvitanovic:2008zz}.

The numerical coupling constants of one massive particle and two tachyons \eqref{eq:TTM_couplingsII} demonstrate that even these couplings of a relatively simple process are not that simple.
The couplings involving two or three massive particles are expected to be considerably more involved. Deriving them in an explicit $SO(D-1)$ covariant form would need sufficient motivation, and would likely require superior technology.

%%%%%%%%%%%%%%%%%%%%%%%%%%%%%%%%%%%%%%%%%%%%%%%%%%%%%%%%%%
%%%%%%%%%%%%%%%%%%%%%%%%%%%%%%%%%%%%%%%%%%%%%%%%%%%%%%%%%%

\section{Target space definition of the string S-matrix at tree level in a flat background}\label{sec:stringdef}

The results obtained above can be gathered into a working definition of the S-matrix of string
theory at tree level in a flat background. As will be elaborated upon in the discussion
section, an explicit and self-contained proof that the set of conditions given below leads to
a fully consistent S-matrix is still lacking. However, from the results obtained so far such
as the Koba-Nielsen amplitude obtained above it is plausible that the produced S-matrix
will be identical to the worldsheet-derived one. This will be checked further by studying
the output of the definition in the open superstring case. The definition can be summarised as follows:

\boxit{
The tree level S-matrix for open strings in a flat background is determined by:
\begin{itemize}
\item unitarity 
\item locality
\item $D$ dimensional (super-)Poincar\'e invariance
\item standard tree level colour-ordering (equation \eqref{eq:stdcolorordering})
\item universal monodromy relations (equation \eqref{eq:monodfund})
\item under (generalised) BCFW shifts of colour-adjacent particles, the amplitude behaves as it does in the corresponding string theory
\item a strict ordering between the location of poles and of roots. In particular there is a unique smallest mass (super)particle. 
\end{itemize}}

The last three requirements are those that are special to string theory. Colour-ordering forces
scattering amplitudes to have poles only in adjacent channels. Locality enters by the requirement
that three point amplitudes are polynomial functions of the external momenta. This translates
by unitarity to `polynomity' of residues of higher point amplitudes in forbidden channel momentum
invariants.  This was a crucial ingredient in extending the Veneziano amplitude to higher
multiplicity \cite{DiVecchia:2007vd, Koba:1969rw}. Note that for four particles the requirement
of BCFW shift behaviour is almost literally the same as Regge behaviour. For two tachyons for
instance, equation \eqref{eq:largeztachyons} holds. 

The ordering requirement for roots and poles amounts to the following: there is a number $x$ such  for any momentum invariant $s_{i \ldots j}$ the roots of the amplitudes as a function of this variable are located at $s_{i \ldots j} < x$ and the poles at $s_{i \ldots j} \geq x$. This number is the mass of the smallest mass (super)particle, which will be taken to be unique.

Closed strings can simply be defined by the KLT relations. The KLT relations are closely related to the monodromy relations. This is already clear from a close reading of the KLT paper: the monodromy relations are used implicitly. A more modern and precise connection is through the `momentum kernel' of \cite{BjerrumBohr:2010hn}. A particularly neat geometrical observation about the relation between monodromy and KLT for the four point amplitude which can also be used to find the roots is made in \cite{Lancaster:1989qc}. A higher point generalisation is unknown. 

A further comment concerns Poincar\'e invariance, which for superstring theory should naturally be enlarged to super-Poincar\'e invariance. As noted above, unitarity forces in both cases the $D\leq D_{\textrm{crit}}$ constraint. The value of the critical dimension is dependent on the theory ($26$ and $10$ for vanilla bosonic or superstrings respectively). 

A final comment is that the possibility of using the monodromy relations as an extension of duality for foundational purposes was already conjectured for five point amplitudes in \cite{D'Adda:1971te}. What is added in this article is a calculational path to make their argument precise for, in principle, any external matter content and any number of particles.

%%%%%

\subsubsection*{On minimality}
The definition given above is certainly sufficient for the bosonic string: the output S-matrix is the same as in the worldsheet approach. As already indicated above, it is highly desirable to obtain known properties of the string theory S-matrix such as unitarity (reality of couplings) without invoking conformal symmetry. On the other side, there is a pressing question if the above definition is minimal. That is, is there perhaps a smaller set of criteria possible? 

There are physics reasons to suspect such a smaller set is indeed possible. The main
motivation for this is the 'folk-theorem' that there are no interacting quantum field
theories with a finite number of particles with spins bigger than two. From an on-shell
perspective, this `theorem' has been discussed in a series of papers \cite{Benincasa:2007xk},
\cite{Benincasa:2011pg} and \cite{Benincasa:2011kn} (see also \cite{Feng:2011jxa}).
Suggestively, the path to a consistent theory suggested in the last two references
involves roots of amplitudes. It is suspected  that combining the above analysis with this line
of reasoning might lead to the elimination of the requirement of imposing monodromy
relations. Similarly, in the close string sector there might be an argument which does
away with the assumption of the KLT relations, perhaps in favour of some form of what
would be called holomorphic factorisation on the worldsheet. The requirement that there
is a unique lowest mass (super)particle might also be unnecessary. 

The above set of conditions is a working definition. As is usual in high energy physics
there are a number of hidden assumptions. One of these is for instance that only standard,
causal, Feynman type propagators are allowed (this feeds into the ``residues at poles
from unitarity'' argument). It would be interesting to reach a definition up to more
rigorous mathematical standards. It will be interesting to see where this differs from
the much more axiomatic approach of \cite{Bahns:2012wj}: as shown above, any physical theory which
contains the Veneziano amplitude has a critical dimension by unitarity.

\subsubsection*{On extendability}
The above definition is tailored to flat backgrounds. Analogs of at least some of the
assumptions can be worked out however in quite generic backgrounds.  Unitarity for
instance should have an analog in any background. Furthermore, monodromy relations for
open strings can in principle be derived in any background, see \cite{Boels:2010bv}.
Crucial here is that it is known that vertex operators in the open string generically
obey a braid relation,
\begin{equation}\label{eq:braidrel}
:V_1: :V_2: = R_{12} :V_2: :V_1: \,.
\end{equation}  
As pointed out first in \cite{Moore:1989ni}, the $R$ factors obey generically the
Yang-Baxter equation. This simply follows from consistency of the three point scattering
amplitude in string theory. Hence analogs of the monodromy relations should exist for
open strings in any background by following the same steps as in \cite{Boels:2010bv}.
Actually, for closed strings one can also repeat the step in deriving a KLT-like
relation between open and closed string amplitudes in any background. This follows
as the KLT paper is basically only concerned with relating the measure of the
integration over the moduli space of the $N$-punctured sphere to that of two
$N$-punctured discs. The braid relation in equation \eqref{eq:braidrel} can then
be inserted for the proper (but very formal) form of the curved background KLT relations. 

It would be extremely interesting if these short observations could be turned into a tool to study scattering amplitudes in non-trivial string backgrounds. This is however far beyond the scope of the present paper.

\subsection{Massless amplitudes in the open superstring: five points}

As a further illustrative example it will be explored below how the conditions posed above can be solved in the superstring case. As the four point case was discussed above, let us focus on five and more particles. Although the approach used previously will also work, for variety here the explicit formula
\begin{multline}\label{eq:5ptsuperdoubleres}
\Res_{s_{12}\to A} \Res_{s_{123}\to B} \left[ \Amp(12345)\right] = (-1)^{A+B} \pi^{-2}  \sin(\pi k_{34}) \\ \left[ \sin(\pi k_{24}) \Amp(14235) + \sin(\pi (k_{23} +k_{24})) \Amp(14325) \right] ,
\end{multline}
will be used. This is obtained by solving the monodromy relations for $\Amp(14235)$ and  $\Amp(14325)$. The amplitudes on the right-hand side have poles as a function of the following variables:
\begin{equation}
\begin{array}{c} A(14235) :\\
k_{14} = B - k_{24} - k_{34} , \\ 
k_{24} ,\\
k_{23} ,\\
k_{35}  = B - A - k_{34} ,\\
k_{15} = k_{23} + k_{24} + k_{34} ,
 \end{array} 
 \qquad 
\begin{array}{c} A(14325) :\\
k_{14} = B - k_{24} - k_{34} ,   \\ 
k_{34} ,\\
k_{23} ,\\
k_{25}  =  A-k_{23} - k_{24} ,\\
k_{15} = k_{23} + k_{24} + k_{34} .
 \end{array} 
 \end{equation}
 
 \subsubsection{Isolating the roots and fixing an ansatz}
Analysing the right-hand side of equation \eqref{eq:5ptsuperdoubleres} gives roots for instance for 
\begin{equation}
k_{24}> 0 , \qquad k_{23} > 0 , \qquad  k_{24} + k_{23} < A , \qquad \{k_{24}  \in \mathbb{N} ,k_{23} \in \mathbb{N} \} ,
\end{equation}
by avoiding the poles of the amplitudes on the right-hand side. These conditions are less strong
compared to the bosonic string case. However, there is much more information left unused in the
above equation. First, on any massless pole the superstring amplitudes factorise into massless
amplitudes with less legs. Since it was already shown the four point amplitude is proportional
to the field theory amplitude $\Amp^F$, the massless residues of the five point string amplitude
are proportional to the massless residues of the field theory amplitude. Hence it is natural to write as an ansatz for the residue,
\begin{equation}\label{eq:ansatz5ptsusy}
\Res_{s_{12}\to A,s_{123}\to B} \left[\Amp(12345)\right] = \Res_{s_{12}\to A,s_{123}\to B} \left[ F_1  \Amp^F(12345) + F_2  \Amp^F(13245) \right] .
\end{equation}
Although this ansatz is natural, it helps to know by the results in \cite{Mafra:2011nv} that it will be enough. Let us furthermore introduce the notation
\be
G_1 = \Res_{s_{12}\to A,s_{123}\to B} F_1 , \qquad G_2 = \Res_{s_{12}\to A,s_{123}\to B} F_2 .
\ee

The right-hand side of equation \eqref{eq:5ptsuperdoubleres} now gives the functions $G_i$ an important property: they have roots for 
\begin{equation}
k_{24}\geq 0 , \qquad k_{23} \geq 0 , \qquad  k_{24} + k_{23} < A , \qquad \{k_{24}  \in \mathbb{N} ,k_{23} \in \mathbb{N} \} ,
\end{equation}
and
\begin{equation}
k_{24}\geq 0 , \qquad k_{34} \geq 0 , \qquad  k_{24} + k_{34} < B , \qquad \{k_{24}  \in \mathbb{N} ,k_{34} \in \mathbb{N} \} .
\end{equation}
Note the appearance of the equality signs. This property can be argued as follows: first consider
$k_{24} \rightarrow 0$, while $A > k_{23}+k_{24} \in \mathbb{N}$. The right-hand side of equation
\eqref{eq:5ptsuperdoubleres} gives a contribution proportional to the massless residue of the 
$\Amp(14235)$ amplitude in the $(4,2)$ channel. This is proportional to the residue of the
corresponding field theory amplitude, $\Amp^F(14235)$ in the $(4,2)$ channel, which can be
expressed in terms of $\Amp^F(12345)$ and $\Amp^F(13245)$ by the field theory BCJ relations.
This picks up the residue of the $\Amp^F(13245)$ amplitude, while $\Amp^F(12345)$ does not
diverge in the limit so that it does not contribute.
This is to be compared to \eqref{eq:ansatz5ptsusy}. $G_{2} \rightarrow 0$ is required to
extract the residue of $\Amp^F(13245)$ and $G_{1} \rightarrow 0$ follows because
$\Amp^F(12345)$ does not vanish by itself in this limit.

For $k_{34} \rightarrow 0$ a similar reasoning gives that with $k_{24}$ a positive integer for which
$k_{34} + k_{24} < B$ one has to demand $G_{1} \rightarrow 0$ and $G_{2} \rightarrow 0$. This isolates
the $k_{34}$ residue in the second amplitude on the right-hand side of equation \eqref{eq:5ptsuperdoubleres}.
The remaining zero for $k_{23}=0$ follows from the solution of the monodromy relations in terms of $\Amp(13425)$ and $\Amp(14325)$,
\begin{multline}\label{eq:5ptsuperdoubleresII} 
 \Res_{s_{123}\to B} \Res_{s_{12}\to A} \left[ \Amp(12345)\right] = (-1)^{A+B} \pi^{-2} \sin(\pi k_{23}) \\  \left[ \sin(\pi k_{24}) \Amp(13425) + \sin(\pi (k_{34} +k_{24})) \Amp(14325)\right] ,
\end{multline}
where the relevant variables are
\begin{equation}
\begin{array}{c} A(13425) :\\
k_{13} = A-B - k_{23} ,  \\ 
k_{34} ,\\
k_{24} ,\\
k_{25}  = A-k_{23} - k_{24} ,\\
k_{15} = k_{23} + k_{24} + k_{34} ,
 \end{array} 
 \qquad 
\begin{array}{c} A(14325) :\\
k_{14} = B - k_{24} - k_{34}  , \\ 
k_{34} ,\\
k_{23} ,\\
k_{25}  =  A-k_{23} - k_{24} ,\\
k_{15} = k_{23} + k_{24} + k_{34} .
 \end{array} 
 \end{equation}

Just as before, there is a maximal spin at each mass level. In the above notation, this is $A+1$ and $B+1$ in the respective channels. Therefore, one would expect that for instance an $\Amp^F(12345)$ amplitude would be multiplied by a polynomial of maximal degree $A+B$, with a similar spin-induced fine-structure of powers of $k_{34}$, $k_{24}$ and $k_{23}$ as elucidated above for the bosonic string.  To be more precise, under a $(1,2)$ channel BCFW shift the residue scales as
\begin{align}\label{eq:largeztachyonsatresidue5ptSUSY}
\lim_{z\rightarrow \infty} \textrm{Res}_{s_{12} \rightarrow A} & \Amp(12345)  \sim z^{A} \Amp^F(12345)(z) \left(f_1\left(\frac{1}{z}\right) \right) ,
\end{align}
while for a $(4,5)$ channel shift 
\begin{align}
\lim_{z\rightarrow \infty} \textrm{Res}_{s_{45} \rightarrow B} & \Amp(12345)  \sim z^{B} \Amp^F(12345)(z)\left(f_1\left(\frac{1}{z}\right) \right) ,
 \end{align}
holds. Note the analogy to \eqref{eq:largeztachyonsatresidue5pt} in the bosonic string case. Just as in that bosonic string case it is natural to use the following basis of polynomials, 
\be \label{eq:superstring5ptpols}
f_a(k_{23},k_{24},k_{34}) = \binom{k_{23}}{A-a} \binom{k_{24}}{a}\binom{k_{34}}{B-a} , \quad 0 \leq a \leq \min(A,B) ,
\ee
which scales as $z^A$ under a $(1,2)$ BCFW shift and as $z^B$ under a $(4,5)$ BCFW shift. 

The analysis of the maximal spin gives for the ansatz of equation \eqref{eq:ansatz5ptsusy} the existence of vectors of numbers $G_1^a$,  $G_2^a$ and  $\tilde{G}_2^a$ such that
\begin{equation}
G_1 = G_1^a f_a(k_{23},k_{24},k_{34}) ,  \qquad G_2 = \left(-k_{23} G_2^a +  \tilde{G}_2^a\right)\, f_a(k_{23},k_{24},k_{34})  .
\end{equation}
Here the possibility of a first order polynomial in $k_{23}$ in the $G_2$ polynomial follows as the $(1,2)$ shift of the field theory amplitude $A(13245)$ is suppressed by one power of $z$ as it is a non-adjacent BCFW shift, while the $(4,5)$ shift of this amplitude is colour-adjacent. Having fixed the complete functional form of the amplitude, it remains to compute the above three vectors of numbers. 

\subsubsection{Solving consistency constraints to obtain full result}
Some constraints follow by the requirement that the combination on the right-hand side of equation \eqref{eq:ansatz5ptsusy} have no poles, while the field theory amplitudes have residual kinematic poles at the residue. This immediately forces 
\begin{equation}
G_{1}^B = 0 , \qquad G_2^0 = 0 , \qquad  \tilde{G}_2^0 =0 ,
\end{equation}
by absence of poles in the $(3,4)$ and $(2,4)$ channels respectively.
To see this, take for instance $k_{34}=0$ in \eqref{eq:ansatz5ptsusy}. The pole in $\Amp^F(12345)$
is not cancelled by a factor of $k_{34}$ in $f_B(k_{23},k_{24},k_{34})$ hence $G_{1}^B = 0$.
For these channels only one of the field theory amplitudes in the basis has a potential pole. In the $(2,3)$ channel both field theory amplitudes develop a pole, leading to the constraint
\begin{equation}\label{eq:nos23pole}
G_{1}^A -G_{2}^A =  0 .
\end{equation}
Here it was used that in this channel both field theory amplitudes in \eqref{eq:ansatz5ptsusy} factorise into the
same lower point amplitudes $\Amp^F(23P) \Amp^F(P451)$.
If $A<B$, then avoiding the pole in the $k_{13}$ channel forces 
\begin{equation}\label{eq:nos13pole}
\tilde{G}_2^c = k_{23} G_{2}^c= (A-B) \, G_{2}^c \qquad \textrm{for   }  A<B .
\end{equation}

The $(1,5)$ channel yields further information as the vanishing of the residue of the pole in the $(1,5)$ channel implies
\begin{equation}\label{eq:noresiduein15channel}
G_1 + \frac{k_{34}}{k_{24}} G_2 = 0 \qquad \textrm{for   }  k_{15} = k_{23} + k_{34} + k_{24} = 0 .
\end{equation}
Here the BCJ relation $k_{34} \Amp^F(P234) + (k_{34}+k_{23}) \Amp^F(P324)=0$ has been used to pull out an overall $\Amp^F(P234)\Amp^F(P51)$.
Evaluating this constraint on the kinematic point $k_{23}  = A - k_{24}, k_{24}=c$ and $k_{34} = - A$ with $c$ an integer $0<c<A$ gives
\begin{equation}
c \, G_1^c - A \,((c-A) \, G_2^c +  \tilde{G}_2^c ) = 0 ,
\end{equation}
while evaluating it on the compatible kinematic point $k_{34} = B - k_{24}, k_{24}=c$ and $k_{23} = - B$ with $c$ an integer $0<c< B$ gives
\begin{equation}
c \, G_1^c + (B-c)\, (B\, G_2^c +  \tilde{G}_2^c ) = 0 .
\end{equation}
These equations can be solved for  $ \tilde{G}_2^c $ and $G_1^c$ to give
\begin{equation}\label{eq:solsforG}
G_1^c = \frac{1}{c} A\,(c-B)\, G_2^c ,  \qquad \tilde{G}_2^c = (A-B) \, G_{2}^c .
\end{equation}

The cases $c=0$ and $c=A$ or $c=B$ are special as they would hit poles of the residue in the $(1,5)$ channel. The correct approach is to first take the kinematic limits, obtaining for instance
\begin{equation}
\frac{\partial G_1}{\partial k_{34}} + \frac{1}{k_{24}} G_2  = 0 \qquad \textrm{for  } k_{34}=0, \,k_{23} + k_{24} = 0 ,
\end{equation}
as well as
\begin{equation}
\frac{1}{k_{34}} G_1 + \frac{\partial G_2}{\partial k_{24}}  = 0   \qquad \textrm{for  } k_{24}=0, \,k_{23} + k_{34} = 0 ,
\end{equation}
by the requirement that the residues at these poles have to vanish.
Here the derivatives single out the terms linear in the corresponding variable since that variable is taken to be zero in both cases.
Again a BCJ relation was used to pull out an overall factor containing three $3$-point amplitudes. From the first 
\begin{equation}
 \sum_{a=0}^{\min(A,B-1)} \frac{G_{1}^a (-1)^{B-a-1} }{B-a} \left(k_{23} \atop A-a \right)\left(- k_{23} \atop a \right) 
-\frac{1}{k_{23}} \left(-k_{23} G_2^B  + \tilde{G}_2^B \right) \left(k_{23} \atop A-B \right) \left(-k_{23} \atop B \right) = 0  ,
\end{equation}
while from the second
\begin{equation}
- \frac{G_1^0}{k_{23}} \left(k_{23} \atop A \right) \left(-k_{23} \atop B \right) + \sum_{a=1}^{\min(A,B)}  \left(-k_{23} G_2^a + \tilde{G}_2^a \right)  \frac{(-1)^{a-1}}{a} \left(k_{23} \atop A-a \right)\left(- k_{23} \atop B- a \right) = 0 ,
\end{equation}
is obtained. Both equations can be solved \emph{uniquely} for $G_{1}^a$ resp.\ $G_2^a$ since the polynomials these coefficients multiply differ by two powers of $k_{23}$. Starting from the maximal power term of degree $A+B$ one can solve for $G_{2}^a$ and  $\tilde{G}_{2}^a$. To read off the solution we use
\bea
\sum_{a=0}^{\min(A,B-1)} \left(k_{23} \atop A-a \right)\left(- k_{23} \atop a \right) &= \frac{B (A-B-k_{23})}{A k_{23}} \left(k_{23} \atop A-B \right)\left(- k_{23} \atop B \right) , \\
\sum_{a=1}^{\min(A,B)}  \left(k_{23} \atop A-a \right)\left(- k_{23} \atop B- a \right) &= \frac{A B}{k_{23} (A-B-k_{23})} \left(k_{23} \atop A \right)\left(- k_{23} \atop B \right) ,
\eea{eq:binsums}
and obtain the solutions
\be
\tilde{G}_2^a = (A-B) \, G_{2}^a , \qquad G_1^a = (-1)^{B-a-1} \frac{(B-a)A}{B}\, G_2^B   ,
\ee
and
\begin{equation}\label{eq:solsforGII}
\tilde{G}_2^a = (A-B) \, G_{2}^a , \qquad G_2^a = (-1)^{a-1} \frac{a}{A\,B}\, G_1^0  ,
\end{equation}
for the two equations.
Note that together with \eqref{eq:solsforG} already more than enough constraints were found to fix the residue up to a constant
and all redundant constraints that were obtained in different ways are compatible.
The result for $G_1$ and $G_2$ is
\bea
G_1 &= \sum\limits_{a=0}^\infty \frac{(B-a)}{B} (-1)^a G_1^0 f_a(k_{23},k_{24},k_{34}) ,\\
G_2 &= \sum\limits_{a=0}^\infty \frac{(A-B-k_{23})a}{A \, B} (-1)^{a-1} G_1^0 f_a(k_{23},k_{24},k_{34}) .
\eea{eq:Gresult}
Hence the ansatz for the $A,B$ residue is fixed up to an overall constant by consistency requirements.  Note that with this solution a factor of $k_{13}$ factors out of the $G_2$ function. 
From this result it is also manifest that there is no pole in the $(1,3)$ channel within the ansatz remaining. 
The remaining constant $G_1^0$ is the string coupling constant times a numerical factor which can only depend on $A$ and $B$. This is easily determined from equation \eqref{eq:5ptsuperdoubleres}.

\begin{comment}

\begin{multline}
 \Amp^F (12345)  = \frac{1}{A B} \left[k_{24} (A-B +k_{34}) \Amp^F(14235) +  \right. \\ \left.  (- B +k_{34})(-A+k_{23}+k_{24}) \Amp^F(14325)  \right]
\end{multline}

\begin{multline}
 \Amp^F (13245)  =  \frac{1}{(B-A +k_{23} ) B} \left[ (-B +k_{24}) (A-B +k_{34})  \Amp^F(14235)   \right. \\ \left.  +k_{34} (-A+k_{23}+k_{24}) \Amp^F(14325)  \right]
\end{multline}
\end{comment}

The result just obtained corresponds indeed to the residue of the known open superstring theory five point amplitude. The simplest form in the literature can be found in \cite{Mafra:2011nv}, for which
\bea
F_1 &= k_{12} k_{34} \int\limits_0^1 \int\limits_0^1 dx dy x^{k_{45}} y^{k_{12}-1} (1-x)^{k_{34}-1} (1-y)^{k_{23}} (1-xy)^{k_{24}} \\
&= \sum\limits_{a=0}^{\infty} \frac{(-1)^{A+B-a} A k_{34}}{(k_{12}+A)(k_{45}+B)} \binom{k_{23}}{A-a} \binom{k_{24}}{a} \binom{k_{34}-1}{B-a-1}\\
&= \sum\limits_{a=0}^{\infty} \frac{(-1)^{A+B-a} A (B-a)}{(k_{12}+A)(k_{45}+B)} \binom{k_{23}}{A-a} \binom{k_{24}}{a} \binom{k_{34}}{B-a} ,
\eea{eq:G1_5pt}
and
\bea
F_2 &= k_{13} k_{24} \int\limits_0^1 \int\limits_0^1 dx dy x^{k_{45}} y^{k_{12}} (1-x)^{k_{34}} (1-y)^{k_{23}} (1-xy)^{k_{24}-1} \\
&= \sum\limits_{a=0}^{\infty} \frac{(-1)^{A+B-a-1} (k_{12}+k_{23}-k_{45}) k_{24}}{(k_{12}+A)(k_{45}+B)} \binom{k_{23}}{A-a-1} \binom{k_{24}-1}{a} \binom{k_{34}}{B-a-1}\\
&= \sum\limits_{a=0}^{\infty} \frac{(-1)^{A+B-a} (k_{12}+k_{23}-k_{45}) a}{(k_{12}+A)(k_{45}+B)} \binom{k_{23}}{A-a} \binom{k_{24}}{a} \binom{k_{34}}{B-a} ,
\eea{eq:G2_5pt}
hold. The residues studied above are easily read off from these equations. As these are correctly obtained, the full result follows by BCFW on-shell recursion.

Note that the derivation above almost exclusively uses physical input such as locality, unitarity and Regge-behaviour. It should be stressed there is a non-trivial step in the above at the point where an ansatz for the five point string theory amplitude is written in terms of field theory amplitudes. Note that by the consistency conditions that the residue is local it is easy to see that an ansatz with a single field theory amplitude would not work. Although the ansatz is very natural, in principle it could be in the above approach that it would ultimately turn out to be insufficient. While it is known \cite{Mafra:2011nv} that this does not occur, it would be interesting to have a target space understanding of this.

\subsection{Massless amplitudes in the open superstring: higher points \& other matters}

For more than five points as well as for fermionic matter the above analysis can be repeated. For
higher points the results are at least easy to sketch: one expects to be able to write the residue in terms
of a $(N-3)!$ element basis of field theory amplitudes, multiplied by a polynomial in the remaining
kinematic variables. At a chosen pole in the multiperipheral channel the maximum powers of these
variables are set by the scaling of the string theory amplitude under the relevant BCFW-like shift
of equation \eqref{eq:largeztachyonsatresidueNpt}. For more than $11$ particles this will require
a formal analytic continuation to higher dimensional amplitudes. The monodromy relations can then
be used to obtain roots of the chosen residue. This, taken together with the cases where the residues
do not vanish but instead involve lower point string amplitudes with massless matter only is fully
expected to  completely fix the residue. Using on-shell recursion, this then fixes the amplitude. 

For guidance, one technical tool needed for this program will be worked out here, which is the result of the shift of $l$ colour-adjacent momenta of equation \eqref{eq:largeztachyonsatresidueNpt}, reproduced here for convenience, 
\begin{equation}\label{eq:extBCFWlikeshift}
k_1 \rightarrow k_1 - (l-1) q z , \quad k_2 \rightarrow k_2 + q z , \quad \ldots , \quad  k_{l} \rightarrow k_l+ q z ,
\end{equation}
where $q$ is orthogonal to all momenta $k_i$ for $i \in \{1,\ldots,l\}$ and itself. 

\subsubsection*{Field theory scaling}
Standard Feynman-'t Hooft type power-counting gives a scaling of $\sim z^{l-1}$ for any amplitude: this follows from graphs with three point couplings only, the $l$ shifted gluons on the outside and at least one external (off-shell) leg. It will be argued here that the color-ordered field theory amplitudes where particles $1$ through $l$ appear adjacent in color-order actually scale like $z$. Note that the particles do not have to appear as an ordered set: there should be no unshifted particles splitting the set in two in colour-ordering. Those amplitudes which have such a split set are suspected to scale suppressed by one power of $z$. 

To demonstrate this one can study Yang-Mills theory in a special gauge: using the light cone vector $q$ as a
gauge choice as first suggested in \cite{ArkaniHamed:2008yf} for the $l=2$ case. For this gauge there are several classes of Feynman diagrams which are singular:
those that have a linear sum of \emph{only} momenta from the set $\{k_1, \ldots, k_l \}$ in one of
their propagators. For a propagator of this type, $q$ is not a valid light cone gauge choice as
\begin{equation}
q \cdot \left(\sum_{i \in \sigma} k_i \right) = 0 .
\end{equation}
Here $\sigma$ is a subset of $\{1,\ldots, l\}$. Technically, this condition blows up a term in the light cone gauge propagator in the graph with only shifted particles on one side. This can be regulated as in \cite{Boels:2010nw} by choosing an auxiliary light cone gauge vector $q + x k_1$ and letting $x$ tend to zero. Note that the (regulated) singular propagators are not orthogonal to $q$. The result is that the singular graphs potentially could contribute positive powers of $z$ scaling. However, any power of $z$ will appear with a $q^{\mu}$ contracted into the remainder of the graph. This remainder must be a three point vertex with a non-shifted momenta: the $q$ must contract with these momenta. 

The singular graphs are classified by the number of connections to the remainder of the graph. There is at least one, but potentially more. First consider the class with one connection. Since the shift leaves all mandelstam invariants from the singular part invariant, any $q$ dependence must contract with outside lines. For the shifted legs, $q \cdot \xi \sim \frac{1}{z}$, so this would lead to subleading contributions. Hence for one connection between singular and other parts, the maximal scaling is $\sim z$, obtained when $q$ contracts into the one singular propagator connecting to the remainder of the graph. 

A graph with more connections to the remainder can be made from the class with one by adding un-shifted external lines in any position. This leads to a decrease by one power of $\frac{1}{z}$ from a non-singular additional hard propagator in the graph\footnote{a potential $z^0$ contribution from the hard propagator is easily seen to cancel out as it would contract two q's into the same three vertex.}. There can also be an additional positive power of $z$ since there can now be two singular propagators ending on a non-singular three vertex: this situation only occurs for four and more shifted particles. This implies the amplitude scales as $\sim z$ under the shift \eqref{eq:extBCFWlikeshift}. We suspect that shifts of field theory amplitudes where the particles $1,\ldots, l$ do not appear adjacent in color-order are suppressed by an additional power of $\frac{1}{z}$, but so far have only checked this for three shifted particles. 

\subsubsection*{String theory scaling}
For the shift \eqref{eq:extBCFWlikeshift} one finds for the string theory amplitude in the multiperipheral channel to scale to leading order in $z$ as
\begin{align}\label{eq:behundergenshift}
\textrm{Res}_{s_{1\ldots l} \rightarrow A_l} & \Amp_N \sim z^{A_l + 1} \left(f_1\left(\frac{1}{z}\right) \right) ,
\end{align}
where $A_l$ is the level at the $s_{1 \ldots l}$ pole. This can be argued for by an extension of the analysis in  \cite{Cheung:2010vn, Boels:2010bv}. Basically, this is a saddle point approximation for the integral of the position of $l-1$ vertex operators, with $y_1$ fixed at zero and $y_{l+1}$ at $-\infty$. Schematically, this integral looks like
\begin{equation}
\int_{-\infty}^{y_{l-1}} dy_l \ldots \int_{y_3}^{0} dy_2 \langle :V(y_{l}): \ldots :V(y_{2})::V(y_{1}=0): \ldots \rangle .
\end{equation}
Calculating the OPE of the vertex operators as done in \cite{Cheung:2010vn, Boels:2010bv} and taking a saddle approximation yields $y_i z \sim 1$ for positions $y_i$ in the large $z$ limit. From the saddle-point gaussian integral one power of $\frac{1}{z}$ follows for each integrated variable within the saddle point approximation. From the OPE one power of $z$ follows for each polarisation vector, as well as a $z^{A_L}$ Regge-type factor. Note that this argument is basically the usual string theory observation that the maximal spin at level $A_l$ is $A_l+1$ in the superstring. To see this, consider the residue of the pole of the superstring amplitude. The left and right amplitudes are functions of the mandelstams, but these are invariant under the shift above. Possible contractions of polarisation vectors with momenta would yield subleading-in-$z$ terms. Hence the maximal possible scaling is set by the momenta appearing in the polarisation sum which contract into the unshifted amplitude. This scaling is simply a count of possible momenta and hence the spin, giving $A_l+1$ by the known spectrum of the string. 

\subsubsection*{Synthesis}
These scaling results translate into constraints on the coefficients in a chosen basis of field theory amplitudes. In the multiperipheral channel it is natural to write an ansatz
\begin{equation}
\Amp(1,2,3,4,\ldots,N)\lfloor_{s_{1,\ldots,l} = A_l} = \sum_{\sigma \in P(2,3,\ldots, N-2)} G_{\sigma} \Amp^{F}(1,\sigma, N-1,N) ,
\end{equation} 
since this is rich enough to capture all points where the multiperipheral residue can be explicitly calculated by tuning to a pole in the cross-channel. The chosen basis contains many colour ordered amplitudes which have particles ordered similarly to the canonical order, which will lead to tighter bounds on the $G$ polynomials. These bounds arise by considering the generalised BCFW scalings of equation \eqref{eq:extBCFWlikeshift}. The constraints on the coefficient function follow from the known shifts of the string theory, e.g. equation \eqref{eq:behundergenshift}, and the shifts of the field theory amplitudes that appear in the ansatz. Note that the only quantity which is really needed in this computation is the ratio of the string-to-field theory scaling under the generalised shift. 

In a next step, one derives further constraints on the functions $G$ by using the monodromy relations.
Similar patterns of roots will be obtained as in the bosonic string case, again at first with
$k_{ij}>0$ conditions instead of $\geq$. Note this in effect fixes the functions $G$ up
to a finite degree polynomial. This leads to a more refined form for the ansatz. It is strongly suspected that the
$G$ polynomials have roots with $k_{ij} \geq 0$-type conditions. The locality conditions (no remaining poles in the residue) are
expected to fix most, if not all of the remaining freedom of the ansatz. Note that there are even more conditions
available from known factorisations of the residue into products of massless particles in
cross-channels to fix coefficients. Hence it is fully expected that the string theory answer in the form written
in \cite{Mafra:2011nv} is reproduced. We leave a full proof of this to future work. 

\subsubsection*{Other matter}
The inclusion of fermionic matter is up to five points trivial as the fermionic matter amplitudes are related to the bosonic ones by the on-shell supersymmetric Ward identities. It would be interesting to explore amplitudes above five points using an on-shell superspace formalism. Note that in the calculation above supersymmetry did not play a prominent role. It is not absent though: it is hidden in the form of the three point amplitude with three massless gluons.

%%%%%%%%%%%%%%%%%%%%%%%%%%%%%%%%%%%%%%%%%%%%%%%%%%%%%%%%%%
%%%%%%%%%%%%%%%%%%%%%%%%%%%%%%%%%%%%%%%%%%%%%%%%%%%%%%%%%%

\section{Discussion and conclusion}

The motivation behind this article is to restart the exploration of string theory from a target space point of view. This is driven
 by recent developments in field theory which harken back to string theory's very roots in the analytic S-matrix program. 
 The encouraging results obtained above on the basis of this motivation fall roughly into two categories: calculational techniques and foundational questions. 

Within the first category it has been shown in examples that the monodromy relations can be used to
calculate residues at singularities directly. This eliminates in effect a sum over the tower of
states known to be present in string theory. Put differently, all the couplings of these higher
states are exquisitely fine-tuned to yield this simple result. 

The residues at singularities may then be combined with on-shell recursion to compute full amplitudes, leaving relatively simple
infinite sums over the levels. In effect, the monodromy relations pick out a unique deformation of Yang-Mills theory amplitudes. 
Note the infinite sums have a function as in the field theory limit
they are intimately connected to the appearance of multiple zeta values. Recently interesting
patterns in this limit for massless superstring amplitudes were pointed out in
\cite{Schlotterer:2012ny}, see also
\cite{Drummond:2013vz, Stieberger:2013wea, Broedel:2013tta, Boels:2013jua}. It would certainly
be interesting to see how those patterns are related to the patterns of roots exposed in this
article. More generally, the field theory limit of the BCFW-type expressions for amplitudes is
an interesting area to be explored. Extending the on-shell calculational strategy to even just one
loop level is another direction of interesting research as here much less is known (see
\cite{Green:2013bza} for recent work in this direction from a worldsheet perspective). In principle
the tree level S-matrix tightly constrains string loop corrections through unitarity, but turning
this observation into a statement of practical utility is not necessarily easy. 

In the foundational category a new, purely target space based definition of the tree level S-matrix
of string theory has been proposed. Since the essential inputs of this definition are known to be
satisfied in string theory, there can be little doubt that its output is consistent, but it would
be very interesting to completely evade resorting to the worldsheet picture. As an example of this
unitarity has been checked above up to a degree. A full target-space based proof of unitarity would
be welcome. As discussed above, the target space approach to string theory may generalise to backgrounds other than flat ones.
It would already be very welcome to have one fully-worked
example of this for a non-trivial curved background. Of course, there might be much to be gained by using
the worldsheet to derive the analogs of the monodromy relations and large BCFW shifts as inputs for a
given target space calculation. 

It would be interesting to see if the set of conditions proposed
above as a definition of the S-matrix is minimal or not. It should be noted that at least morally
speaking our proposal for a definition of the S-matrix is similar to that of \cite{Moore:1993ns}.
In that article a particular kind of generalised monodromy conditions were derived which in contrast
to equation \eqref{eq:monodfund} do depend on particle content. To calculate a given amplitude one
still needs to perform a worldsheet computation in that approach. Deriving the relations proposed
in \cite{Moore:1993ns} from \eqref{eq:monodfund} might be interesting as it is a possible route
to check overall consistency of the approach. Moreover, it would be interesting to find out what
the minimal set of conditions in target space is for which the answer is ``all known string theories''. 

A definition of closed string theory through the KLT relations leaves much to be desired. Better
would be to find a more intrinsic definition which would yield KLT as an output. This would be in the
realm of a string theory extension of colour-kinematics duality.  More down to earth, an understanding
of the location of the roots of the closed string amplitudes could perhaps be obtained in a different
way.  In general the story of the roots of amplitudes certainly deserves further study. 

It should be pointed out that one of our initial motivations to study unitarity was to obtain
explicitly $SO(D-1)$ covariant three point amplitudes for arbitrary matter content. The forms
in the literature commonly involve only manifest $SO(D-2)$ covariance, with some exception for
maximal spin \cite{Schlotterer:2010kk}. However, from the results in Section \ref{sec:unit} for
the two tachyons case it is clear that even these in their current formulation are not simple.
We suspect that in a sense to be made precise they are not written in the right variables.
Finding these right variables probably would lead one to a form of an `on-shell string field'
as an analog to the more well-known on-shell superfields, but this remains a direction to be explored. 

It is rather remarkable that the on-shell intuition which originally led to Veneziano's amplitude can be made into a computational engine for any number of points, bypassing the later-found worldsheet picture. This suggests the worldsheet point of view and especially conformal symmetry may not be as central to string theory as always thought. Exploring this different viewpoint on the theory should prove fruitful beyond the results obtained here.

\acknowledgments
RB would like to thank the Institute for Advanced Study for a visit and Nima Arkani-Hamed for discussions which provided the inspiration for the present article. This work was supported by the German Science Foundation (DFG) within the Collaborative Research Center 676 ``Particles, Strings and the Early Universe''.

%%%%%%%%%%%%%%%%%%%%%%%%%%%%%%%%%%%%%%%%%%%%%%%%%%%%%%%%%%
%%%%%%%%%%%%%%%%%%%%%%%%%%%%%%%%%%%%%%%%%%%%%%%%%%%%%%%%%%

\appendix

\section{Multi-residues of tachyon amplitudes from the worldsheet}

This appendix contains an explicit derivation of the multiple residue of the Koba-Nielsen amplitude.  With the conventional gauge fixing
\begin{equation}
z_1 = 0, \quad z_{N-1} = 1, \quad z_N \to \infty\;,
\end{equation}
the Koba-Nielsen formula reads
\begin{align}
\Amp_N =  \prod\limits_{u=2}^{N-2} \int\limits_{0}^{z_{u+1}} \dd z_{u}  \prod\limits_{v=2}^{N-2} z_v^{k_{1v}} \prod\limits_{l=2}^{N-2} (1- z_l)^{k_{l,N-1}} \prod\limits_{\substack{i,j\\1< i<j< N-1}} (z_j-z_i)^{k_{ij}}  .
\end{align}
Using binomial expansion\footnote{Alternatively, one could use Mellin-Barnes representations here, see e.g. \cite{Yuan:2014gva} for a systematic approach.}
\be
(z_j-z_i)^{k_{ij}}  = \sum\limits_{a_{ij}=0}^\infty \binom{k_{ij}}{a_{ij}}(-1)^{a_{ij}} z_i^{a_{ij}} z_j^{k_{ij}-a_{ij}} ,
\ee
the amplitude becomes
\begin{align}
\Amp_N = \sum\limits_{a_{23},\ldots,a_{N-2,N-1}=0}^\infty 
\prod\limits_{\substack{i,j\\1<i<j<N}} (-1)^{a_{ij}} \binom{k_{ij}}{a_{ij}}
 \prod\limits_{u=2}^{N-2} \int\limits_{0}^{z_{u+1}} \dd z_{u}
  z_i^{a_{ij}}
  \prod\limits_{v=2}^{N-2} z_v^{k_{1v}}
 \prod\limits_{\substack{s,t\\1< s<t<N-1}} z_t^{k_{st}-a_{st}} .
\end{align}
Doing the integrals one by one, one finds that the (l+1)th integral gives the factor
\be
\frac{1}{\alpha' (k_1+\ldots+k_l)^2 + \sum\limits_{\substack{1<u \leq l\\l<v<N}} a_{uv} - 1} .
\ee
Now compute the $N$-point tachyon amplitudes with all internal particles on-shell $-\alpha' (k_1+\ldots+k_l)^2=A_l-1$. In our other notation these mass levels correspond to $A_2=A, A_3=B$ and so on. Doing the integral using binomial expansion, one can see a way to write the result in general
\bea
&\left( \prod\limits_{l=2}^{N-2} \Res_{s_{1\ldots l}\to A_l-1} \right) \Amp_N 
= (-1)^{N-3} \sum\limits_{a_{23},\ldots,a_{N-2,N-1}=0}^\infty 
\prod\limits_{\substack{i,j\\1<i<j<N}} (-1)^{a_{ij}} \binom{k_{ij}}{a_{ij}} \prod\limits_{l=2}^{N-2} \delta_{A_l,\sum\limits_{\substack{1<u \leq l\\l<v<N}} a_{uv}} .
\eea{eq:A_N_result}
For example the double residue of the $5$-tachyon amplitude is
\bea
&\Res_{s_{12}\to A_2-1} \Res_{s_{123}\to A_3-1} \Amp_5 \\
={}& \sum\limits_{a_{23},a_{24},a_{34}=0}^\infty (-1)^{a_{23}+a_{24}+a_{34}}
\binom{k_{23}}{a_{23}} \binom{k_{24}}{a_{24}} \binom{k_{34}}{a_{34}}  \delta_{A_2, a_{23}+a_{24}} \delta_{A_3, a_{24}+a_{34}} \\
={}& \sum\limits_{a_{24}=0}^{\min(A_2,A_3)}  (-1)^{A_2+A_3-a_{24}} \binom{k_{23}}{A_2 - a_{24}} \binom{k_{24}}{a_{24}}  \binom{k_{34}}{A_3 - a_{24}} .
\eea{eq:A_5_result}
Analogously for $N=6$
\bea
& \Res_{s_{12}\to A_2-1} \Res_{s_{123}\to A_3-1} \Res_{s_{1234}\to A_4-1} \Amp_6 \\
={}& -\sum\limits_{a_{23},\ldots,a_{45}=0}^\infty (-1)^{a_{23}+a_{24}+a_{25}+a_{34}+a_{35}+a_{45}}
\binom{k_{23}}{a_{23}} \binom{k_{24}}{a_{24}} \binom{k_{25}}{a_{25}} \binom{k_{34}}{a_{34}} \binom{k_{35}}{a_{35}} \binom{k_{45}}{a_{45}}  \\
& \cdot \delta_{A_2, a_{23}+a_{24}+a_{25}} \delta_{A_3, a_{24}+a_{25}+a_{34}+a_{35}} \delta_{A_4, a_{25}+a_{35}+a_{45}} \\
={}& -\sum\limits_{a_{24}=0}^{\min(A_2,A_3)} \sum\limits_{a_{25}=0}^{\min(A_2,A_3,A_4)} \sum\limits_{a_{35}=0}^{\min(A_3,A_4)} (-1)^{A_2+A_3+A_4-a_{24}-2a_{25}-a_{35}}\\
& \cdot \binom{k_{23}}{A_2 - a_{24}-a_{25}} \binom{k_{24}}{a_{24}} \binom{k_{25}}{a_{25}} \binom{k_{34}}{A_3 - a_{24}-a_{25}-a_{35}} \binom{k_{35}}{a_{35}} \binom{k_{45}}{A_4 - a_{25}-a_{35}} .
\eea{eq:A_6_result}

%%%%%%%%%%%%%%%%%%%%%%%%%%%%%%%%%%%%%%%%%%%%%%%%%%%%%%%%%%
%%%%%%%%%%%%%%%%%%%%%%%%%%%%%%%%%%%%%%%%%%%%%%%%%%%%%%%%%%

\section{On-shell space of kinematic variables}
\label{sec:on-shell_space}

In Section \ref{sec:unit} the Koba-Nielsen amplitudes were factored into $3$-point
amplitudes by putting all the Mandelstams $s_{12\ldots}$ on the mass shell.
It is shown in this appendix how in this configuration any $k_{ij}$ can be
expressed in terms of the remaining $\frac{(N-2)(N-3)}{2}$ independent variables $k_{ij}$ with $1<i<j<N$.
After obtaining a result which can contain any $k_{ij}$ the following rules can be used to remove spurious kinematic variables.
First momentum conservation is used to remove the variables $k_{iN}$
\be
k_{iN} = \sum\limits_{j=1}^{N-1} -k_{ij} .
\ee
The $N-3$ conditions to put $k_1+k_2$ up to $k_1+k_2+\ldots+k_{N-2}$ on-shell can be used to eliminate $k_{12},\ldots,k_{1,N-2}$
\bea
-\alpha' (k_1+k_2)^2 &= A_2-1 ,\\
-\alpha' (k_1+k_2+k_3)^2 &= A_3-1 ,\\
\vdots\\
-\alpha' (k_1+k_2+\ldots+k_{N-2})^2 &= A_{N-2}-1 .
\eea{eq:on-shell_conditions}
Finally, there is always one additional condition that eliminates $k_{1,N-1}$. This condition is found by removing the $1$ using momentum conservation and then $k_{N-1,N}$ using the last on-shell condition again (using momentum conservation to get $-\alpha' (k_{N-1}+k_N)^2 = A_{N-2}-1$)
\be
k_{1,N-1} = \sum\limits_{j=2}^{N} -k_{j,N-1}  = \sum\limits_{j=2}^{N-1} -k_{j,N-1} + A_{N-2}+1  .
\ee
Now only those invariants are left that appear in the residues of the $N$-tachyon amplitude \eqref{eq:A_N_max_residue_ansatz}.

%%%%%%%%%%%%%%%%%%%%%%%%%%%%%%%%%%%%%%%%%%%%%%%%%%%%%%%%%%
%%%%%%%%%%%%%%%%%%%%%%%%%%%%%%%%%%%%%%%%%%%%%%%%%%%%%%%%%%

\section{Cyclicity as alternative input for fixing the residue coefficients}

For the $4$ and $5$ point tachyon amplitudes in the bosonic string the coefficients for the basis elements can also be fixed just by the assumption that the amplitudes are cyclic. It is likely that a generalisation to $N$ points is possible.

\subsubsection*{4 points}
\label{sec:cyclicity_fixes_coefficients_4point}

With the residues derived before, the $4$-point amplitude is
\be
\sum\limits_{A=0}^\infty \binom{k_{23}}{A} \frac{h_A}{k_{12}+A+1} ,
\ee
and the coefficients $h_A$ are to be determined. Due to momentum conservation $k_{34}=k_{12}$ holds and so cyclic invariance yields
\be
\sum\limits_{A=0}^\infty \binom{k_{23}}{A} \frac{h_A}{k_{12}+A+1} =
\sum\limits_{B=0}^\infty \binom{k_{12}}{B} \frac{h_B}{k_{23}+B+1} .
\ee
to calculate the constants $h_A$ consider the line $k_{12}=k_{23}-A'$ (with $A' \in \mathbb{N}$) in the space of kinematic variables
and  multiply both sides with $(k_{23}+1)$
\be
(k_{23}+1)\sum\limits_{A=0}^\infty \binom{k_{23}}{A} \frac{h_A}{k_{23}-A'+A+1} =
(k_{23}+1)\sum\limits_{B=0}^\infty \binom{k_{23}-A'}{B} \frac{h_B}{k_{23}+B+1} .
\ee
Now set $k_{23}=-1$ and obtain
\be
 (-1)^{A'} h_{A'} = h_0 .
\ee

\subsubsection*{5 points}
\label{sec:cyclicity_fixes_coefficients_5point}

The $5$-point amplitude is in terms of the basis derived above \eqref{eq:ResResA5_poly_basis}
\be
\sum\limits_{A,B=0}^\infty \sum\limits_{a=0}^{\min(A,B)} \binom{k_{23}}{A-a} \binom{k_{34}}{B-a} \binom{k_{24}}{a} \frac{h_{A,B,a}}{(k_{12}+A+1) (k_{45}+B+1)} .
\ee
It is useful to change to a set of variables that is mapped to itself under a cyclic relabelling of the external particles. For this, exchange $k_{24}$ for $k_{51}$ using
$k_{24} = k_{51}-k_{23}-k_{34}-1$
\be
\sum\limits_{A,B=0}^\infty \sum\limits_{a=0}^{\min(A,B)} \binom{k_{23}}{A-a} \binom{k_{34}}{B-a} \binom{k_{51}-k_{23}-k_{34}-1}{a} \frac{h_{A,B,a}}{(k_{12}+A+1) (k_{45}+B+1)} .
\ee
Consider the cyclic permutation by two positions
\be
\sum\limits_{A,B,a} h_{A,B,a} \frac{\binom{k_{23}}{A-a} \binom{k_{34}}{B-a} \binom{k_{51}-k_{23}-k_{34}-1}{a}}{(k_{12}+A+1) (k_{45}+B+1)}
= \sum\limits_{C,D,b} h_{C,D,b} \frac{\binom{k_{45}}{C-b} \binom{k_{51}}{D-b} \binom{k_{23}-k_{45}-k_{51}-1}{b}}{(k_{34}+C+1) (k_{12}+D+1)} .
\ee
This time restrict to $k_{45}=k_{34}-B'$ and multiply by $(k_{12}+A'+1)(k_{34}+C' + 1)$
\bea
& (k_{12}+A'+1)(k_{34}+C' + 1) \sum\limits_{A,B} \sum\limits_{a=0}^{\min(A,B)} h_{A,B,a} \frac{\binom{k_{23}}{A-a} \binom{k_{34}}{B-a} \binom{k_{51}-k_{23}-k_{34}-1}{a}}{(k_{12}+A+1) (k_{34}-B'+B+1)} \\
={}& (k_{12}+A'+1)(k_{34}+C' + 1) \sum\limits_{C,D} \sum\limits_{b=0}^{\min(C,D)}h_{C,D,b} \frac{\binom{k_{34}-B'}{C-b} \binom{k_{51}}{D-b} \binom{k_{23}-k_{34}+B'-k_{51}-1}{b}}{(k_{34}+C+1) (k_{12}+D+1)} .
\eea{eq:A5_coeff_relation1}
Set $k_{12}=-A'-1$ and $k_{34} =-C' -1$
\bea
& \sum\limits_{a=0}^{\min(A',B'+C')} h_{A',B'+C',a} \binom{k_{23}}{A'-a} \binom{-C' -1}{B'+C'-a} \binom{k_{51}-k_{23}+C'}{a} \\
={}& \sum\limits_{b=0}^{\min(C',A')} h_{C',A',b} \binom{-B'-C'-1}{C'-b} \binom{k_{51}}{A'-b} \binom{k_{23}+B'+C'-k_{51}}{b} .
\eea{eq:A5_coeff_relation2}
Now set $C'=0$
\be
 \sum\limits_{a=0}^{\min(A',B')} h_{A',B',a} \binom{k_{23}}{A'-a} \binom{-1}{B'-a} \binom{k_{51}-k_{23}}{a} =  h_{0,A',0}  \binom{k_{51}}{A'}  .
\label{eq:A5_coeff_relation3}
\ee
One can choose $k_{23}=k_{51}\not \in \mathbb{Z}$ (where the integers are avoided to make sure not to hit a zero) to gain
\be
 h_{A',B',0}  \binom{-1}{B'}  =  h_{0,A',0}   .
\ee
Applying this formula twice to $h_{0,0,0}$ gives us all coefficients with $a =0$
\be
 h_{A',B',0}   = (-1)^{A'+B'}  h_{0,0,0}   .
\label{eq:A5_coeff_a0}
\ee
To calculate the other coefficients, go back to \eqref{eq:A5_coeff_relation3} and set $k_{23} = -1$ and $k_{51}=a'-1$ with $a' \in \mathbb{N}, 1 \leq a' \leq\min(A',B')$ 
\be
 0 = \sum\limits_{a=0}^{a'} h_{A',B',a} \binom{-1}{A'-a} \binom{-1}{B'-a} \binom{a'}{a} = (-1)^{A'+B'}\sum\limits_{a=0}^{a'} h_{A',B',a} \binom{a'}{a}   .
\ee
These are enough equations to fix all $h_{A',B',a'}$ and given that the alternating sum of binomial coefficients vanishes, the solution is
\be
h_{A',B',a'}= (-1)^{a'} h_{A',B',0} .
\ee
Together with \eqref{eq:A5_coeff_a0} the result is
\be
h_{A',B',a'}= (-1)^{A'+B'+a'} h_{0,0,0} .
\ee

%%%%%%%%%%%%%%%%%%%%%%%%%%%%%%%%%%%%%%%%%%%%%%%%%%%%%%%%%%
%%%%%%%%%%%%%%%%%%%%%%%%%%%%%%%%%%%%%%%%%%%%%%%%%%%%%%%%%%

\section{Couplings of two tachyons and one massive particle}
\label{sec:one_massive_coeffs}

In this appendix \eqref{A4fact} is used to compute the general $3$-point coupling of two tachyons and one arbitrary on-shell particle. 
The right-hand side of \eqref{A4fact}  consists of contractions of the terms $(k_1 - k_2)^{\mu}$ and $(k_3 - k_4)^{\mu}$, i.e.\
\be
\left.  \frac{\alpha'}{2} (k_1 - k_2)^{\mu} \xi_{\mu}^{I} \xi_{\nu}^{I}(k_3 - k_4)^{\nu} \right|_{s_{12}=A-1} = s_{23} + \frac{A+3}{2} ,
\label{eq:A4_single_crossing}
\ee
and
\be
 \left.  \frac{\alpha'}{2} (k_1 - k_2)^{\mu} \xi_{\mu}^{I} \xi_{\nu}^{I}(k_1 - k_2)^{\nu} \right|_{s_{12}=A-1}
=  \left.  \frac{\alpha'}{2} (k_3 - k_4)^{\mu} \xi_{\mu}^{I} \xi_{\nu}^{I}(k_3 - k_4)^{\nu} \right|_{s_{12}=A-1}
= \frac{A+3}{2} .
\label{eq:A4_single_trace}
\ee
Start by writing the residues of the Veneziano amplitude as a function of the polynomial \eqref{eq:A4_single_crossing}. Then all couplings 
$c_{A,\alpha}$ that appear as part of the $3$-point amplitudes on the right-hand side of \eqref{A4fact} are computed by matching up the coefficients of these polynomials on both sides.

The residues of the Veneziano amplitude at mass level $A \in {\mathbb N}_0$ are 
\begin{equation}
-\Res_{s_{12}\to A-1} \Amp_4(s_{12},s_{23})
= \frac{1}{A!} 
\prod\limits_{i=1}^A(s_{23} +  1 + i) .
\end{equation}
These residues can be expressed as linear combinations of the terms \eqref{eq:A4_single_crossing}. For even $A$, 
\begin{equation}
\begin{aligned}
-\Res_{s_{12}\to A-1} \Amp_4(s_{12},s_{23}) &= \frac{1}{A!} 
\prod\limits_{i=1}^\frac{A}{2}\left\{ \left(s_{23} + \frac{A+3}{2}\right)^2 - \left(i - \frac{1}{2}\right)^2 \right\} \\
&=  \sum\limits_{k=0}^\frac{A}{2} V_{k, A\, \text{even}} \left(s_{23} + \frac{A+3}{2}\right)^{A - 2k} ,
\end{aligned}
\label{veneziano_pole_even}
\end{equation}
holds, with
\begin{equation}
V_{0, A\, \text{even}} = \frac{1}{A!}, \quad V_{k, A\, \text{even}} = \frac{(-1)^k}{A!}  \sum\limits_{j_1=1}^\frac{A}{2} \left(j_1 - \frac{1}{2}\right)^2
 \sum\limits_{j_2=j_1+1}^\frac{A}{2} \hspace{-.4em} \left(j_2 - \frac{1}{2}\right)^2 
\ldots
 \sum\limits_{j_k=j_{k-1}+1}^\frac{A}{2} \hspace{-.4em} \left(j_k - \frac{1}{2}\right)^2 .
\end{equation}
Similarly, for odd $A$
\begin{align}
-\Res_{s_{12}\to A-1} \Amp_4(s_{12},s_{23}) &= \frac{1}{A!} 
\left(s_{23} + \frac{A+3}{2}\right)\prod\limits_{i=1}^\frac{A-1}{2}\left\{ \left(s_{23} + \frac{A+3}{2}\right)^2 - i^2 \right\} \nonumber\\
&= \sum\limits_{k=0}^\frac{A-1}{2} V_{k, A\, \text{odd}} \left(s_{23} + \frac{A+3}{2}\right)^{A - 2k} ,
\label{veneziano_pole_odd}
\end{align}
with
\begin{equation}
V_{0, A\, \text{odd}} = \frac{1}{A!}, \qquad V_{k, A\, \text{odd}} =\frac{(-1)^k}{A!} \sum\limits_{j_1=1}^\frac{A-1}{2} j_1^2 
 \sum\limits_{j_2=j_1+1}^\frac{A-1}{2} j_2^2 
\ldots
 \sum\limits_{j_k=j_{k-1}+1}^\frac{A-1}{2} j_k^2 .
\end{equation}
$V_{k, A\, \text{even}}$ and $V_{k, A\, \text{odd}}$ are essentially the central factorial numbers $t(A,k)$ and $t_2(A,k)$ %see \href{oeis.org/A008955} and \href{oeis.org/A008956})
\begin{align}
V_{k, A\, \text{even}} &=\frac{(-1)^k}{A! 4^k} t_2\left(\frac{A}{2}, k\right) , &0 \leq k \leq \lfloor \frac{A}{2} \rfloor ,\\
V_{k, A\, \text{odd}} &= \frac{(-1)^k}{A!} t\left(\frac{A-1}{2}, k\right) , &0 \leq k \leq \lfloor \frac{A}{2} \rfloor ,
\end{align}
where 
\begin{align}
t(n, 0) &= 1,\nonumber \\
t(n, n) &= (n!)^2, \label{eq:cfn1}\\
t(n, k) &= n^2 t(n - 1, k - 1) + t(n - 1, k) ,\nonumber \\
t_2(n, 0) &= 1, \nonumber \\
t_2(n, n) &= ((2n - 1)!!)^2 , \label{eq:cfn2} \\
t_2(n, k) &= (2n - 1)^2 t_2(n - 1, k - 1) + t_2(n - 1, k) .\nonumber 
\end{align}

The interesting part on the right-hand side of \eqref{A4fact} are the projectors
$\mathbf{P}_{\alpha}$ to the ${SO(D-1)}$ or ${SO(D-2)}$ irrep $\alpha$.
Since the $3$-point amplitudes with two tachyons are already fully symmetric
\eqref{TTM_sym}, only 
the projector which projects out the trace part of the product of two symmetric
$|\alpha|$-tensors is needed. This is derived below in Section \ref{sec:tl_projector} and reads
\bea
\mathbf{P}^{'}_{I_1 \ldots I_{|\alpha|}, J_1 \ldots J_{|\alpha|}} 
= \sum\limits_{k=0}^{\lfloor \frac{|\alpha|}{2} \rfloor}  W_{|\alpha|,k}  \prod\limits_{a=1}^k \delta_{I_{2a-1}, I_{2a}} \delta_{J_{2a-1}, J_{2a}} \prod\limits_{b=2k+1}^{|\alpha|} \delta_{I_b,J_b} ,
\eea{eq:tracefree_projector}
where $\lfloor \frac{|\alpha|}{2} \rfloor$ is the greatest integer less or equal $\frac{|\alpha|}{2}$ and
\begin{equation}
W_{|\alpha|, 0} = 1, \qquad W_{|\alpha|,k} = (-1)^k \frac{|\alpha|!}{(|\alpha|-2k)! 2^k k!}  \prod\limits_{a=1}^k \frac{1}{d+2|\alpha|-2a-2} ,
\end{equation}
with $d=D-1$.
Since only symmetric tracefree irreps appear it is unambiguous to label the couplings $c_{A,\alpha}$ by $c_{A,|\alpha|}$ in this section.
Inserting the projector into \eqref{A4fact} and using  \eqref{eq:A4_single_crossing} and \eqref{eq:A4_single_trace} yields
\bea
& \sum\limits_{|\alpha|=0}^A c_{A,|\alpha|}^2
 \sum\limits_{k=0}^{\lfloor \frac{|\alpha|}{2} \rfloor}  W_{|\alpha|,k}
 \left( \frac{A+3}{2} \right)^{2k} \left( s_{23} + \frac{A+3}{2} \right)^{|\alpha|-2k} \\
={}& c_{A,A}^2  \left(s_{23} + \frac{A+3}{2}\right)^{A}
+ \left[ c_{A,A}^2 W_{A,1} \left( \frac{A+3}{2} \right)^{2}  + c_{A,A-2}^2 \right]  \left(s_{23} + \frac{A+3}{2}\right)^{A-2} \\
&+ \left[ c_{A,A}^2 W_{A,2} \left( \frac{A+3}{2} \right)^{4} + c_{A,A-2}^2 W_{A,1} \left( \frac{A+3}{2} \right)^{2} + c_{A,A-4}^2 \right]  \left(s_{23} + \frac{A+3}{2}\right)^{A-4} + \ldots .
\eea{eq:RHS_A4fact}
Here it was already used that only even or odd powers of the polynomial in $s_{23}$ appear exclusively in (\ref{veneziano_pole_even},\ref{veneziano_pole_odd}). This implies that all $c_{A,|\alpha|}^2$
 with $(A-|\alpha|)$ odd are zero which is expected as explained in Section \ref{sec:unit_ttm}. Now a recursive formula for $c_{A,|\alpha|}^2$ can be read off by matching up (\ref{veneziano_pole_even},\ref{veneziano_pole_odd}) and \eqref{eq:RHS_A4fact}
\begin{equation}
c_{A,A}^2 = V_{0,A}, \qquad c_{A,A-2k}^2 = V_{k,A} - \sum\limits_{l=1}^{k} c_{A,A-2k+2l}^2 W_{A,l}  \left(\frac{A+3}{2}\right)^{2l} .
\label{veneziano_coefficients_recursive}
\end{equation}
Observing that each term which multiplies the number $V_{k,A}$ in \eqref{veneziano_coefficients_recursive} contains the same power of $\left( \frac{A+3}{2}\right)$,
the recursion relation can be cast into the form
\begin{equation}
c_{A,A-2k}^2 = \sum\limits_{l=0}^{k} V_{k-l,A} \left(\frac{A+3}{2} \right)^{2l} M^{A,k}_l ,
\label{eq:veneziano_coefficients}
\end{equation}
with
\begin{equation}
M^{A,k}_0 = 1 , \qquad  M^{A,k}_l = - \sum\limits_{j=1}^{l} W_{A-2k+2l, j}  M^{A,k}_{l-j} .
\end{equation}
This can be expressed in a closed form. Start simplifying with the observation that (with $|\alpha|=A-2k$ and $\lambda = (\lambda_1, \ldots, \lambda_m)$ denotes a partition of $l$) each term in $M^{A,k}_l$
consists of a product 
\be
\prod\limits_{i=1}^m W_{|\alpha|+2(l-\sum_{k<i} \lambda_k), \lambda_i} .
\ee
A common factor can be pulled out of all of these products
\bea
 \prod\limits_{i=1}^m W_{|\alpha|+2(l-\sum_{k<i} \lambda_k), \lambda_i} 
= &\prod\limits_{i=1}^m \frac{(-1)^{\lambda_i} 
(|\alpha|+2(l-\sum_{k<i} \lambda_k))_{2\lambda_i} }{2^{2\lambda_i}} \tilde W_{|\alpha|+2(l-\sum_{k<i} \lambda_k), \lambda_i} \\
= &\frac{(-1)^{l} 
(|\alpha|+1)^{(2l)} }{2^{2l}} 
\prod\limits_{i=1}^m  \tilde W_{|\alpha|+2(l-\sum_{k<i} \lambda_k), \lambda_i} , \\
\text{where} \quad \tilde W_{|\alpha|, j} = & \frac{1}{j! (\frac{d}{2} + |\alpha| - 2)_j} ,
\eea{eq:wprime_appears}
and the raising and falling factorials $(x)^{(l)} = (x+l-1)_l = x (x+1) (x+2)\ldots (x+l-1)$ were used.
$M^{A,k}_l$ is proportional to this overall factor
\begin{equation}
M^{A,k}_l = \frac{(-1)^{l} (|\alpha|+1)^{(2l)} }{2^{2l}} \tilde M^{A,k}_l, \qquad \tilde M^{A,k}_0 = 1 , \qquad  \tilde M^{A,k}_l = - \sum\limits_{j=1}^{l} \tilde W_{|\alpha|+2l, j} \tilde M^{A,k}_{l-j} .
\end{equation}
Next  by induction it can be proven that
\begin{equation}
\tilde M^{A,k}_l = \frac{(-1)^l}{ l! (\frac{d}{2}+|\alpha|)^{(l)}  } .
\label{eq:M_tilde}
\end{equation}
The statement is true for $l=0$ . Plugging in $\tilde M^{A,k}_{l-j}$ into the recursive definition for the induction step yields
\bea
\tilde M^{A,k}_l &= - \sum\limits_{j=1}^{l} \frac{1}{j! (\frac{d}{2} + |\alpha| + 2l - 2)_j} \frac{(-1)^{l-j}}{ (l-j)! (\frac{d}{2}+|\alpha|)^{(l-j)}  }  \\
&= \frac{1}{(\frac{d}{2}+|\alpha|)^{(2l-1)}} \sum\limits_{j=1}^{l}  \frac{(-1)^{l-j+1} (\frac{d}{2}+|\alpha|+l-j)^{(l-1)}  }{j! (l-j)!} .
\eea{eq:M_induction_step}
To show that this equals \eqref{eq:M_tilde} use the identity
\be
\sum\limits_{j=0}^{l}  \frac{(-1)^{l-j+1} (\frac{d}{2}+|\alpha|+l-j)^{(l-1)}  }{j! (l-j)!} = 0 ,
\ee
which can be proved using computer algebra.
This yields the expression
\begin{equation}
M^{A,k}_l = \frac{(|\alpha|+1)^{(2l)} }{2^{2l} l! (\frac{d}{2}+|\alpha|)^{(l)}  } ,
\end{equation}
which inserted into \eqref{eq:veneziano_coefficients} yields the final result
\begin{align}
c_{A,|\alpha|}^2 = \begin{cases}
{\displaystyle \sum\limits_{l=0}^{\frac{A-|\alpha|}{2}} V_{\frac{A-|\alpha|}{2}-l,A} \left(\frac{A+3}{4}  \right)^{2l} \frac{(|\alpha|+1)^{(2l)}}{l!(\frac{d}{2}+|\alpha|)^{(l)} } } \qquad &A-|\alpha|\ \text{even}, \\[1em]
0 \qquad &A-|\alpha|\ \text{odd}.
\end{cases}
\label{eq:TTM_couplings}
\end{align}
This is the result quoted in the main text.

\subsection{Projector from symmetric to traceless symmetric tensors}
\label{sec:tl_projector}

In this section the projector from symmetric tensors to traceless symmetric tensors
and its simplified form that can be used when contracting with symmetric tensors 
from both sides are derived.
Be $T^m = T^{I_1 \ldots I_m}$ a symmetric $m$-tensor over $\mathbb{R}^d$ and $T^m_k$ its contraction
with $k$ Kronecker deltas
\begin{equation}
T^m_k = T^{I_1 \ldots I_m} \delta_{I_{1} I_{2}} \ldots \delta_{I_{2k - 1} I_{2k}} .
\end{equation}
Furthermore, let $\delta^l T^m_k$ be the symmetrised product of $T^m_k$ and $l$ Kronecker deltas.
For example
\bea
\delta^1 T^4_1 = &\left( \delta^{I_1 I_2} T^{I_3 I_4 J_1 J_2} + \delta^{I_1 I_3} T^{I_2 I_4 J_1 J_2} +\delta^{I_1 I_4} T^{I_2 I_3 J_1 J_2} \right. \\
& \left. +\delta^{I_2 I_3} T^{I_1 I_4 J_1 J_2} + \delta^{I_2 I_4} T^{I_1 I_3 J_1 J_2} +\delta^{I_3 I_4} T^{I_1 I_2 J_1 J_2} \right) \delta_{J_1 J_2} .
\eea{eq:deltaT_example}
The number of terms contained in $\delta^l T^m_k$ is
\begin{equation}
\#^{k,l} = \frac{(m-2k+2l)!}{(m-2k)! 2^l l!} .
\label{P_def}
\end{equation}
Our ansatz for constructing the projector is to subtract all terms that have the correct number of indices and are manifestly symmetric
\begin{align}
\mathbf{P} T^m &= T^m - \frac{1}{Q_1} \left\{ \delta^1 T^m_1 -\frac{1}{Q_2} \left\{  \delta^2 T^m_2 - \frac{1}{Q_3} \left\{  \delta^3 T^m_3 - \ldots\right\}\right\}\right\} .
\label{eq:tl_ansatz}
\end{align}
The coefficients $Q_i$ are determined by solving
\be
0 = \delta^{JK} \mathbf{P} T^{J K I_3 \ldots I_m}  .
\label{eq:tl_condition}
\ee
In order to solve this equation first analyse how often terms with a given distribution of these indices over the Kronecker deltas and the tensor $T^m_k$ appear in $\delta^l T^m_k$.
To this end the following notation is introduced,
\bea               
  (\delta^l T^m_k)^{JK}_{(2,0,0)} &= \sum \delta^{JK} \delta^{\ldots} \ldots T^{\ldots} ,\\
  (\delta^l T^m_k)^{JK}_{(1,1,0)} &= \sum \delta^{J\ldots} \delta^{K\ldots} \ldots T^{\ldots} ,\\
  (\delta^l T^m_k)^{JK}_{(1,0,1)} &= \sum \delta^{J\ldots} \ldots T^{K \ldots} , \\
  (\delta^l T^m_k)^{JK}_{(0,1,1)} &= \sum \delta^{K\ldots} \ldots T^{J \ldots} , \\
  (\delta^l T^m_k)^{JK}_{(0,0,2)} &= \sum \delta^{\ldots} \ldots T^{JK \ldots} ,
\eea{eq:deltaT_notation}
where
\begin{equation}
\delta^l T^m_k = (\delta^l T^m_k)^{JK}_{(2,0,0)} + (\delta^l T^m_k)^{JK}_{(1,1,0)} + (\delta^l T^m_k)^{JK}_{(1,0,1)} + (\delta^l T^m_k)^{JK}_{(0,1,1)} + (\delta^l T^m_k)^{JK}_{(0,0,2)}.
\label{split_delta_T}
\end{equation}
Analogously to the overall number of terms \eqref{P_def} the number of terms in the sums in \eqref{eq:deltaT_notation} are
\begin{eqnarray}
  \#^{k,l}_{(2,0,0)} =& \displaystyle \frac{(m-2k +2l -2)!}{(m-2k)! 2^{l-1}(l-1)!} &= \#^{k,l-1}  \label{dkTmk2}\\
  \#^{k,l}_{(1,1,0)} =& \displaystyle  \frac{(m-2k +2l -2)!}{(m-2k)! 2^{l-2}(l-2)!} &= \#^{k,l-1} 2 (l-1) \label{dkTmk11}\\
  \#^{k,l}_{(1,0,1)} = \#^{k,l}_{(0,1,1)} =&\displaystyle  \frac{(m-2k +2l -2)!}{(m-2k-1)! 2^{l-1}(l-1)!} &= \#^{k,l-1} (m - 2k)\label{dkTmk1}\\
  \#^{k,l}_{(0,0,2)} =&\displaystyle  \frac{(m-2k +2l -2)!}{(m-2k-2)! 2^{l}l!} &= \#^{k+1,l} \label{dkTmk0}
\end{eqnarray}
As a consistency check, one can show
\begin{equation}
\#^{k,l} = \#^{k,l}_{(2,0,0)} + \#^{k,l}_{(1,1,0)} + \#^{k,l}_{(1,0,1)} + \#^{k,l}_{(0,1,1)} + \#^{k,l}_{(0,0,2)} .
\end{equation}
When contracted with $\delta^{JK}$, the first four lines of \eqref{eq:deltaT_notation} each turn into the $\#^{k,l-1}$ terms $\delta^{l-1} T^m_k$ times an integer factor
which can be read of from the right-hand side of (\ref{dkTmk2}-\ref{dkTmk1}).
\bea
& \delta^{JK}\left((\delta^l T^m_k)^{JK}_{(2,0,0)} + (\delta^l T^m_k)^{JK}_{(1,1,0)} + (\delta^l T^m_k)^{JK}_{(1,0,1)} + (\delta^l T^m_k)^{JK}_{(0,1,1)} \right) \\
={}& (d + 2 (l-1) + 2(m - 2k) ) \delta^{l-1} T^m_k\\
\equiv{} & R_{k,l} \delta^{l-1} T^m_k 
\eea{eq:contract_delta_T_1}
The last line of \eqref{eq:deltaT_notation} contracted with $\delta^{JK}$ becomes
\begin{equation}
  \delta^{JK}(\delta^l T^m_k)^{JK}_{(0,0,2)} = \delta^{l} T^m_{k+1} .
\label{eq:contract_delta_T_2}
\end{equation}
Now insert \eqref{split_delta_T} and (\ref{eq:contract_delta_T_1}-\ref{eq:contract_delta_T_2}) into \eqref{eq:tl_condition}
\begin{align}
0 = \delta^{JK} \mathbf{P} T^m =& T^m_1 - \frac{1}{Q_1} \Bigg\{ R_{1,1} T^m_1 + \delta^1 T^m_2 -\frac{1}{Q_2} \Big\{ R_{2,2}  \delta^1 T^m_2 + \delta^2 T^m_3\\
& {} - \frac{1}{Q_3} \left\{ R_{3,3} \delta^2 T^m_3 + \delta^3 T^m_4 - \ldots\right\}\Big\}\Bigg\} ,
\end{align}
and conclude
\begin{equation}
Q_i = R_{i,i} = d + 2(m - i - 1) .
\end{equation}
This proves
\begin{align}
\mathbf{P}  T^m =& \sum\limits_{k=0}^{\lfloor \frac{m}{2} \rfloor} (-1)^k \left(\prod\limits_{j=1}^{k} \frac{1}{d+2m-2j-2} \right) \delta^k T^m_k .
\label{tracefree_tensor}
\end{align}
If $\mathbf{P}$ is contracted to symmetric tensors on both sides every term in $\delta^k T^m_k$ yields the same contribution. 
Define the simplified projector $\mathbf{P}^{'}$ where $\delta^k T^m_k$ is replaced by one of its terms times the number of terms
which is given by \eqref{P_def}
\bea
\mathbf{P}^{'}_{I_1 \ldots I_{m}, J_1 \ldots J_{m}} 
= \sum\limits_{k=0}^{\lfloor \frac{m}{2} \rfloor}  W_{m,k}  \prod\limits_{a=1}^k \delta_{I_{2a-1}, I_{2a}} \delta_{J_{2a-1}, J_{2a}} \prod\limits_{b=2k+1}^{m} \delta_{I_b,J_b} ,
\eea{eq:tracefree_projector_2}
where
\begin{equation}
W_{m, 0} = 1, \qquad W_{m,k} = (-1)^k \frac{m!}{(m-2k)! 2^k k!}  \prod\limits_{a=1}^k \frac{1}{d+2m-2a-2} .
\end{equation}
This simplified projector satisfies for two symmetric tensors $T_L^m, T_R^m$
\begin{equation}
T^m_L \mathbf{P} T^m_R = T^m_L \mathbf{P}^{'} T^m_R .
\end{equation}

\bibliographystyle{JHEP}

\bibliography{stringtargetspace}

\end{document}

%% file: bt/A5-mp.pdf_t
\begin{picture}(0,0)%
\includegraphics{A5-mp.pdf}%
\end{picture}%
\setlength{\unitlength}{3947sp}%
\begingroup\makeatletter\ifx\SetFigFont\undefined%
\gdef\SetFigFont#1#2#3#4#5{%
  \reset@font\fontsize{#1}{#2pt}%
  \fontfamily{#3}\fontseries{#4}\fontshape{#5}%
  \selectfont}%
\fi\endgroup%
\begin{picture}(3620,2055)(1486,-2701)
\put(1501,-2461){\makebox(0,0)[lb]{\smash{{\SetFigFont{25}{30.0}{\familydefault}{\mddefault}{\updefault}{\color[rgb]{0,0,0}2}%
}}}}
\put(3226,-2686){\makebox(0,0)[lb]{\smash{{\SetFigFont{25}{30.0}{\familydefault}{\mddefault}{\updefault}{\color[rgb]{0,0,0}3}%
}}}}
\put(1501,-961){\makebox(0,0)[lb]{\smash{{\SetFigFont{25}{30.0}{\familydefault}{\mddefault}{\updefault}{\color[rgb]{0,0,0}1}%
}}}}
\put(4876,-961){\makebox(0,0)[lb]{\smash{{\SetFigFont{25}{30.0}{\familydefault}{\mddefault}{\updefault}{\color[rgb]{0,0,0}5}%
}}}}
\put(4876,-2461){\makebox(0,0)[lb]{\smash{{\SetFigFont{25}{30.0}{\familydefault}{\mddefault}{\updefault}{\color[rgb]{0,0,0}4}%
}}}}
\end{picture}%

%% file: bt/AN-mp.pdf_t
\begin{picture}(0,0)%
\includegraphics{AN-mp.pdf}%
\end{picture}%
\setlength{\unitlength}{3947sp}%
\begingroup\makeatletter\ifx\SetFigFont\undefined%
\gdef\SetFigFont#1#2#3#4#5{%
  \reset@font\fontsize{#1}{#2pt}%
  \fontfamily{#3}\fontseries{#4}\fontshape{#5}%
  \selectfont}%
\fi\endgroup%
\begin{picture}(6480,2163)(1486,-2809)
\put(1501,-2461){\makebox(0,0)[lb]{\smash{{\SetFigFont{25}{30.0}{\familydefault}{\mddefault}{\updefault}{\color[rgb]{0,0,0}2}%
}}}}
\put(3226,-2686){\makebox(0,0)[lb]{\smash{{\SetFigFont{25}{30.0}{\familydefault}{\mddefault}{\updefault}{\color[rgb]{0,0,0}3}%
}}}}
\put(1501,-961){\makebox(0,0)[lb]{\smash{{\SetFigFont{25}{30.0}{\familydefault}{\mddefault}{\updefault}{\color[rgb]{0,0,0}1}%
}}}}
\put(7951,-2461){\makebox(0,0)[lb]{\smash{{\SetFigFont{25}{30.0}{\familydefault}{\mddefault}{\updefault}{\color[rgb]{0,0,0}$N-1$}%
}}}}
\put(7951,-961){\makebox(0,0)[lb]{\smash{{\SetFigFont{25}{30.0}{\familydefault}{\mddefault}{\updefault}{\color[rgb]{0,0,0}$N$}%
}}}}
\put(5776,-2686){\makebox(0,0)[lb]{\smash{{\SetFigFont{25}{30.0}{\familydefault}{\mddefault}{\updefault}{\color[rgb]{0,0,0}$N-2$}%
}}}}
\end{picture}%

%% file: bt/glue_projector.pdf_t
\begin{picture}(0,0)%
\includegraphics{glue_projector.pdf}%
\end{picture}%
\setlength{\unitlength}{3947sp}%
\begingroup\makeatletter\ifx\SetFigFont\undefined%
\gdef\SetFigFont#1#2#3#4#5{%
  \reset@font\fontsize{#1}{#2pt}%
  \fontfamily{#3}\fontseries{#4}\fontshape{#5}%
  \selectfont}%
\fi\endgroup%
\begin{picture}(8680,2428)(1586,-1575)
\put(5626,-511){\makebox(0,0)[lb]{\smash{{\SetFigFont{34}{40.8}{\familydefault}{\mddefault}{\updefault}{\color[rgb]{0,0,0}$\mathbf{P}_\alpha$}%
}}}}
\put(3301,-436){\makebox(0,0)[lb]{\smash{{\SetFigFont{34}{40.8}{\familydefault}{\mddefault}{\updefault}{\color[rgb]{0,0,0}$\mathbf{P}^{\perp}_{k}$}%
}}}}
\put(7951,-436){\makebox(0,0)[lb]{\smash{{\SetFigFont{34}{40.8}{\familydefault}{\mddefault}{\updefault}{\color[rgb]{0,0,0}$\mathbf{P}^{\perp}_{k}$}%
}}}}
\end{picture}%

%% file: bt/k1perp.pdf_t
\begin{picture}(0,0)%
\includegraphics{k1perp.pdf}%
\end{picture}%
\setlength{\unitlength}{3947sp}%
\begingroup\makeatletter\ifx\SetFigFont\undefined%
\gdef\SetFigFont#1#2#3#4#5{%
  \reset@font\fontsize{#1}{#2pt}%
  \fontfamily{#3}\fontseries{#4}\fontshape{#5}%
  \selectfont}%
\fi\endgroup%
\begin{picture}(5487,2426)(5829,-2474)
\put(8836,-1351){\makebox(0,0)[lb]{\smash{{\SetFigFont{34}{40.8}{\familydefault}{\mddefault}{\updefault}{\color[rgb]{0,0,0}$\mathbf{P}^{\perp}_{k_1+k_2}$}%
}}}}
\put(6676,-1411){\makebox(0,0)[lb]{\smash{{\SetFigFont{34}{40.8}{\familydefault}{\mddefault}{\updefault}{\color[rgb]{0,0,0}$k_1$}%
}}}}
\end{picture}%

%% file: bt/k2perp.pdf_t
\begin{picture}(0,0)%
\includegraphics{k2perp.pdf}%
\end{picture}%
\setlength{\unitlength}{3947sp}%
\begingroup\makeatletter\ifx\SetFigFont\undefined%
\gdef\SetFigFont#1#2#3#4#5{%
  \reset@font\fontsize{#1}{#2pt}%
  \fontfamily{#3}\fontseries{#4}\fontshape{#5}%
  \selectfont}%
\fi\endgroup%
\begin{picture}(5487,2426)(5829,-2474)
\put(8836,-1351){\makebox(0,0)[lb]{\smash{{\SetFigFont{34}{40.8}{\familydefault}{\mddefault}{\updefault}{\color[rgb]{0,0,0}$\mathbf{P}^{\perp}_{k_1+k_2}$}%
}}}}
\put(6451,-1411){\makebox(0,0)[lb]{\smash{{\SetFigFont{34}{40.8}{\familydefault}{\mddefault}{\updefault}{\color[rgb]{0,0,0}$-k_2$}%
}}}}
\end{picture}%

%% file: bt/k1k2perp.pdf_t
\begin{picture}(0,0)%
\includegraphics{k1k2perp.pdf}%
\end{picture}%
\setlength{\unitlength}{3947sp}%
\begingroup\makeatletter\ifx\SetFigFont\undefined%
\gdef\SetFigFont#1#2#3#4#5{%
  \reset@font\fontsize{#1}{#2pt}%
  \fontfamily{#3}\fontseries{#4}\fontshape{#5}%
  \selectfont}%
\fi\endgroup%
\begin{picture}(5487,2426)(5829,-2474)
\put(8836,-1351){\makebox(0,0)[lb]{\smash{{\SetFigFont{34}{40.8}{\familydefault}{\mddefault}{\updefault}{\color[rgb]{0,0,0}$\mathbf{P}^{\perp}_{k_1+k_2}$}%
}}}}
\put(6151,-1411){\makebox(0,0)[lb]{\smash{{\SetFigFont{34}{40.8}{\familydefault}{\mddefault}{\updefault}{\color[rgb]{0,0,0}$k_1-k_2$}%
}}}}
\end{picture}%

%% file: bt/k1k2.pdf_t
\begin{picture}(0,0)%
\includegraphics{k1k2.pdf}%
\end{picture}%
\setlength{\unitlength}{3947sp}%
\begingroup\makeatletter\ifx\SetFigFont\undefined%
\gdef\SetFigFont#1#2#3#4#5{%
  \reset@font\fontsize{#1}{#2pt}%
  \fontfamily{#3}\fontseries{#4}\fontshape{#5}%
  \selectfont}%
\fi\endgroup%
\begin{picture}(3237,2294)(5829,-2408)
\put(6151,-1411){\makebox(0,0)[lb]{\smash{{\SetFigFont{34}{40.8}{\familydefault}{\mddefault}{\updefault}{\color[rgb]{0,0,0}$k_1-k_2$}%
}}}}
\end{picture}%

%% file: bt/A_TTM_diagram.pdf_t
\begin{picture}(0,0)%
\includegraphics{A_TTM_diagram.pdf}%
\end{picture}%
\setlength{\unitlength}{3947sp}%
\begingroup\makeatletter\ifx\SetFigFont\undefined%
\gdef\SetFigFont#1#2#3#4#5{%
  \reset@font\fontsize{#1}{#2pt}%
  \fontfamily{#3}\fontseries{#4}\fontshape{#5}%
  \selectfont}%
\fi\endgroup%
\begin{picture}(4390,3274)(5976,-3251)
\put(8476,-3061){\makebox(0,0)[lb]{\smash{{\SetFigFont{34}{40.8}{\familydefault}{\mddefault}{\updefault}{\color[rgb]{0,0,0}$A,\alpha$}%
}}}}
\put(10126,-436){\makebox(0,0)[lb]{\smash{{\SetFigFont{34}{40.8}{\familydefault}{\mddefault}{\updefault}{\color[rgb]{0,0,0}$\mu_1$}%
}}}}
\put(10126,-2311){\makebox(0,0)[lb]{\smash{{\SetFigFont{34}{40.8}{\familydefault}{\mddefault}{\updefault}{\color[rgb]{0,0,0}$\mu_{|\alpha|}$}%
}}}}
\put(6526,-1411){\makebox(0,0)[lb]{\smash{{\SetFigFont{34}{40.8}{\familydefault}{\mddefault}{\updefault}{\color[rgb]{0,0,0}$k_1-k_2$}%
}}}}
\end{picture}%

%% file: bt/A_TTM_diagram_plain.pdf_t
\begin{picture}(0,0)%
\includegraphics{A_TTM_diagram_plain.pdf}%
\end{picture}%
\setlength{\unitlength}{3947sp}%
\begingroup\makeatletter\ifx\SetFigFont\undefined%
\gdef\SetFigFont#1#2#3#4#5{%
  \reset@font\fontsize{#1}{#2pt}%
  \fontfamily{#3}\fontseries{#4}\fontshape{#5}%
  \selectfont}%
\fi\endgroup%
\begin{picture}(3990,2448)(5976,-2485)
\put(6526,-1411){\makebox(0,0)[lb]{\smash{{\SetFigFont{34}{40.8}{\familydefault}{\mddefault}{\updefault}{\color[rgb]{0,0,0}$k_1-k_2$}%
}}}}
\end{picture}%

%% file: bt/A_TTM_sym.pdf_t
\begin{picture}(0,0)%
\includegraphics{A_TTM_sym.pdf}%
\end{picture}%
\setlength{\unitlength}{3947sp}%
\begingroup\makeatletter\ifx\SetFigFont\undefined%
\gdef\SetFigFont#1#2#3#4#5{%
  \reset@font\fontsize{#1}{#2pt}%
  \fontfamily{#3}\fontseries{#4}\fontshape{#5}%
  \selectfont}%
\fi\endgroup%
\begin{picture}(5190,2448)(5976,-2485)
\put(6526,-1411){\makebox(0,0)[lb]{\smash{{\SetFigFont{34}{40.8}{\familydefault}{\mddefault}{\updefault}{\color[rgb]{0,0,0}$k_1-k_2$}%
}}}}
\end{picture}%

%% file: bt/4point.pdf_t
\begin{picture}(0,0)%
\includegraphics{4point.pdf}%
\end{picture}%
\setlength{\unitlength}{3947sp}%
\begingroup\makeatletter\ifx\SetFigFont\undefined%
\gdef\SetFigFont#1#2#3#4#5{%
  \reset@font\fontsize{#1}{#2pt}%
  \fontfamily{#3}\fontseries{#4}\fontshape{#5}%
  \selectfont}%
\fi\endgroup%
\begin{picture}(8649,3214)(1125,-3551)
\put(1675,-1726){\makebox(0,0)[lb]{\smash{{\SetFigFont{34}{40.8}{\familydefault}{\mddefault}{\updefault}{\color[rgb]{0,0,0}$k_1-k_2$}%
}}}}
\put(5026,-3361){\makebox(0,0)[lb]{\smash{{\SetFigFont{34}{40.8}{\familydefault}{\mddefault}{\updefault}{\color[rgb]{0,0,0}$A,\alpha$}%
}}}}
\put(5101,-1711){\makebox(0,0)[lb]{\smash{{\SetFigFont{34}{40.8}{\familydefault}{\mddefault}{\updefault}{\color[rgb]{0,0,0}$\mathbf{P}_\alpha$}%
}}}}
\put(7876,-1711){\makebox(0,0)[lb]{\smash{{\SetFigFont{34}{40.8}{\familydefault}{\mddefault}{\updefault}{\color[rgb]{0,0,0}$k_3-k_4$}%
}}}}
\end{picture}%

%% file: bt/ex21_1234.pdf_t
\begin{picture}(0,0)%
\includegraphics{ex21_1234.pdf}%
\end{picture}%
\setlength{\unitlength}{3947sp}%
\begingroup\makeatletter\ifx\SetFigFont\undefined%
\gdef\SetFigFont#1#2#3#4#5{%
  \reset@font\fontsize{#1}{#2pt}%
  \fontfamily{#3}\fontseries{#4}\fontshape{#5}%
  \selectfont}%
\fi\endgroup%
\begin{picture}(4846,1844)(2978,-2183)
\put(3226,-1411){\makebox(0,0)[lb]{\smash{{\SetFigFont{34}{40.8}{\familydefault}{\mddefault}{\updefault}{\color[rgb]{0,0,0}$k_1-k_2$}%
}}}}
\put(6226,-1411){\makebox(0,0)[lb]{\smash{{\SetFigFont{34}{40.8}{\familydefault}{\mddefault}{\updefault}{\color[rgb]{0,0,0}$k_3-k_4$}%
}}}}
\end{picture}%

%% file: bt/ex22_A12.pdf_t
\begin{picture}(0,0)%
\includegraphics{ex22_A12.pdf}%
\end{picture}%
\setlength{\unitlength}{3947sp}%
\begingroup\makeatletter\ifx\SetFigFont\undefined%
\gdef\SetFigFont#1#2#3#4#5{%
  \reset@font\fontsize{#1}{#2pt}%
  \fontfamily{#3}\fontseries{#4}\fontshape{#5}%
  \selectfont}%
\fi\endgroup%
\begin{picture}(2788,1844)(6278,-2183)
\put(6526,-1411){\makebox(0,0)[lb]{\smash{{\SetFigFont{34}{40.8}{\familydefault}{\mddefault}{\updefault}{\color[rgb]{0,0,0}$k_1-k_2$}%
}}}}
\end{picture}%

%% file: bt/ex22_A34.pdf_t
\begin{picture}(0,0)%
\includegraphics{ex22_A34.pdf}%
\end{picture}%
\setlength{\unitlength}{3947sp}%
\begingroup\makeatletter\ifx\SetFigFont\undefined%
\gdef\SetFigFont#1#2#3#4#5{%
  \reset@font\fontsize{#1}{#2pt}%
  \fontfamily{#3}\fontseries{#4}\fontshape{#5}%
  \selectfont}%
\fi\endgroup%
\begin{picture}(2788,1844)(5336,-2183)
\put(6526,-1411){\makebox(0,0)[lb]{\smash{{\SetFigFont{34}{40.8}{\familydefault}{\mddefault}{\updefault}{\color[rgb]{0,0,0}$k_3-k_4$}%
}}}}
\end{picture}%

%% file: bt/A_TMM_diagram.pdf_t
\begin{picture}(0,0)%
\includegraphics{A_TMM_diagram.pdf}%
\end{picture}%
\setlength{\unitlength}{3947sp}%
\begingroup\makeatletter\ifx\SetFigFont\undefined%
\gdef\SetFigFont#1#2#3#4#5{%
  \reset@font\fontsize{#1}{#2pt}%
  \fontfamily{#3}\fontseries{#4}\fontshape{#5}%
  \selectfont}%
\fi\endgroup%
\begin{picture}(8655,5219)(1711,-4733)
\put(1726,-436){\makebox(0,0)[lb]{\smash{{\SetFigFont{34}{40.8}{\familydefault}{\mddefault}{\updefault}{\color[rgb]{0,0,0}$\mu_1$}%
}}}}
\put(10126,-436){\makebox(0,0)[lb]{\smash{{\SetFigFont{34}{40.8}{\familydefault}{\mddefault}{\updefault}{\color[rgb]{0,0,0}$\nu_1$}%
}}}}
\put(6226,-736){\makebox(0,0)[lb]{\smash{{\SetFigFont{34}{40.8}{\familydefault}{\mddefault}{\updefault}{\color[rgb]{0,0,0}$q$}%
}}}}
\put(10126,-2311){\makebox(0,0)[lb]{\smash{{\SetFigFont{34}{40.8}{\familydefault}{\mddefault}{\updefault}{\color[rgb]{0,0,0}$\nu_{|\beta|}$}%
}}}}
\put(6376,-4336){\makebox(0,0)[lb]{\smash{{\SetFigFont{34}{40.8}{\familydefault}{\mddefault}{\updefault}{\color[rgb]{0,0,0}$-k_T$}%
}}}}
\put(5401,-4336){\makebox(0,0)[lb]{\smash{{\SetFigFont{34}{40.8}{\familydefault}{\mddefault}{\updefault}{\color[rgb]{0,0,0}$k_T$}%
}}}}
\put(1726,-2236){\makebox(0,0)[lb]{\smash{{\SetFigFont{34}{40.8}{\familydefault}{\mddefault}{\updefault}{\color[rgb]{0,0,0}$\mu_{|\alpha|}$}%
}}}}
\put(8476,-3061){\makebox(0,0)[lb]{\smash{{\SetFigFont{34}{40.8}{\familydefault}{\mddefault}{\updefault}{\color[rgb]{0,0,0}$B,\beta$}%
}}}}
\put(3076,-3061){\makebox(0,0)[lb]{\smash{{\SetFigFont{34}{40.8}{\familydefault}{\mddefault}{\updefault}{\color[rgb]{0,0,0}$A,\alpha$}%
}}}}
\end{picture}%

%% file: bt/5point.pdf_t
\begin{picture}(0,0)%
\includegraphics{5point.pdf}%
\end{picture}%
\setlength{\unitlength}{3947sp}%
\begingroup\makeatletter\ifx\SetFigFont\undefined%
\gdef\SetFigFont#1#2#3#4#5{%
  \reset@font\fontsize{#1}{#2pt}%
  \fontfamily{#3}\fontseries{#4}\fontshape{#5}%
  \selectfont}%
\fi\endgroup%
\begin{picture}(16700,5219)(2475,-3833)
\put(13411,-451){\makebox(0,0)[lb]{\smash{{\SetFigFont{34}{40.8}{\familydefault}{\mddefault}{\updefault}{\color[rgb]{0,0,0}$\mathbf{P}^{\perp}_{k_4+k_5}$}%
}}}}
\put(15301,-511){\makebox(0,0)[lb]{\smash{{\SetFigFont{34}{40.8}{\familydefault}{\mddefault}{\updefault}{\color[rgb]{0,0,0}$\mathbf{P}_\beta$}%
}}}}
\put(17275,-526){\makebox(0,0)[lb]{\smash{{\SetFigFont{34}{40.8}{\familydefault}{\mddefault}{\updefault}{\color[rgb]{0,0,0}$k_4-k_5$}%
}}}}
\put(6886,-451){\makebox(0,0)[lb]{\smash{{\SetFigFont{34}{40.8}{\familydefault}{\mddefault}{\updefault}{\color[rgb]{0,0,0}$\mathbf{P}^{\perp}_{k_1+k_2}$}%
}}}}
\put(5701,-511){\makebox(0,0)[lb]{\smash{{\SetFigFont{34}{40.8}{\familydefault}{\mddefault}{\updefault}{\color[rgb]{0,0,0}$\mathbf{P}_\alpha$}%
}}}}
\put(3025,-526){\makebox(0,0)[lb]{\smash{{\SetFigFont{34}{40.8}{\familydefault}{\mddefault}{\updefault}{\color[rgb]{0,0,0}$k_1-k_2$}%
}}}}
\put(10876,164){\makebox(0,0)[lb]{\smash{{\SetFigFont{34}{40.8}{\familydefault}{\mddefault}{\updefault}{\color[rgb]{0,0,0}$q$}%
}}}}
\put(10051,-3436){\makebox(0,0)[lb]{\smash{{\SetFigFont{34}{40.8}{\familydefault}{\mddefault}{\updefault}{\color[rgb]{0,0,0}$k_3$}%
}}}}
\put(11026,-3436){\makebox(0,0)[lb]{\smash{{\SetFigFont{34}{40.8}{\familydefault}{\mddefault}{\updefault}{\color[rgb]{0,0,0}$-k_3$}%
}}}}
\put(6301,-2161){\makebox(0,0)[lb]{\smash{{\SetFigFont{34}{40.8}{\familydefault}{\mddefault}{\updefault}{\color[rgb]{0,0,0}$A,\alpha$}%
}}}}
\put(14476,-2161){\makebox(0,0)[lb]{\smash{{\SetFigFont{34}{40.8}{\familydefault}{\mddefault}{\updefault}{\color[rgb]{0,0,0}$B,\beta$}%
}}}}
\end{picture}%

%% file: bt/ex22_perp1.pdf_t
\begin{picture}(0,0)%
\includegraphics{ex22_perp1.pdf}%
\end{picture}%
\setlength{\unitlength}{3947sp}%
\begingroup\makeatletter\ifx\SetFigFont\undefined%
\gdef\SetFigFont#1#2#3#4#5{%
  \reset@font\fontsize{#1}{#2pt}%
  \fontfamily{#3}\fontseries{#4}\fontshape{#5}%
  \selectfont}%
\fi\endgroup%
\begin{picture}(4330,1512)(1886,-2858)
\put(2551,-2611){\makebox(0,0)[lb]{\smash{{\SetFigFont{34}{40.8}{\familydefault}{\mddefault}{\updefault}{\color[rgb]{0,0,0}$k_1+k_2$}%
}}}}
\put(4276,-2611){\makebox(0,0)[lb]{\smash{{\SetFigFont{34}{40.8}{\familydefault}{\mddefault}{\updefault}{\color[rgb]{0,0,0}$k_1+k_2$}%
}}}}
\end{picture}%

%% file: bt/ex22_perp2.pdf_t
\begin{picture}(0,0)%
\includegraphics{ex22_perp2.pdf}%
\end{picture}%
\setlength{\unitlength}{3947sp}%
\begingroup\makeatletter\ifx\SetFigFont\undefined%
\gdef\SetFigFont#1#2#3#4#5{%
  \reset@font\fontsize{#1}{#2pt}%
  \fontfamily{#3}\fontseries{#4}\fontshape{#5}%
  \selectfont}%
\fi\endgroup%
\begin{picture}(4330,1512)(1886,-3576)
\put(2551,-2611){\makebox(0,0)[lb]{\smash{{\SetFigFont{34}{40.8}{\familydefault}{\mddefault}{\updefault}{\color[rgb]{0,0,0}$k_1+k_2$}%
}}}}
\put(4276,-2611){\makebox(0,0)[lb]{\smash{{\SetFigFont{34}{40.8}{\familydefault}{\mddefault}{\updefault}{\color[rgb]{0,0,0}$k_1+k_2$}%
}}}}
\end{picture}%

%% file: bt/ex22_perp3.pdf_t
\begin{picture}(0,0)%
\includegraphics{ex22_perp3.pdf}%
\end{picture}%
\setlength{\unitlength}{3947sp}%
\begingroup\makeatletter\ifx\SetFigFont\undefined%
\gdef\SetFigFont#1#2#3#4#5{%
  \reset@font\fontsize{#1}{#2pt}%
  \fontfamily{#3}\fontseries{#4}\fontshape{#5}%
  \selectfont}%
\fi\endgroup%
\begin{picture}(4330,1844)(1886,-3908)
\put(2551,-2611){\makebox(0,0)[lb]{\smash{{\SetFigFont{34}{40.8}{\familydefault}{\mddefault}{\updefault}{\color[rgb]{0,0,0}$k_1+k_2$}%
}}}}
\put(4276,-2611){\makebox(0,0)[lb]{\smash{{\SetFigFont{34}{40.8}{\familydefault}{\mddefault}{\updefault}{\color[rgb]{0,0,0}$k_1+k_2$}%
}}}}
\put(2551,-3661){\makebox(0,0)[lb]{\smash{{\SetFigFont{34}{40.8}{\familydefault}{\mddefault}{\updefault}{\color[rgb]{0,0,0}$k_1+k_2$}%
}}}}
\put(4276,-3661){\makebox(0,0)[lb]{\smash{{\SetFigFont{34}{40.8}{\familydefault}{\mddefault}{\updefault}{\color[rgb]{0,0,0}$k_1+k_2$}%
}}}}
\end{picture}%

%% file: bt/ex22_A3q0.pdf_t
\begin{picture}(0,0)%
\includegraphics{ex22_A3q0.pdf}%
\end{picture}%
\setlength{\unitlength}{3947sp}%
\begingroup\makeatletter\ifx\SetFigFont\undefined%
\gdef\SetFigFont#1#2#3#4#5{%
  \reset@font\fontsize{#1}{#2pt}%
  \fontfamily{#3}\fontseries{#4}\fontshape{#5}%
  \selectfont}%
\fi\endgroup%
\begin{picture}(3730,3721)(5336,-3758)
\put(6451,-3361){\makebox(0,0)[lb]{\smash{{\SetFigFont{34}{40.8}{\familydefault}{\mddefault}{\updefault}{\color[rgb]{0,0,0}$k_3$}%
}}}}
\put(7426,-3361){\makebox(0,0)[lb]{\smash{{\SetFigFont{34}{40.8}{\familydefault}{\mddefault}{\updefault}{\color[rgb]{0,0,0}$-k_3$}%
}}}}
\end{picture}%

%% file: bt/ex22_A3q1.pdf_t
\begin{picture}(0,0)%
\includegraphics{ex22_A3q1.pdf}%
\end{picture}%
\setlength{\unitlength}{3947sp}%
\begingroup\makeatletter\ifx\SetFigFont\undefined%
\gdef\SetFigFont#1#2#3#4#5{%
  \reset@font\fontsize{#1}{#2pt}%
  \fontfamily{#3}\fontseries{#4}\fontshape{#5}%
  \selectfont}%
\fi\endgroup%
\begin{picture}(3730,3721)(5336,-3758)
\put(6451,-3361){\makebox(0,0)[lb]{\smash{{\SetFigFont{34}{40.8}{\familydefault}{\mddefault}{\updefault}{\color[rgb]{0,0,0}$k_3$}%
}}}}
\put(7426,-3361){\makebox(0,0)[lb]{\smash{{\SetFigFont{34}{40.8}{\familydefault}{\mddefault}{\updefault}{\color[rgb]{0,0,0}$-k_3$}%
}}}}
\end{picture}%

%% file: bt/ex22_A3q2.pdf_t
\begin{picture}(0,0)%
\includegraphics{ex22_A3q2.pdf}%
\end{picture}%
\setlength{\unitlength}{3947sp}%
\begingroup\makeatletter\ifx\SetFigFont\undefined%
\gdef\SetFigFont#1#2#3#4#5{%
  \reset@font\fontsize{#1}{#2pt}%
  \fontfamily{#3}\fontseries{#4}\fontshape{#5}%
  \selectfont}%
\fi\endgroup%
\begin{picture}(3730,2448)(5336,-2485)
\end{picture}%

%% file: bt/ex22_perp4.pdf_t
\begin{picture}(0,0)%
\includegraphics{ex22_perp4.pdf}%
\end{picture}%
\setlength{\unitlength}{3947sp}%
\begingroup\makeatletter\ifx\SetFigFont\undefined%
\gdef\SetFigFont#1#2#3#4#5{%
  \reset@font\fontsize{#1}{#2pt}%
  \fontfamily{#3}\fontseries{#4}\fontshape{#5}%
  \selectfont}%
\fi\endgroup%
\begin{picture}(4330,1512)(1886,-2858)
\put(2551,-2611){\makebox(0,0)[lb]{\smash{{\SetFigFont{34}{40.8}{\familydefault}{\mddefault}{\updefault}{\color[rgb]{0,0,0}$k_4+k_5$}%
}}}}
\put(4276,-2611){\makebox(0,0)[lb]{\smash{{\SetFigFont{34}{40.8}{\familydefault}{\mddefault}{\updefault}{\color[rgb]{0,0,0}$k_4+k_5$}%
}}}}
\end{picture}%

%% file: bt/ex22_perp5.pdf_t
\begin{picture}(0,0)%
\includegraphics{ex22_perp5.pdf}%
\end{picture}%
\setlength{\unitlength}{3947sp}%
\begingroup\makeatletter\ifx\SetFigFont\undefined%
\gdef\SetFigFont#1#2#3#4#5{%
  \reset@font\fontsize{#1}{#2pt}%
  \fontfamily{#3}\fontseries{#4}\fontshape{#5}%
  \selectfont}%
\fi\endgroup%
\begin{picture}(4330,1512)(1886,-3576)
\put(2551,-2611){\makebox(0,0)[lb]{\smash{{\SetFigFont{34}{40.8}{\familydefault}{\mddefault}{\updefault}{\color[rgb]{0,0,0}$k_4+k_5$}%
}}}}
\put(4276,-2611){\makebox(0,0)[lb]{\smash{{\SetFigFont{34}{40.8}{\familydefault}{\mddefault}{\updefault}{\color[rgb]{0,0,0}$k_4+k_5$}%
}}}}
\end{picture}%

%% file: bt/ex22_perp6.pdf_t
\begin{picture}(0,0)%
\includegraphics{ex22_perp6.pdf}%
\end{picture}%
\setlength{\unitlength}{3947sp}%
\begingroup\makeatletter\ifx\SetFigFont\undefined%
\gdef\SetFigFont#1#2#3#4#5{%
  \reset@font\fontsize{#1}{#2pt}%
  \fontfamily{#3}\fontseries{#4}\fontshape{#5}%
  \selectfont}%
\fi\endgroup%
\begin{picture}(4330,1844)(1886,-3908)
\put(2551,-3661){\makebox(0,0)[lb]{\smash{{\SetFigFont{34}{40.8}{\familydefault}{\mddefault}{\updefault}{\color[rgb]{0,0,0}$k_4+k_5$}%
}}}}
\put(4276,-2611){\makebox(0,0)[lb]{\smash{{\SetFigFont{34}{40.8}{\familydefault}{\mddefault}{\updefault}{\color[rgb]{0,0,0}$k_4+k_5$}%
}}}}
\put(4276,-3661){\makebox(0,0)[lb]{\smash{{\SetFigFont{34}{40.8}{\familydefault}{\mddefault}{\updefault}{\color[rgb]{0,0,0}$k_4+k_5$}%
}}}}
\put(2551,-2611){\makebox(0,0)[lb]{\smash{{\SetFigFont{34}{40.8}{\familydefault}{\mddefault}{\updefault}{\color[rgb]{0,0,0}$k_4+k_5$}%
}}}}
\end{picture}%

%% file: bt/ex22_A45.pdf_t
\begin{picture}(0,0)%
\includegraphics{ex22_A45.pdf}%
\end{picture}%
\setlength{\unitlength}{3947sp}%
\begingroup\makeatletter\ifx\SetFigFont\undefined%
\gdef\SetFigFont#1#2#3#4#5{%
  \reset@font\fontsize{#1}{#2pt}%
  \fontfamily{#3}\fontseries{#4}\fontshape{#5}%
  \selectfont}%
\fi\endgroup%
\begin{picture}(2788,1844)(5336,-2183)
\put(6526,-1411){\makebox(0,0)[lb]{\smash{{\SetFigFont{34}{40.8}{\familydefault}{\mddefault}{\updefault}{\color[rgb]{0,0,0}$k_4-k_5$}%
}}}}
\end{picture}%

%% file: bt/6point.pdf_t
\begin{picture}(0,0)%
\includegraphics{6point.pdf}%
\end{picture}%
\setlength{\unitlength}{3947sp}%
\begingroup\makeatletter\ifx\SetFigFont\undefined%
\gdef\SetFigFont#1#2#3#4#5{%
  \reset@font\fontsize{#1}{#2pt}%
  \fontfamily{#3}\fontseries{#4}\fontshape{#5}%
  \selectfont}%
\fi\endgroup%
\begin{picture}(14221,10996)(2528,-3608)
\put(9301,-361){\makebox(0,0)[lb]{\smash{{\SetFigFont{34}{40.8}{\familydefault}{\mddefault}{\updefault}{\color[rgb]{0,0,0}$\mathbf{P}_\beta$}%
}}}}
\put(10546,-226){\makebox(0,0)[lb]{\smash{{\SetFigFont{34}{40.8}{\familydefault}{\mddefault}{\updefault}{\color[rgb]{0,0,0}$\mathbf{P}^{\perp}_{k_1+k_2+k_3}$}%
}}}}
\put(6721,-226){\makebox(0,0)[lb]{\smash{{\SetFigFont{34}{40.8}{\familydefault}{\mddefault}{\updefault}{\color[rgb]{0,0,0}$\mathbf{P}^{\perp}_{k_1+k_2+k_3}$}%
}}}}
\put(15151,-3211){\makebox(0,0)[lb]{\smash{{\SetFigFont{34}{40.8}{\familydefault}{\mddefault}{\updefault}{\color[rgb]{0,0,0}$-k_4$}%
}}}}
\put(14176,-3211){\makebox(0,0)[lb]{\smash{{\SetFigFont{34}{40.8}{\familydefault}{\mddefault}{\updefault}{\color[rgb]{0,0,0}$k_4$}%
}}}}
\put(14626,3989){\makebox(0,0)[lb]{\smash{{\SetFigFont{34}{40.8}{\familydefault}{\mddefault}{\updefault}{\color[rgb]{0,0,0}$\mathbf{P}_\gamma$}%
}}}}
\put(3526,-3211){\makebox(0,0)[lb]{\smash{{\SetFigFont{34}{40.8}{\familydefault}{\mddefault}{\updefault}{\color[rgb]{0,0,0}$k_3$}%
}}}}
\put(4501,-3211){\makebox(0,0)[lb]{\smash{{\SetFigFont{34}{40.8}{\familydefault}{\mddefault}{\updefault}{\color[rgb]{0,0,0}$-k_3$}%
}}}}
\put(5026,389){\makebox(0,0)[lb]{\smash{{\SetFigFont{34}{40.8}{\familydefault}{\mddefault}{\updefault}{\color[rgb]{0,0,0}$q$}%
}}}}
\put(14251,389){\makebox(0,0)[rb]{\smash{{\SetFigFont{34}{40.8}{\familydefault}{\mddefault}{\updefault}{\color[rgb]{0,0,0}$r$}%
}}}}
\put(3976,3989){\makebox(0,0)[lb]{\smash{{\SetFigFont{34}{40.8}{\familydefault}{\mddefault}{\updefault}{\color[rgb]{0,0,0}$\mathbf{P}_\alpha$}%
}}}}
\put(12751,3539){\makebox(0,0)[lb]{\smash{{\SetFigFont{34}{40.8}{\familydefault}{\mddefault}{\updefault}{\color[rgb]{0,0,0}$C,\gamma$}%
}}}}
\put(9226,-2236){\makebox(0,0)[lb]{\smash{{\SetFigFont{34}{40.8}{\familydefault}{\mddefault}{\updefault}{\color[rgb]{0,0,0}$B,\beta$}%
}}}}
\put(5701,3539){\makebox(0,0)[lb]{\smash{{\SetFigFont{34}{40.8}{\familydefault}{\mddefault}{\updefault}{\color[rgb]{0,0,0}$A,\alpha$}%
}}}}
\put(3601,6014){\makebox(0,0)[lb]{\smash{{\SetFigFont{34}{40.8}{\familydefault}{\mddefault}{\updefault}{\color[rgb]{0,0,0}$k_1-k_2$}%
}}}}
\put(15601,6014){\makebox(0,0)[rb]{\smash{{\SetFigFont{34}{40.8}{\familydefault}{\mddefault}{\updefault}{\color[rgb]{0,0,0}$k_5-k_6$}%
}}}}
\put(14401,2639){\makebox(0,0)[lb]{\smash{{\SetFigFont{34}{40.8}{\familydefault}{\mddefault}{\updefault}{\color[rgb]{0,0,0}$\mathbf{P}^{\perp}_{k_5+k_6}$}%
}}}}
\put(3751,2639){\makebox(0,0)[lb]{\smash{{\SetFigFont{34}{40.8}{\familydefault}{\mddefault}{\updefault}{\color[rgb]{0,0,0}$\mathbf{P}^{\perp}_{k_1+k_2}$}%
}}}}
\end{picture}%